\documentclass[11pt]{article}
\usepackage{graphicx,amsmath,amsthm,amssymb}
\usepackage{boxedminipage}
\usepackage{natbib}
\usepackage{amsfonts}
\usepackage{float}
\usepackage[usenames,dvipsnames,table]{xcolor}
\usepackage{colortbl}
\newtheorem{theorem}{Theorem}
\newtheorem{lemma}{Lemma}
\newtheorem{coro}{Corollary}

\newcommand{\be}{\begin{eqnarray}}
\newcommand{\ee}{\end{eqnarray}}
\newcommand{\ben}{\begin{eqnarray*}}
\newcommand{\een}{\end{eqnarray*}}
\newcommand{\bes}{\begin{eqnarray*}}
\newcommand{\ees}{\end{eqnarray*}}
\newcommand{\utwi}[1]{\mbox{\boldmath $ #1$}}
\newcommand{\ba}{{\utwi{a}}}

\newcommand{\bff}{{\utwi{f}}}
\newcommand{\bg}{{\utwi{g}}}

\newcommand{\bi}{{\utwi{i}}}
\newcommand{\bj}{{\utwi{j}}}

\newcommand{\bu}{{\utwi{u}}}
\newcommand{\bv}{{\utwi{v}}}

\newcommand{\bx}{{\utwi{x}}}

\newcommand{\bA}{{\utwi{A}}}

\newcommand{\bE}{{\utwi{E}}}
\newcommand{\bF}{{\utwi{F}}}
\newcommand{\bG}{{\utwi{G}}}

\newcommand{\bI}{{\utwi{I}}}

\newcommand{\bM}{{\utwi{M}}}

\newcommand{\bP}{{\utwi{P}}}

\newcommand{\bU}{{\utwi{U}}}
\newcommand{\bV}{{\utwi{V}}}
\newcommand{\bW}{{\utwi{W}}}
\newcommand{\bX}{{\utwi{X}}}

\newcommand{\bZ}{{\utwi{Z}}}

\newcommand{\hbU}{\hat{\utwi{U}}}

\newcommand{\bLambda}{{\utwi{\mathnormal\Lambda}}}

\newcommand{\bSig}{{\utwi{\mathnormal\Sigma}}}

\newcommand{\bvarepsilon}{{\utwi{\mathnormal\varepsilon}}}
\newcommand{\bzeta}{{\utwi{\mathnormal\zeta}}}

\newcommand{\bTheta}{{\utwi{\mathnormal\Theta}}}
\newcommand{\bGamma}{{\utwi{\mathnormal\Gamma}}}

\newcommand{\bPhi}{{\utwi{\mathnormal\Phi}}}
\newcommand{\bxi}{{\utwi{\mathnormal\xi}}}

\newcommand{\lam}{{\lambda}}

\newcommand{\bDelta}{{\utwi{\mathnormal\Delta}}}

\newcommand{\cM}{{\cal M}}

\newcommand{\cE}{{\cal E}}
\newcommand{\cF}{{\cal F}}
\newcommand{\cG}{{\cal G}}
\newcommand{\cZ}{{\cal Z}}
\newcommand{\cR}{{\cal R}}
\newcommand{\cS}{{\cal S}}
\newcommand{\cX}{{\cal X}}
\newcommand{\cY}{{\cal Y}}
\newcommand{\cW}{{\cal W}}
\newcommand{\R}{\mathbb{R}}

\renewcommand{\vec}{\mathrm{vec}}

\definecolor{dollarbill}{rgb}{0.52, 0.73, 0.4}

\newcommand\chzM[1]{{#1}}

\def\bbDelta{{\overline \bDelta}}

\newcommand{\bel}{\begin{eqnarray}\label}
\newcommand{\eel}{\end{eqnarray}}
\def\E{\mathbb{E}}

\def\mat{\hbox{\rm mat}}

\def\Kds{{d_1\times\cdots\times d_K}}
\def\timesKAs{{\times_{k=1}^K \bA_k}}
\def\timesKAsKAs{{\times_{k=1}^{2K} \bA_k}}
\def\timesKPs{{\times_{k=1}^K \bP_k}}
\def\timesKPsKPs{{\times_{k=1}^{2K} \bP_k}}
\def\timesKhPs{{\times_{k=1}^K{\widehat{\bP_k}}}}
\def\timesKhPsKhPs{{\times_{k=1}^{2K}{\widehat{\bP_k}}}}

\def\timesKPAsKPAs{{\times_{k=1}^{2K}{\bP_k\bA_k}}}

\def\cV{{\cal V}}

\def\hbPk{\widehat{\bP_k}}

\def\hbu{\widehat{\bu}}
\def\mat{\hbox{\rm mat}}
\def\barE{\overline\E}

\begin{document}
\centerline{{\Large\textbf{Factor Models for High-Dimensional Tensor
      Time Series}}}

\bigskip \centerline{Rong Chen, Dan Yang and \chzM{Cun-Hui} Zhang\footnote{
Rong Chen is Professor at Department of
    Statistics, Rutgers University, Piscataway, NJ 08854. E-mail:
    rongchen@stat.rutgers.edu.
 Dan Yang
    is Associate Professor, Faculty of Business and Economics, The
    University of Hong Kong, Hong Kong. E-mail:
    dyanghku@hku.hk.
    \chzM{Cun-Hui} Zhang is Professor at
    Department of Statistics, Rutgers University, Piscataway, NJ
    08854. E-mail: czhang@stat.rutgers.edu.
Rong Chen is the
    corresponding author. Chen's research is supported
in part by National Science Foundation
grants DMS-1503409, DMS-1737857 and IIS-1741390.
Yang's research is
supported in part under NSF grant IIS-1741390.
Zhang's research is supported in part by NSF \chzM{grants DMS-1721495, IIS-1741390 and CCF-1934924.}
}} \medskip \centerline{Rutgers University and The University of Hong Kong}

\bigskip

\smallskip

\centerline{\bf{Abstract}}

\smallskip

\noindent
Large tensor (multi-dimensional array) data are now
routinely collected in a wide range of applications,
due to modern data collection capabilities. Often such
observations are taken over time, forming tensor time series.
In this paper
we present a factor model approach for analyzing high-dimensional dynamic
tensor time series and multi-category dynamic transport networks.
Two estimation procedures
along
with their theoretical properties and simulation results are presented.
Two applications are used to illustrate the model and its interpretations.

\vspace{3em}

\noindent%

\pagenumbering{arabic}
\setcounter{page}{1}

\section{Introduction}

\begin{figure}
\centering
\includegraphics[width=\linewidth,height=\textheight,keepaspectratio=true]{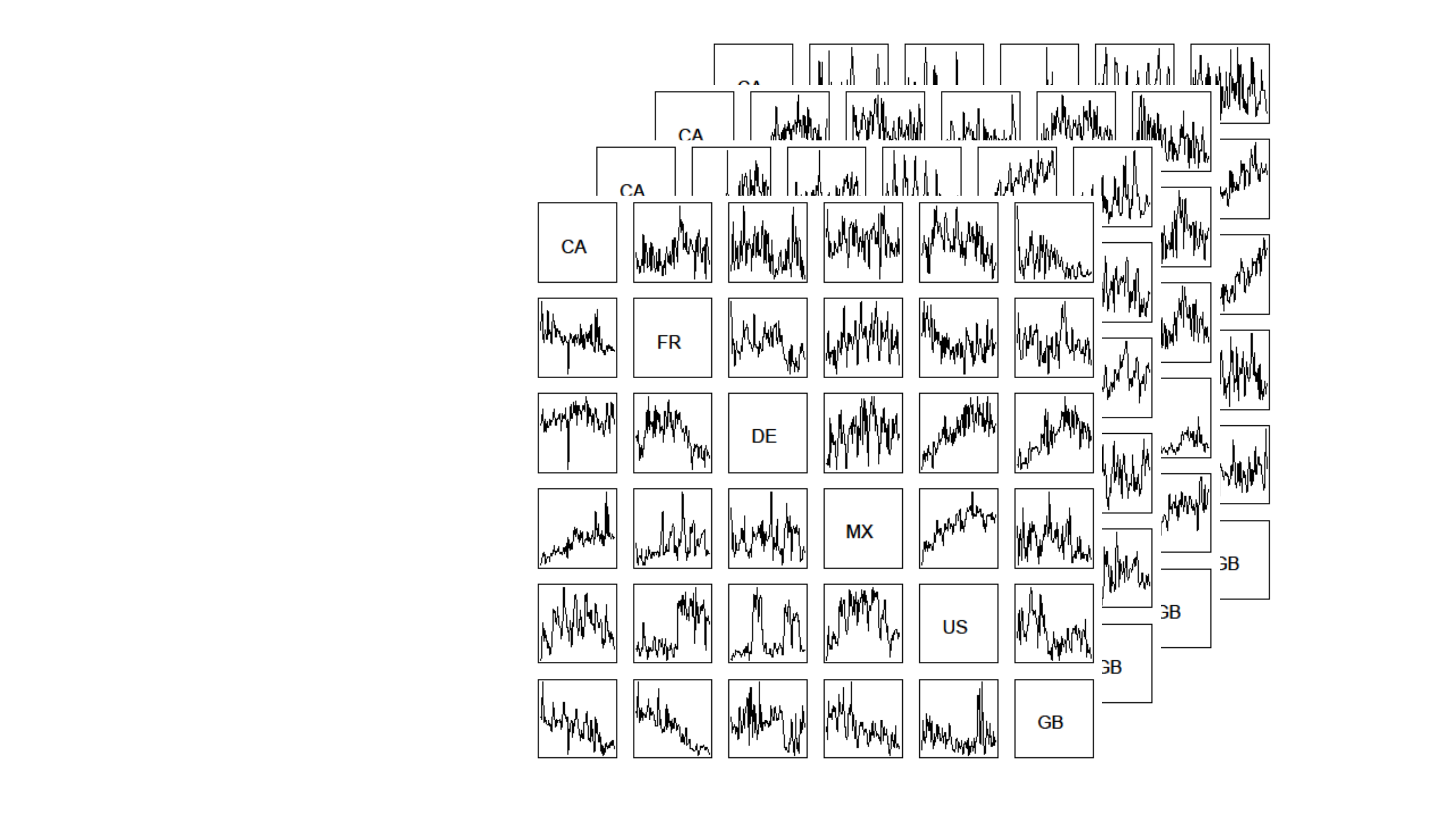}
\caption{Monthly import-export volume time series of
four categories of products (Chemical, Food, Machinery and Electronic, and
Footwear and Headwear) among six countries (US, Canada, Mexico, Germany, UK
and France) from January 2001 to December 2017.}
\label{fig1}
\end{figure}

Modern data collection capability has led to massive quantity of
time series. High dimensional time series observed in
tensor form are becoming more and
more commonly seen in various fields such as economics, finance,
engineering, environmental sciences, medical research and others.
For example, Figure \ref{fig1}
shows the monthly import-export volume time series of
four categories of products (Chemical, Food, Machinery and Electronic, and
Footwear and Headwear) among six countries (US, Canada, Mexico, Germany, UK
and France) from January 2001 to December 2016. At each time point,
the observations can be arranged into a three-dimensional tensor, with
the diagonal elements for each product category unavailable.
This is
part of a larger data set with 15 product categories and 22 countries which
we will study in detail in Section~\ref{example1}.
Univariate time series deals with one item in the
tensor (e.g. Food export series of US to Canada).
Panel time series analysis focuses on the co-movement of
one row (fiber) in the tensor (e.g. Food export of US to all other
countries). Vector
time series analysis also focuses on the co-movement of
one fiber in the tensor (e.g. Export of US to Canada in all product
categories). \cite{Wang&2019, Chen&Chen2019} and \cite{Chen&2019} studied
matrix time series. Their analysis deals with a matrix slice of the tensor
(e.g. the import-export activities between all the countries in
one product category). In this paper we develop a factor model for the
analysis of the entire tensor time series simultaneously.

The import-export network  belongs to the general class of dynamic
transport (traffic) network. The focus of such a network is the volume
of traffic on the links between the nodes on the network.
The availability of complex and diverse
network data, recorded over periods of time and in very large scales,
brings new opportunities with
challenges \citep{aggarwal2014evolutionary}.
For example, weekly activities in different forms (e.g. text messages, email,
phone conversations, and
personal interactions) and on different topics
(politics, food, travel, photo,  emotions, etc)
among friends on a social network form a transport network similar to the
import-export network, but as a four-dimensional tensor time series.
The number of passengers flying between
a group of cities with a group of airlines in different classes (economy or
business)
on different days of the week
can  be represented as a five-dimensional tensor time series.
In Section~\ref{example2} we will present a second example on
taxi traffic patterns in New York city. With
the city being divided into 69 zones, we study the volume of passenger
pickups and drop-offs by taxis among the zones, at different hours during the
day as a daily time series of a $69\times 69\times 24$ tensor.

Note that most developed
statistical inference methods in network analysis
are often confined to static network data such as social network
\citep{Goldenberg&2010,snijders2006statistical,hanneke2010discrete,kolaczyk2014statistical,ji&jin2016,Zhao&2012,Phan&Airoldi2015}.
Of course most
networks are dynamic in nature. One important challenge is to
develop stochastic
models/processes that capture the dynamic dependence and dynamic changes of
 a network.


Besides dynamic traffic networks, tensor time series
are also observed in many other applications.
For example, in economics, many economic indicators such as GDP, unemployment
rate and inflation index are reported quarterly by many countries, forming a
matrix-valued time series.
Functional MRI produces a sequence of 3-dimensional brain images (forming
3-dimensional tensors) that changes with different stimulants.
Temperature and salinity levels observed
at a regular grid of locations and a set of different
depth in the ocean form 3-dimensional tensors and are observed over time.

Such tensor systems are often very large. Thirty economic indicators from 30
countries yield total
900 individual time series. Import-export volume of 15 product
categories among 20 countries makes up almost 6,000 individual
time series. FMRI images
often consist of hundreds of thousands of voxels observed over time.


The aim of this paper is to develop a factor model
to systematically study the dynamics of tensor systems by
jointly modeling the entire tensor
simultaneously, while preserving the tensor structure and the time series
structure.
This is different from the more
conventional time series analysis which deals with scalar or vector
observations \citep{Box&Jenkins76, Brockwell&Davis91,
Shumway&stoffer02,Tsay05,Tong90,Fan&Yao03,Hard97,Tsay&Chen2018}
and multivariate time series analysis
\citep{Hannan70,Lutkepohl93}, panel time series analysis
\citep{Baltagi05,Hsiao03,Geweke-1977,Sargent-Sims-1977}
and spatial-temporal modelling
\citep{Bennett79,Cressie93,Stein99,Stroud&01,Woolrich&04,Handcock&Wallis94,Mardia&98,Wikle&Cressie99,Wikle&98,Irwin00}.

We mainly focus
on the cases when the tensor dimension is large.
When dealing with
many time series simultaneously, dimension reduction is one of main
approaches
to extract common information from the data without being overwhelmed
by the idiosyncratic variations. One of the most powerful tools for
dimension
reduction in time series analysis is the dynamic factor model in which
'common' information is summarized into a small number of factors
and the co-movement of the time series is assumed to be driven by these
factors and their inherited dynamic structures
\citep{Bai-2003,Forni-Hallin-Lippi-Reichline-2000,
Stock-Watson-2012, Bai-Ng-2008,connor&2012,
Chamberlain-1983, pena1987,Pan-Yao-2008}.  We will follow this approach
in our development.

The tensor factor model in this paper is similar to the
matrix factor model studied in \cite{Wang&2019}.
Specifically, we use a Tucker decomposition type of formation
to relate the high-dimensional tensor observations to a low-dimensional
latent tensor factor that is assumed to vary over time.
Two estimation approaches, named TIPUP and TOPUP, are studied.
Asymptotic properties of the estimators are
investigated, which provides a comparison between the two estimation
methods.
The estimation procedure used in \cite{Wang&2019} in the matrix setting
is essentially the TOPUP procedure.
We show that the convergence rate they obtained
for the TOPUP can be improved. On the other hand,
the TIPUP has a faster rate than the TOPUP, under
a mildly more restrictive condition on the level of signal cancellation.
The developed  theoretical properties also cover the cases where
  the dimensions of the tensor factor increase with the dimension of
  the observed tensor time series.

The paper is organized as follows. Section 2 contains some
preliminary information on the approach of factor models that we
will adopt and the basic notations of tensor analysis. Section 3 introduces
a general framework of factor models for large tensor
time series, which is assumed to be the sum of a signal part and a
noise part. The signal part has a multi-linear factor form, consisting
of a low-dimensional tensor that varies over time, and a set of
fixed loading matrices in a Tucker decomposition form.
Section 4 discusses two general estimation procedures. Their
theoretical properties are shown in Section 5.
In section 6 we present some simulation studies to
demonstrate the performance of the estimation procedures. Two
applications are presented in Section 7 to illustrate the model and its
interpretations.

\section{Preliminary: dynamic factor models and foundation of tensor}\label{sec2}

In this section we briefly review the linear factor model approach to
panel time series data
and tensor data analysis.
Both serve as a foundation of our approach to tensor time series.

Let $\{(x_{i,t})_{d\times T}\}$ be a set of panel time series. Dynamic factor
model assumes
\begin{equation}\label{vector_factor}
\bx_t=\bA\bff_t+\bvarepsilon_t,
\mbox{\ or equivelently \ }
x_{it} = a_{i1}f_{1t}+\ldots+a_{ir}f_{rt}+\varepsilon_{it}
 \mbox{\ for } i=1, \ldots, d,
\end{equation}
where $\bff_{t}=(f_{1t},\ldots, f_{rt})^\top$ is a set of unobserved latent
factor time series with dimension
$r\ll d$; The row vector $\ba_{i}=(a_{i1},\ldots, a_{ir})$, treated as unknown and deterministic, is
called factor loading of the $i$-th series.
The collection of all $\ba_{i}$ is called the loading matrix $\bA$.
The idiosyncratic noise $\bvarepsilon_t$ is assumed to
be uncorrelated with the factors $\bff_t$ in all leads and lags.
Both $\bA$ and $\bff_t$ are unobserved hence some further model assumptions are needed.
Two different types of model assumptions are adopted
 in the literature.
One type of models assumes that a
common factor must have impact on `most' (defined asymptotically)
of the time series, but allows the idiosyncratic noise to have weak cross-correlations and
weak autocorrelations \citep{Geweke-1977,Sargent-Sims-1977,
Forni-Hallin-Lippi-Reichline-2000,
Stock-Watson-2012, Bai-Ng-2008, Stock-Watson-2006,Bai&Ng2002,Hallin&Liska2007,
Chamberlain-1983, Chamberlain-Rothschild-1983, connor&2012,
Connor&Linton2007, Fan&2016, FanWangZhong2019, Pena&Poncela2006, Bai&Li2012}.
Under such sets of assumptions,
principle component analysis (PCA) of the sample covariance matrix is typically used to
estimate the space spanned by the columns of the loading matrix, with various extensions.
Another type of models assumes that
the factors accommodate all dynamics, making the idiosyncratic noise
`white' with no autocorrelation but allowing substantial
contemporary cross-correlation among the error process
\citep{pena1987,Pan-Yao-2008,Lam&2011,Lam&Yao2012, Chang2018}. The estimation of the
loading space is done by an eigen analysis based on
the non-zero lag autocovariance matrices.
In this paper we adopt the second approach in our model development.

The key feature of the factor model is that
all co-movements of the data are driven by the common factor $\bff_t$
and the factor loading $\ba_i$ provides a link between the
underlying factors and the $i$-th series $x_{it}$.
This approach has three major benefits: (i)
It achieves great reduction in model complexity (i.e. the number of parameters)
as the autocovariance
matrices are now determined by the loading matrix
$\bA$ and the much smaller autocovariance matrix of the
factor process $\bff_t$;
(ii) The hidden dynamics (the co-movements) become
transparent, leading to clearer and more insightful understanding. This
is especially important when the co-movement of the time series
is complex and difficult to discover without proper modeling of the full panel;
(iii) The estimated factors can be used as input and
instrumental variables in models in downstream data analyses,
providing summarized and parsimonious information of the whole series.

In the following we briefly review
tensor data analysis without involving time series (equivalently at a fixed time point),
mainly for the purpose of
fixing the notation in our later discussion. For more detailed information,
see \cite{KoldaBader09}.

A tensor is a multidimensional array, a generalization of a matrix. The order
of a tensor is the number of dimensions, also known as the number of
modes. Fibers of a tensor are the higher order analogue of matrix rows and columns, which can be obtained by fixing all but one of the modes. For example, a matrix is a tensor of order $2$,
and a matrix column is a mode-1 fiber and a matrix row is a mode-2 fiber.

Consider an order-$K$ tensor $\cX\in\R^{d_1\times\cdots\times d_K}$.
Following \cite{KoldaBader09},
the $k$-mode product of
$\cX$ with a matrix $\bA\in\R^{\tilde{d_k}\times d_k}$ is an order-$K$ tensor of size $d_1\times\cdots\times d_{k-1}\times \tilde{d_k}\times d_{k+1}\times...\times d_K$ and will be denoted by $\cX\times_k \bA$. Elementwise, $(\cX\times_k \bA)_{i_1\cdots i_{k-1}j i_{k+1}\cdots i_K} = \sum_{i_k=1}^{d_k}
x_{i_1\cdots i_k\cdots i_K}a_{ji_k}$.
Similarly, the $k$-mode product of an order-K tensor
with a vector $\ba\in\R^{d_k}$ is an order-$(K-1)$ tensor of size $d_1\times...\times d_{k-1}\times d_{k+1}\times...\times d_K$ and denoted by $\cX\times_k \ba$. Elementwise,
$(\cX\times_k \ba)_{i_1\cdots i_{k-1} i_{k+1}\cdots i_K}
= \sum_{i_k=1}^{d_k} x_{i_1\cdots i_k\cdots i_K}a_{i_k}$. Let $d=d_1\dots d_K$ and $d_{-k}=d/d_k$.
The mode-k unfolding matrix
$\mat_k(\cX)$ is a $d_k\times d_{-k}$ matrix by assembling all $d_{-k}$ mode-k
fibers as columns of the matrix. One may also  stack a tensor into a vector. Specifically,
$\vec(\cX)$ is a vector in $\R^d$ formed by stacking
mode-$1$ fibers of $\cX$ in the order of modes $2,\ldots,K$.

The CP decomposition
\citep{CarrollChang1970,Harshman1970} and Tucker decomposition
 \citep{Tucker1963,Tucker1964,Tucker1966} are two major extensions of the matrix singular value
decomposition (SVD) to tensors of higher order.
Recall that the SVD of a matrix $\bX\in \R^{d_1\times d_2}$  of rank $r$
has two equivalent forms: $\bX= \sum_{l=1}^r \lambda_l\bu^{(1)}_l\bu_l^{(2)\top}$,
which decomposes a matrix into a sum of $r$ rank-one matrices,
and $\bX= \bU_1\bLambda_r\bU_2^\top$,
where $\bU_1$ and $\bU_2$ are orthonormal matrices of size $d_1\times r$ and
$d_2\times r$ spanning the column and row spaces of $\bX$
respectively, and $\bLambda_r$ is an $r\times r$ diagonal matrix with $r$
positive singular values on its diagonal.
In parallel, CP decomposes an order-$K$ tensor $\cX$ into a
sum of rank one tensors,
$\cX=\sum_{l=1}^r \lambda_l \bu_l^{(1)} \otimes \bu_l^{(2)}\otimes\cdots\otimes\bu_l^{(K)}\in \R^{d_1\times\cdots\times d_k}$,
where ``$\otimes$'' represents the tensor product.
The vectors ${\bu}_l^{(k)}\in\R^{d_k}, l=1,2,...,r$, are not necessarily orthogonal to each other, which differs from the matrix SVD.
The Tucker decomposition boils down to $K$ orthonormal matrices
$\bU_k\in\R^{d_k\times r_k}$
containing basis vectors spanning $k$-mode fibers of the tensor,
a potentially much smaller `core'
tensor $\cG\in\R^{r_1\times r_2\times\cdots\times r_K}$ and the relationship
\be
\label{eq:tucker}
\cX= \cG\times_1\bU_1\times_2\bU_2\times_3\cdots\times_K\bU_K
=\cG\times_{k=1}^K\bU_k.
\ee
Note that the core tensor $\cG$ is
similar to the $\bLambda_r$ in the middle of matrix SVD but
now it is not necessarily diagonal.

\section{A Tensor Factor Model}

In tensor times series, the observed tensors would depend on $t=1,\ldots,T$ and be denoted by $\cX_t\in\R^{d_1\times\cdots\times d_K}$ as a series of order-$K$ tensors.
By absorbing time, we may stack $\cX_t$ into an order-$(K+1)$ tensor
$\cY\in\R^{d_1\times\cdots\times d_K\times T}$, with time $t$ as the
$(K+1)$-th mode, referred to as the time-mode.
We assume the following decomposition
\be
\cY=\cS+\cR, \mbox{\ \ or equivalently \ \ } \cX_t = \cM_t + \cE_t,
\label{second}
\ee
where $\cS$ is the dynamic signal component and $\cR$ is a white noise part. In
the second expression (\ref{second}),
$\cM_t$ and $\cE_t$ are the corresponding signal
and noise components of $\cX_t$, respectively.
We assume that the noise $\cE_t$ are uncorrelated (white)
across time, following \cite{Lam&Yao2012}.

In this model, all dynamics are contained in the signal component $\cM_t$.
We assume that $\cM_t$ is in a lower-dimensional space and has certain multilinear decomposition.
We further assume that any component in this multilinear decomposition that involves the
time-mode is random and dynamic, and will be called a factor component
(depending on its order, it will be called a scalar factor $f_t$,
a vector factor $\bff_t$, a matrix factor $\bF_t$, or a tensor factor $\cF_t$),
which when concatenated along the time-mode
forms a higher order object,
such as $\bg,\bG,\cG$. Any
components of $\cM_t$ other than $\cF_t$ are assumed to be deterministic
and will be called the loading components.

Although it is tempting to directly model $\cS$ with standard tensor
decomposition approaches to find its lower dimensional structure,
the dynamics and dependency in the time direction (auto-dependency)
are important and should be treated differently.
Traditional tensor decomposition using tensor SVD/PCA
on $\cS$ ignores the special role of the time-mode and the covariance
structure in the time direction,
and treats the signal $\cS$ as deterministic
\citep{MontanariRichard2014,Anandkumar2014,Hopkins+2015,Sun+2016}.
Such a direct approach often leads to inferior inference results
as our preliminary results have demonstrated \citep{Wang&2019}.
In our approach, the component in the time direction is considered as
latent and random.
As a result, our model assumptions and interpretations, and
their corresponding estimation procedures and theoretical properties are
significantly different.

In the following we propose a specific model for tensor time series,
based on a decomposition similar to Tucker decomposition. Specifically, we
assume that
\begin{equation}
\cS= \cG\times_1\bA_1\times_2\ldots\times_K\bA_K
\mbox{ \ or equivalently \ }
\cM_t=\cF_t\times_1\bA_1\times_2\ldots\times_K\bA_K
\label{eq:tucker-model-2}
\end{equation}
where $\cF_t$ is itself a tensor times series of dimension
$r_1\times\ldots \times r_K$ with
small $r_k\ll d_k$ and $\bA_k$ are $d_k\times r_k$ loading matrices.
We assume without loss of generality in the sequel that $\bA_k$ is of rank $r_k$.

Model (\ref{eq:tucker-model-2}) resembles a Tucker-type decomposition similar to (\ref{eq:tucker})
where the core tensor $\cG\in\R^{r_1\times\ldots \times r_K\times T}$ is the factor
term and the loading matrices $\bA_k\in\R^{d_k\times r_k}$
are constant matrices, whose column spaces are identifiable.
The core tensor $\cF_t$ is usually much smaller than $\cX_t$
in dimension.
It drives all the comovements of
individual time series in $\cX_t$.
For matrix time series, model (\ref{eq:tucker-model-2})
becomes $\bM_t=\bF_t\times_1\bA_1\times_2\bA_2=\bA_1\bF_t\bA_2^\top$.
The matrix version of Model (\ref{eq:tucker-model-2}) was considered
in \cite{Wang&2019}, which also provided several model
interpretations. Most of their interpretations can be extended to the
  tensor factor model. In this paper we consider more general model settings and  more powerful
estimation procedures.

As $\cF_t \to \cX_t = \cF_t\timesKAs+ \cE_t$ is a linear mapping from $\R^{r_1\times\cdots\times r_K}$ to $\R^{d_1\times\cdots\times d_K}$. It can be written as a matrix acting on vectors as in
\[
\vec(\cX_t)=
{\rm Kronecker}\left(\bA_K,\ldots,\bA_1\right)
\vec(\cF_t)
+\vec(\cE_t),
\]
where ${\rm Kronecker}\left(\bA_K,\ldots,\bA_1\right)\in \R^{d\times r}$ is the Kronecker product,
$d = \prod_{k=1}^Kd_k$,
$r = \prod_{k=1}^Kr_k$,
and $\vec(\cdot)$ is the tensor stacking operator as described in
Section~\ref{sec2}. While $\otimes$ is often used to denote the Kronecker product, we shall avoid this usage as $\otimes$ is preserved to denote the tensor product in this paper. For example, in the case of $K=2$ with observation $\bX_t\in\R^{d_1\times d_2}$,  $\bX_{t-h}\otimes \bX_t$ is a $d_1\times d_2\times d_1\times d_2$ tensor of order four, not a matrix of dimension $d_1^2\times d_2^2$, as we would need to consider the model-2 unfolding of $\bX_{t-h}\otimes \bX_t$ as a $d_2\times(d_1^2d_2)$ matrix.
The Kronecker expression
exhibits the same form as in the factor model for panel time series
except that the loading matrix of size $d\times r$
in the vector
factor model is assumed to have a Kronecker product structure of $K$ matrices
of much smaller sizes $d_i\times r_i$ ($i=1,\ldots,K$).
Hence the tensor factor model reduces the number of parameters in the
loading matrices from $dr = d_1r_1\ldots d_Kr_K$
in the stacked vector version
to $d_1r_1+\ldots+d_Kr_K$, a very significant dimension reduction.
The dimension reduction comes from the assumption imposed on the
loading matrices.

It would be tempting to assume the orthonormality of $\bA_k\in\R^{d_k\times r_k}$
as in SVD and Tucker decomposition. However, in the high-dimensional setting, this may not be compatible in general with the assumption that $\cF_t$ is a ``regular" factor series with unit order of magnitude, which we may also want to impose; The magnitude of $\cX_t$ would have to be absorbed into either $\cF_t$ or $\{\bA_k\}$ in model \eqref{eq:tucker-model-2}. In addition, the orthonormality of $\bA_k$ would be incompatible with the expression of the strength of the factor in terms of the norm of $\bA_k$ as in the literature \citep{Bai-Ng-2008, Lam&Yao2012, Wang&2019}. Thus, we shall consider general $\bA_k$ to preserve flexibility.

Let $\bA_k = \bU_k\bLambda_k\bV_k^\top$ be the SVD of $\bA_k$. The tensor time series in \eqref{eq:tucker-model-2} can be then written as $\cX_t = \big(\cF_t\times_{k=1}^K(\bLambda_k\bV_k^\top)\big)\times_{k=1}^K \bU_k$. In the special case where the $r_1\times\cdots\times r_K$ dimensional series  $\big(\cF_t\times_{k=1}^K(\bLambda_k\bV_k^\top)\big)/\lam$ can be viewed as a properly normalized factor for some signal strength parameter $\lam > 0$, we may absorb $\bLambda_k$, $\bV_k^\top$ and $1/\lam$ into  $\cF_t$ and write \eqref{eq:tucker-model-2} as
\bel{lam-F-U}
\cX_t = \lam\big( \cF_t \times_1\bU_1\times_2\cdots\times_K\bU_k\big) + \cE_t
\eel
with orthonormal $\bU_k\in \R^{d_k\times r_k}$ and properly normalized $\cF_t$. This means $\bA_k=c_k\bU_k$ in \eqref{eq:tucker-model-2} with constants $c_k$ and $\lam = \prod_{k=1}^Kc_k$. For example, in the one-factor model where $r_1=\cdots=r_K=1$, $\cX_t = \big(\lam f_t\big)\bu_1\otimes \cdots\otimes \bu_K$ as in \eqref{example-1a} would be discussed in detail below Theorems \ref{th-1} and \ref{th-2}.

\vspace{0.2in}

\noindent{\bf Remark:\ }
In our theoretical development, we do not impose any specific structure for the dynamics of the relatively low-dimensional factor process $\cF_t\in \R^{r_1\times\cdots\times r_K}$, except conditions on the spectrum norm and singular values of certain matrices in the unfolding of the average of the cross-product $(T-h)^{-1}(\sum_{t=h+1}^T \cM_{t-h}\otimes \cM_t)$. As $\cM_t=\cF_t\times_{k=1}^K \bA_k$, these conditions on $\cM_t$ would hold when the condition numbers of $\bA_k^\top\bA_k$ are bounded, e.g. model \eqref{lam-F-U}, and parallel conditions on the spectrum norm and singular values in the unfolding of $(T-h)^{-1}\sum_{t=h+1}^T \cF_{t-h}\otimes \cF_t$ hold
through the consistency of the averages.
For fixed $r_1,\ldots,r_k$, such consistency for the low-dimensional $\cF_t$ has been extensively
studied in the literature with many options such as various mixing conditions.

\vspace{0.2in}

\noindent{\bf Remark:\ }
The above tensor factor model does
not assume any structure on the noises except that the noise process is white.
The estimation
procedures we use
do not require any additional
structure. But in many cases it benefits to allow specific
structures for the contemporary cross-correlation of the elements
of $\cE_t$. For example, 
one may assume $\cE_t=\cZ_t\times_1 \bSig_1^{1/2}
\times_2 \bSig_2^{1/2}\times_3 \ldots \times_K \bSig_K^{1/2}$ where all
elements in $\cZ_t$ are i.i.d. $N(0,1)$. Hence each of the
$\bSig_i$ can be viewed as the common covariance matrix of mode-$i$ fiber
in the tensor $\cE_t$. More efficient estimators may be
constructed to utilize such a structure but is out of the scope of this
paper.


\section{Estimation procedures}

Low-rank tensor approximation is a delicate task.
To begin with, the best rank-$r$ approximation to a tensor may not exist
\citep{Silva2008}
or NP hard to compute \citep{HillarLim2013}.
On the other hand, despite such inherent difficulties, many heuristic
techniques are
widely used and often enjoy great successes in practice.
\cite{MontanariRichard2014} and \cite{Hopkins+2015}, among others,
have considered a rank-one spiked tensor model $\cS+\cR$ as a vehicle to
investigate the requirement of signal-to-noise ratio for consistent estimation
under different constraints of computational resources,
where $\cS = \lam\bu_1\otimes\bu_2\otimes\bu_3$
for some deterministic unit vectors $\bu_k\in\R^{d_k}$ and
all entries of $\cR$ are iid standard normal.
As shown by \cite{MontanariRichard2014}, in the symmetric case
where $d_1=d_2=d_3=d$,
$\cS$ can be estimated consistently by the MLE when $\sqrt{d}/\lambda =o(1)$.
Similar to the case of spiked PCA
\citep{koltchinskii+2011, NegahbanWainwright2011},
it can be shown that the rate achieved by
the MLE is minimax among all estimators when $\cS$ is treated as
deterministic.
However, at the same time it is also unsatisfactory
as the MLE of $\cS$ is NP hard to compute even in this simplest rank one case.
Additional discussion of this and some other key differences between
matrix and tensor estimations can be found in recent studies of related tensor completion problems
\citep{barak2016noisy, yuan2016tensor, yuan2017incoherent, xia2017polynomial, zhang2019cross}.

A commonly used heuristic to overcome this computational difficulty is
tensor unfolding.
In the following we proposed two estimation methods that
are based on a marriage of tensor unfolding
and the use of
lagged cross-product, the tensor version of the autocovariance.
This is due to the dynamic and random
nature of the latent factor process, and the whiteness assumption on the
error process.

As in all factor models,
due to ambiguity, we will only
estimate the linear spaces spanned by the loading matrices with
an orthonormal representation of the loading spaces,
or equivalently, only estimate the orthogonal projection matrix to such spaces.

The lagged cross-product 
operator, which we denote by $\bSig_h$, can be viewed as
the $(2K)$-tensor
\bes
\bSig_h = \E\bigg[\sum_{t=h+1}^T \frac{\cX_{t-h}\otimes \cX_t}{T-h}\bigg]
= \E\bigg[\sum_{t=h+1}^T \frac{\cM_{t-h}\otimes \cM_t}{T-h}\bigg]
\in \R^{\Kds\times \Kds},
\ees
$h=1,\ldots,h_0$. We consider two estimation methods based on the
sample version of $\bSig_h$,
\bel{new-1}
{\overline \bSig}_h =
\sum_{t=h+1}^T \frac{\cX_{t-h}\otimes \cX_t}{T-h},\quad h=1,\ldots,h_0.
\eel
The orthogonal projection to the column space of $\bA_k$ is
\bel{P_k}
\bP_k = \bA_k\big(\bA_k^\top\bA_k\big)^{-1}\bA_k^\top.
\eel
It is the $k$-th principle space
of the tensor time series $\cM_t = \cF_t\timesKAs$ in \eqref{eq:tucker-model-2}.
As $\cM_t  = \cM_t \timesKPs$ for all $t$,
\bes
{\bSig}_h
= {\bSig}_h\timesKPsKPs
= \E\bigg[\sum_{t=h+1}^T \frac{\cF_{t-h}\otimes\cF_t}{T-h}\bigg]\timesKPAsKPAs.
\ees
with the notation $\bA_k = \bA_{k-K}$ and $\bP_k=\bP_{k-K}$ for $k>K$.
Once consistent estimates $\hbPk$ are obtained for $\bP_k$,
the estimation of other aspects of ${\bSig}_h$ can be carried out based on the low-rank
projection of \eqref{new-1},
\bes
{\overline \bSig}_h\timesKhPsKhPs
= \sum_{t=h+1}^T \frac{(\cX_{t-h}\timesKhPs)\otimes (\cX_t\timesKhPs)}{T-h},
\ees
as if the low-rank tensor time series
$\cX_t\timesKhPs$ is observed.
For the estimation of $\bP_k$,
we propose two methods, and both methods can be written in terms of the mode-$k$ matrix unfolding
$\mat_k(\cX_t)$ of $\cX_t$ as follows.

\noindent
{\bf (i) TOPUP method:}
We define a order-5 tensor as
\bel{TOPUP-k}
{\rm TOPUP}_k
= \bigg(\sum_{t=h+1}^T \frac{\mat_k(\cX_{t-h})\otimes \mat_k(\cX_t)}{T-h},  h=1,\ldots,h_0\bigg)
\eel
where $\otimes$ is the tensor product and $h$ is the index for the 5-th mode.
Let $d_{-k}=d/d_k$ with $d= \prod_{k=1}^K d_k$.
As $\mat_k(\cX_t)$ is a $d_k \times d_{-k}$ matrix,
${\rm TOPUP}_k$ is of dimension
$d_k \times d_{-k} \times d_k \times d_{-k}  \times h_0$,
so that $\mat_1\big({\rm TOPUP}_k\big)$ is a $d_k\times (d^2h_0/d_k)$ matrix.
Let
\bel{TOPUP-P_km}
\widehat{\bP_{k,m}} =
\hbox{\rm PLSVD$_m$}\Big(
\mat_1\Big({\rm TOPUP}_k\Big)\Big),
\eel
where PLSVD$_m$ stands for the orthogonal projection to the span of the first $m$
left singular vectors of a matrix.
We estimate the projection $\widehat{\bP_k}$ by $\widehat{\bP_{k,\widehat{r_k}}}$
with a proper $\widehat{r_k}$. When $r_k$ is given,
\bes
\widehat{\bP_k} = \widehat{\bP_{k,r_k}}.
\ees
The above method is expected to yield consistent estimates of $\bP_k$
under proper conditions on
the dimensionality, signal strength and noise level
since \eqref{TOPUP-k} and \eqref{eq:tucker-model-2} imply
\bel{new-3}
\lefteqn{\E\Big[\mat_1\Big({\rm TOPUP}_k\Big)\Big]}
\cr &=& \mat_1\Big(\hbox{$\sum_{t=h+1}^T$} \E\big(\mat_k(\cM_{t-h})\otimes \mat_k(\cM_t)\big)/(T-h),  h=1,\ldots,h_0 \Big)
\cr &=& \mat_k\Big(\Big\{\hbox{$\sum_{t=h+1}^T$} \E\big(\cF_{t-h}\otimes \cF_t\big)/(T-h)\Big\}\timesKAsKAs,  h=1,\ldots,h_0\Big)
\cr &=& \bA_k \mat_k\Big(\hbox{$\sum_{t=h+1}^T$} \E\big(\cF_{t-h}\otimes \cF_t\big)/(T-h)\Big\}
\times_{\ell=1}^{k-1}\bA_\ell \times_{\ell=k+1}^{2K}\bA_{\ell},  h=1,\ldots,h_0\Big).
\eel
This is a product of two matrices, with $\bA_k = \bP_k\bA_k$ on the left.

We note that the left singular vectors of $\mat_1\big({\rm TOPUP}_k\big)$
are the same as the eigenvectors in the PCA of the $d_k\times d_k$ nonnegative-definite matrix
\bel{new-4}
\widehat{\bW}_k = \mat_1\Big({\rm TOPUP}_k\Big)\mat_1^\top\Big({\rm TOPUP}_k\Big),
\eel
which can be viewed as the sample version of
\bel{W_k}\label{eq:WA}
\bW_k = \mat_1\big(\E\big({\rm TOPUP}_k\big)\big)
\mat_1^\top\big(\E\big({\rm TOPUP}_k\big)\big).
\eel
It follows from \eqref{new-3} that $\bW_k$ has a sandwich formula with
$\bA_k$ on the left and $\bA_k^\top$ on the right.

As $\bA_k$ is assumed to be of rank $r_k$, its column space is identical to
that of $\E\big[\mat_1\big({\rm TOPUP}_k\big)\big]$ in \eqref{new-3}
or that of $\bW_k$ in \eqref{eq:WA} as long as they are also of rank $r_k$.
Thus $\bP_k$ is identifiable from the population version of TOPUP$_k$.
However, further identification of the lagged cross-product operator by the TOPUP would involve parameters
specific to the TOPUP approach. For example, if we
write $\bP_k = \bU_k\bU_k^\top$ where $\bU_k = (\bu_{k,1},\ldots,\bu_{k,r_k})$
is orthonormal, the TOPUP estimator \eqref{TOPUP-P_km} is designed to estimate
$(\bu_{k,1},\ldots,\bu_{k,m})$
as the left singular matrix of $\E\big[\mat_1\big({\rm TOPUP}_k\big)\big]$.
Even then, the singular vector  $\bu_{k,m}$ is identifiable only up to the sign through
the projections $\bP_{k,m} = \sum_{j=1}^m \bu_{k,j}\bu_{k,j}^\top$ and $\bP_{k,m-1}$ provided a sufficiently large gap
between the $(m-1)$-th, the $m$-th and the $(m+1)$-th singular values of the matrix relative
to the TOPUP estimation error.

For the ease of discussion, we consider
for example the case of $k=1$ and $K=3$ with stationary factor $\cF_t$ where
$\cX_t\in \R^{d_1\times d_2\times d_3}$ is a 3-way tensor,
and its lag-$h$ ($h>0$) autocovariance $\bSig_h$ is a 6-way
tensor with dimensions
$d_1\times d_2\times d_3\times d_1\times d_2\times d_3$ and elements
$\sigma^{(h)}_{i_1, j_1,k_1, i_2, j_2,k_2}=cov(x_{i_1j_1k_1,t-h},
x_{i_2j_2k_2,t})$.
For the estimation of the column space of the loading matrix $\bA_1$, we write
\begin{eqnarray}
\bW_1 &=& \sum_{h=1}^{h_0}\,\sum_{j_1,k_1}\,\sum_{i_2, j_2,k_2}
\Big(\E\Big[\bx_{\cdot,j_1,k_1,t-h}^{(1)}x_{i_2,j_2,k_2,t}\Big]\Big)
\Big(\E\Big[\bx_{\cdot,j_1,k_1,t-h}^{(1)}x_{i_2,j_2,k_2,t}\Big]\Big)^\top  \nonumber \\
&=& \sum_{h=1}^{h_0}\,\sum_{j_1,k_1}\,\sum_{j_2,k_2}
\Big(\E\Big[\bx_{\cdot,j_1,k_1,t-h}\bx_{\cdot,j_2,k_2,t}^{\top}\Big]\Big)
\Big(\E\Big[\bx_{\cdot,j_1,k_1,t-h}\bx_{\cdot,j_2,k_2,t}^{\top}\Big]\Big)^\top \nonumber \\
&=& \bA_1\bigg(\sum_{h=1}^{h_0}\,\sum_{j_1,k_1}\,\sum_{j_2,k_2}
\bGamma_{j_1,k_1,j_2,k_2,h}\bGamma_{j_1,k_1,j_2,k_2,h}^\top\bigg)\bA_1^\top \label{sandwich}
\end{eqnarray}
in view of \eqref{eq:WA},
where $\bGamma_{j_1,k_1,j_2,k_2,h}=
\rm{cov}(\cF_{t-h},\cF_{t}) \times_2\bA_{2,j_1\cdot}\times_3\bA_{3,k_1\cdot}\times_4 \bA_1\times_5\bA_{2,j_2\cdot}\times_6\bA_{3,k_2\cdot}\in\R^{r_1\times d_1}$.
This is a non-negative definite matrix sandwiched by $\bA_1$ and $\bA_1^\top$.
Hence the column space of $\bA_1$ and the column space of ${\bW_1}$ are the
same, if the matrix between $\bA_1$ and $\bA_1^\top$ in (\ref{sandwich})
is of full rank.

The TOPUP can be then described in terms of the PCA as follows.
Replacing ${\bW_1}$ with its sample version ${\widehat{\bW}_{1}}$ and
through eigenvalue decomposition, we can
estimate the top $r_k$-eigenvectors of
${\widehat{\bW}_{1}}$,
which form a representative of the estimated space spanned by
$\bA_1$.
Representative sets of eigenvectors of $\bA_2$ and $\bA_3$ can be obtained similarly.
This procedure uses the outer-product of all (time shifted)
 mode-1 fibers of the observed tensor
$\cY \in\R^{d_1\times d_2\times d_3\times T}$. Then, after taking the squares,
it sums over the other modes.
By considering positive lags $h>0$, we explicitly utilize
the assumption that the noise process is white, hence avoiding having to deal
with the contemporary covariance structure of $\cE_t$, as it disappears
in $\bGamma_{j_1,k_1,j_2,k_2,h}$ for all $h>0$.
We also note that while the PCA of $\widehat{\bW}_k$ in (\ref{new-4}) is
equivalent to
the SVD in \eqref{TOPUP-P_km} for the estimation of $\bP_k$,
it can be computationally more efficient to perform the SVD directly in many
cases.

We call this {\bf TOPUP} ({\bf T}ime series {\bf O}uter-{\bf P}roduct
{\bf U}nfolding {\bf P}rocedure)
as the tensor product in the matrix unfolding in \eqref{TOPUP-k}
is a direct extension of the vector outer product, which
is actually used in the equivalent formulation in \eqref{sandwich}.
This reduces to the algorithm in \cite{Wang&2019} for matrix time series.

\vspace{0.1in}

\noindent
{\bf (ii) TIPUP method:}
The {\bf TIPUP} ({\bf T}ime series {\bf I}nner-{\bf P}roduct
{\bf U}nfolding {\bf P}rocedure) can be simply described as the replacement of
the tensor product in \eqref{TOPUP-k} with the inner product:
\bel{TIPUP-k}
{\rm TIPUP}_k
= \bigg(\sum_{t=h+1}^T \frac{\mat_k(\cX_{t-h})\mat_k^\top(\cX_t)}{T-h},  h=1,\ldots,h_0\bigg),
\eel
which is treated as a matrix of dimension $d_k\times (d_kh_0)$.
The estimator $\widehat{\bP_{k,m}}$ is then defined as
\bel{TIPUP-P_km}
\widehat{\bP_{k,m}} =
\hbox{\rm PLSVD$_m$}\big({\rm TIPUP}_k\big).
\eel
Again TIPUP is expected to yield consistent estimates of $\bP_k$ in
\eqref{P_k} as
\bel{new-2}
\lefteqn{\E\big[{\rm TIPUP}_k\big]}
\cr &=& \Big(\big\langle \bSig_h, \bI_{k,k+K}\big\rangle_{\{k,k+K\}^c}, h=1,\ldots,h_0\Big)
\\ \nonumber &=&\bA_k {\Big(}\big\langle \E\big[\hbox{$\sum_{t=h+1}^T$}(\cF_{t-h}\otimes \cF_t)/(T-h)\big]
\times_{\ell\neq k, 1\le \ell\le 2K}\bA_\ell,
\bI_{k,k+K}\big\rangle_{\{k,k+K\}^c},  h \le h_0\Big),
\eel
where $\bI_{k,k+K}$ is the $(2K)$-tensor with
elements $(\bI_{k,k+K})_{\bi,\bj} = I\{ \bi_{-k}=\bj_{-k}\}$ at
$\bi=(i_1,\ldots,i_K)$ and $\bj=\{j_1,\ldots,j_K)$, and
$\langle\cdot,  \cdot \rangle_{\{k,k+K\}^c}$ is the inner product summing over
indices other than $\{k,k+K\}$.

We use the superscript $^*$ to indicate the TIPUP counterpart of TOPUP quantities, e.g.
\bel{eq:hat:WA-star}
\widehat{\bW}_k^* = \big({{\rm TIPUP}_k}\big)\big({{\rm TIPUP}_k}\big)^\top
\eel
is the sample version of
\bel{eq:WA-star}
\bW_k^* = \E\big[{\rm TIPUP}_k\big] \E\big[{\rm TIPUP}_k^\top\big].
\eel
We note that by \eqref{new-2} $\bW_k^*$ is again sandwiched between $\bA_k$ and $\bA_k^\top$.
For $k=1$ and $K=3$,
\bel{eq:WA-star2}
\bW_1^* = \bA_1\left[\sum_{h=1}^{h_0}
\left(\sum_{j,k}\bGamma_{j,k,j,k,h}\right)
    \left(\sum_{j,k}\bGamma_{j,k,j,k,h}\right)^\top\right]\bA_1^{\top}
\eel
with $\bGamma_{j_1,k_1,j_2,k_2,h}$ being that in \eqref{sandwich}.
If the middle term in (\ref{eq:WA-star2}) is of full rank, then
the column space of $\bW_1^*$ is the same as that of $\bA_1$.

As in the case of the TOPUP,
for the estimation of the auto-covariance operator beyond $\bP_k$,
the TIPUP would only identify parameters
specific to the approach. For example,
the TIPUP estimator \eqref{TIPUP-P_km} aims to estimate
$\bP^*_{k,m} = \sum_{j=1}^m \bu_{k,j}^*(\bu_{k,j}^*)^\top$ with $\bU_k^*$ being
the left singular matrix of $\E\big[{{\rm TIPUP}_k}\big]$, e.g.
the eigen-matrix of $\bW_1^*$ in \eqref{eq:WA-star2}.
This is evidently different from the projection $\bP_{k,m}$ to the rank $m$
eigen-space of $\bW_k$ in \eqref{eq:WA}, in view of \eqref{sandwich} and \eqref{eq:WA-star2}.


\vspace{0.1in}

\noindent
{\bf Remark:} The differences between the TOPUP and TIPUP are two folds. First,
the TOPUP for estimating
the column space of $\bA_1$ uses the auto-cross-covariance between
{\bf all} the mode-1 fibers in $\cX_{t-h}$ and {\bf all} the mode-1 fibers in
$\cX_{t}$,
with all possible combinations of $\{j_1,k_1,j_2,k_2\}$,
while the TIPUP only uses the auto-cross-covariance between the
mode-1 fibers in $\cX_{t-h}$ and their corresponding mode-1 fibers in
$\cX_{t}$, with all combinations of $\{j,k,j,k\}$ only. Hence
the TIPUP uses less cross-covariance terms in the estimation.
Second, the TOPUP 'squares'
every auto-cross-covariance matrices first (i.e.
$\bGamma_{j_1,k_1,j_2,k_2,h}\bGamma_{j_1,k_1,j_2,k_2,h}^\top$ in \eqref{sandwich})
before the summation, while the TIPUP does the summation of
$\bGamma_{j,k,j,k,h}$
first, before taking the square as in (\ref{eq:WA-star2}). Because
the TOPUP takes the squares first, every term in the summation of
the middle part of (\ref{sandwich}) is semi-positive definite. Hence
if the sum of a subset of them is full rank, then the middle part is full rank and
the column space of $\bA_1$ and ${\bW_1}$ will be the same. On the other
hand, the TIPUP takes the summation first, hence runs into the possibility that
some of the auto-covariance matrices cancel out each other,
making the sum not full rank.
However, the summation first approach
averages out more noises in the sample version
while the TOPUP accumulates more noises by taking the squares first.
The TOPUP also has more terms -- although it amplifies the signal,
it amplifies the noise as well.
The detailed asymptotic convergence rates of both methods represented
in Section 5 reflect the differences. In Section 6 we show a case in
which some of the auto-covariance matrices cancel each other. We note that
complete cancellation does not occur often and can often be avoided by using
a larger $h_0$ in estimation, though partial cancellation can still have impact on the performance of TIPUP in finite samples.

\vspace{0.1in}

\noindent
{\bf Remark: $i$TOPUP and $i$TIPUP:} One can construct iterative
procedures based on the TOPUP and TIPUP respectively.
Note that if  a version of $\bU_2\in \R^{d_2\times r_2}$ and $\bU_3\in \R^{d_3\times r_3}$
are given with $\bP_k = \bU_k\bU_k^\top$, $\bP_1$ can be estimated
via the TOPUP or TIPUP using $\tilde\cX_t^{(1)}=
\cX_t\times_2\bU_2^\top\times_3\bU_3^\top\in\R^{d_1\times r_2\times r_3}$.
Intuitively the performance improves since $\tilde\cX_t^{(1)}$
is of much lower dimension than $\cX_t$ as $r_2 \ll d_2$ and $r_3\ll d_3$.
With the results of the TOPUP and TIPUP
as the starting points, one can
alternate the estimation of $\bP_k$ given other
estimated loading matrices until convergence.
They have similar flavor as tensor power methods. Numerical
experiments show that the iterative procedures do indeed outperform the
simple implementation of the TOPUP and TIPUP. However, their asymptotic
properties require more detailed analysis and are out of the scope of this
paper. The benefit of such iteration has been shown in tensor completion
\citep{xia2017polynomial} among others.

\vspace{0.2in}

\section{Theoretical Results}

Here we present some results of the
theoretical properties of the proposed estimation methods.
Recall that the loading matrix $\bA_k$ is not orthonormal in general, and
our aim is to estimate the projection $\bP_k$ in \eqref{P_k} to the column space of $\bA_k$.
We shall consider the theoretical properties of the estimators under the
following two conditions:

\vspace{0.1in}

\noindent
{\bf Condition A:} {\it
$\cE_t$ are independent Gaussian tensors conditionally on the entire process of
$\{\cF_t\}$. 
In addition, we assume that for some constant $\sigma>0$, we have
\bel{cond-1}
\barE(\bu^\top\vec(\cE_t))^2\le \sigma^2 \|\bu\|_2^2,\quad \bu \in \R^{d},
\eel
where $\barE$ is the conditional expectation given $\{\cF_t, 1\le t\le T\}$
and $d=\prod_{k=1}^Kd_k$}.

Condition A, which holds with equality when $\cE_t$ has iid
$N(0,\sigma^2)$ entries, allows the entries of $\cE_t$
to have a range of dependency structures and different covariance
structures for different $t$.
Under Condition A, we develop a general theory
to describe the ability of the TOPUP and TIPUP estimators to average out the noise $\cE_t$.
It guarantees consistency and provides convergence rates in the estimation of
the principle space of the signal $\cM_t$, or equivalently the projection $\bP_k$,
under proper conditions on the magnitude and certain singular value of the lagged cross-product of $\cM_t$,
allowing the ranks $r_k$ to grow as well as $d_k$ in a sequence of experiments with $T\to\infty$.

We then apply our general theory in two specific scenarios.
The first scenario, also the simpler, is described in the following
condition on the factor series $\cF_t$.



\vspace{0.1in}

\noindent
{\bf Condition B:} {\it
The process $\cF_t\in \R^{r_1\times\dots\times r_K}$ is weakly stationary,
with fixed $r_1,\ldots,r_K$ and fixed expectation for the lagged cross-products
$\cF_{t-h}\otimes \cF_t$, such that
$(T-h)^{-1}\sum_{t=h+1}^T\cF_{t-h}\otimes \cF_t$
converges to $\E\big[\cF_{T-h}\otimes \cF_T\big]$ in probability. }

In the second scenario, described in Conditions C-1 and C-2 below,
conditions on the signal process $\cM_t$ are expressed in terms of
certain factor strength or related quantities.
For the vector factor model (\ref{vector_factor}), \cite{Lam&2011} showed
that the convergence rate of the corresponding
TOPUP estimator is $d/(\lambda^2T^{1/2})$, when
$\lambda\asymp\,$singular$(\bA)=O(d^{(1-\delta')/2})$ and
$0\le \delta'\le 1$. Here
singular$(\bA)$ denotes (any and all) positive singular values of $\bA$,
and $\delta'$ is often referred to as the strength of the factors
\citep{Bai&Ng2002,Doz&2011,Lam&2011}. It reflects the
signal to noise ratio in the factor model.
When $\delta'=0$, singular$(\bA)=O(d^{1/2})$
hence the information contained in the signal
$\bA\bff_t$ increases linearly with the dimension $d$. In this case
the factors are often said to be 'strong' and the convergence rate is
$T^{-1/2}$. When $0<\delta'\le 1$ (weak factors),
the information in the signal increases more slowly than the dimension. In this
case, one needs larger $T$ (longer time series) to compensate in order
to have consistent estimation of the loading spaces.

\vspace{0.1in}

Again, let $d= \prod_{k=1}^K d_k$, $d_{-k}=d/d_k$, $r=\prod_{k=1}^K r_k$ and $r_{-k}=r/r_k$.
Define
\bel{Phi-Theta}
\bPhi_{k,h} = \sum_{t=h+1}^{T}
\frac{\mat_k(\cF_{t-h})\otimes \mat_k(\cF_t)}{T-h},\ \
\bTheta_{k,h} = \sum_{t=h+1}^{T}
\frac{\mat_k(\cM_{t-h})\otimes \mat_k(\cM_t)}{T-h},
\eel
as 4-way tensors respectively of dimensions $r_k\times r_{-k}\times r_k\times r_{-k}$ and $d_k\times d_{-k}\times d_k\times d_{-k}$.
It follows from \eqref{TOPUP-k} that the TOPUP procedure is based on the lagged cross-product
\bes
\bV_{k,h} = \sum_{t=h+1}^T \frac{\mat_k(\cX_{t-h})\otimes \mat_k(\cX_t)}{T-h}
\in \R^{d_k\times d_{-k}\times d_k\times d_{-k}}.
\ees
In fact, it follows from
\eqref{second}, \eqref{eq:tucker-model-2}, \eqref{TOPUP-k}, \eqref{Phi-Theta} and Condition A that
\bel{new-0}
\barE\big[ \bV_{k,h} \big] &=& \bTheta_{k,h}
= \bPhi_{k,h}\times_1\bA_k\times_2\bA_{-k}\times_3\bA_k\times_4\bA_{-k},
\cr \barE\big[\mat_1\big(\hbox{\rm TOPUP}_k\big)\big]
&=& \mat_1\big(\bTheta_{k,h}, 1\le h\le h_0\big)
\\ \nonumber
&=& \mat_1\big(\bPhi_{k,1:h_0}\times_1\bA_k\times_2\bA_{-k}\times_3\bA_k\times_4\bA_{-k}\big),
\eel
where
$\bA_{-k} = {\rm Kronecker}(\bA_{K},\ldots,\bA_{k+1},\bA_{k-1},\ldots,\bA_1) \in \R^{d_{-k}\times r_{-k}}$
and $\bPhi_{k,1:h_0}=\big(\bPhi_{k,1},\ldots,\bPhi_{k,h_0}\big)
\in \R^{r_k\times r_{-k}\times r_k\times r_{-k}\times h_0}$.
Recall that $\mat_k(\cX_t)$ is the mode $k$ unfolding of $\cX_t$ into a
$d_k\times d_{-k}$ matrix.
In connection to the PCA, \eqref{new-4} and \eqref{eq:WA} give
\bes
\widehat{\bW}_k = \sum_{h=1}^{h_0} \mat_1\big(\bV_{k,h}\big)\mat_1^\top\big(\bV_{k,h}\big),\quad
\bW_k = \sum_{h=1}^{h_0} \mat_1\big(\E\, \bTheta_{k,h}\big)\mat_1^\top\big(\E\, \bTheta_{k,h}\big).
\ees

For the TIPUP, define matrices
\bel{Phi-Theta-star}
\bPhi_{k,h}^* = \sum_{t=h+1}^{T}
\frac{\mat_k(\cF_{t-h})
\mat_k^\top(\cF_t)}{T-h},\ \
\bTheta_{k,h}^* = \sum_{t=h+1}^{T}
\frac{\mat_k(\cM_{t-h})\mat_k^\top(\cM_t)}{T-h},
\eel
respectively of dimensions $r_k\times r_k$ and $d_k\times d_k$.
As in \eqref{TIPUP-k}, the TIPUP
procedure is based on
\bes
\bV^*_{k,h} = \sum_{t=h+1}^T \frac{\mat_k(\cX_{t-h})\mat_k^\top(\cX_t)}{T-h} \in \R^{d_k\times d_k},
\ees
which can be viewed as an estimate of $\barE\big[\bV^*_{k,h}\big]= \bTheta^*_{k,h}$.
In model \eqref{lam-F-U},
\bel{new-5}
\barE\big[{\rm TIPUP}_k\big]
= \big(\bTheta^*_{k,1},\ldots,\bTheta^*_{k,h_0}\big)
= \lam^2\bU_k \bPhi^*_{k,1:h_0}(\bU_k,\ldots,\bU_k)^\top
\eel
with $\bPhi^*_{k,1:h_0}=\big(\bPhi^*_{k,1},\ldots,\bPhi^*_{k,h_0}\big)$.
By (\ref{eq:hat:WA-star}) and (\ref{eq:WA-star}),
the above quantities are connected to PCA via
\bel{W_k-star}
\widehat{\bW}^*_k = \sum_{h=1}^{h_0} \bV^*_{k,h}\big(\bV^*_{k,h}\big)^\top,\quad
\bW^*_k = \sum_{h=1}^{h_0} \big(\E\big[\bTheta^*_{k,h}\big]\big)
\big(\E\big[\bTheta^*_{k,h}\big]\big)^\top.
\eel

Our analysis involves the norms of the $d_k\times d_{-k}\times d_k\times d_{-k}$ tensor
$\bTheta_{k,0}$ and
the $d_k\times d_k$ matrix $\bTheta^*_{k,0}$, and the elements the singular values
of $\barE\big[\mat_1({\rm TOPUP}_k)\big]$ and $\barE\big[{\rm TIPUP}_k\big]$.

The first norm is the operator norm of $\bTheta_{k,0}$ as a linear mapping
in $\R^{d_k\times d_{-k}}$:
\bel{new-norm-1}
\big\|\bTheta_{k,0}\big\|_{\rm op} = \max\bigg\{\sum_{i_1,j_1,i_2,j_2}
u_{i_1,j_1}u_{i_2,j_2}\big(\bTheta_{k,0}\big)_{i_1,j_1,i_2,j_2}: \big\|\bU\big\|_{\rm F}=1\bigg\}
\eel
where $\bU$ denotes a $d_k\times d_{-k}$ matrix with elements $u_{i,j}$.
The second is the spectrum norm of $\bTheta^*_{k,0}$ with elements
$\sum_{j =1}^{d_{-k}}\big(\bTheta_{k,0}\big)_{i_1,j,i_2,j}$ and also in the inner-product form
as in \eqref{Phi-Theta-star}:
\bel{new-norm-2}
\big\|\bTheta^*_{k,0}\big\|_{\rm S} = \max_{\|\bu\|_2=1} \bu^\top\bTheta^*_{k,0}\bu
= \max_{\|\bu\|_2=1}\sum_{i_1,i_2,j} u_{i_1}u_{i_2}\big(\bTheta_{k,0}\big)_{i_1,j,i_2,j}.
\eel
\chzM{In fact, treating $\bTheta_{k,0}$ as an $\R^d\to\R^d$ mapping, we have
\bel{norm-relations}
r_k\big\|\bTheta^*_{k,0}\big\|_{\rm S}
\ge r_k^{1/2}\big\|\bTheta^*_{k,0}\big\|_{\rm F}
\ge \text{trace}\big(\bTheta^*_{k,0}\big)
= \text{trace}\big(\bTheta_{k,0}\big)
\le r^{1/2}\big\|\bTheta_{k,0}\big\|_{\rm HS}
\le r \big\|\bTheta_{k,0}\big\|_{\rm op},
\eel
as $\bTheta_{k,0}$ and $\bTheta^*_{k,0}$ have respective ranks $r$ and $r_k$.
See \eqref{new-cond-1} and \eqref{new-cond-2} below for additional discussion.}

Our error bounds for the TOPUP also involve
\bel{new-norm-3a}
\tau_{k,m} = \hbox{the $m$-th largest singular value of }\barE\big[\mat_1\big({\rm TOPUP}_k\big)\big].
\eel
By \eqref{TOPUP-k} and \eqref{new-0},
$\barE\big[\mat_1\big({\rm TOPUP}_k\big)\big]
= \big(\mat_1\big(\bTheta_{k,1}\big),\ldots,\mat_1\big(\bTheta_{k,h_0}\big)\big)\in\R^{d_k\times(d_{-k}dh_0)}$,
so that $\tau_{k,m}^2$ is the $m$-th eigenvalue of
$\sum_{h=1}^{h_0}\mat_1\big(\bTheta_{k,h}\big)\mat_1^\top\big(\bTheta_{k,h}\big)$, a sum of $h_0$
nonnegative-definite matrices. Thus, as $\barE\big[\mat_1\big({\rm TOPUP}_k\big)\big]$ is a second order
process with the left-most factor $\bA_k\in \R^{d_k\times r_k}$ of rank $r_k$,
we characterize the signal strength as
\bel{new-norm-3b}
\lam_k = \sqrt{h_0^{-1/2}\tau_{k,r_k}}
\eel
for the estimation of the orthogonal projection
$\bP_k = \bA_k\big(\bA_k^\top\bA_k\big)^{-1}\bA_k^\top$ in \eqref{P_k}.
For the estimation of the mode-$k$ principle space of a general rank $m$,
the eigen-gap $\tau_{k,m}-\tau_{k,m+1}$ would be involved.

We note that \chzM{by \eqref{new-0}, \eqref{new-norm-1} and Cauchy-Schwarz
\bel{lam_k-relation}
\lam_k^4 = \frac{\tau_{k,r_k}^2}{h_0}
\le \frac{\big\|\barE\big[\mat_1\big(\hbox{\rm TOPUP}_k\big)\big]\big\|_{\rm F}^2}{h_0r_k}
= \sum_{h=1}^{h_0}\frac{\big\|\bTheta_{k,h}\big\|_{\rm HS}^2}{h_0r_k}
\le \frac{\big\|\bTheta_{k,0}\big\|_{\rm HS}^2}{r_k(1-h_0/T)^2}.
\eel
Thus, $\lam_k^2 \le C_1r_{-k}^{-1/2}\big\|\bTheta^*_{k,0}\big\|_{\rm S}/(1-h_0/T)$ when
$r^{1/2}\big\|\bTheta_{k,0}\big\|_{\rm HS} \le C_1 r_k\big\|\bTheta^*_{k,0}\big\|_{\rm S}$
as expected from \eqref{norm-relations}.}
When the condition numbers of $\bA_k^\top\bA_k$
are bounded and \chzM{$\prod_{k=1}^K\|\bA_k\|_{\rm S} = \lam$,}
\bel{tau_k-2}
& \big\|\bTheta_{k,0}\big\|_{\rm op} \asymp \lam^2 \big\|\bPhi_{k,0}\big\|_{\rm op},\quad
\big\|\bTheta^*_{k,0}\big\|_{\rm S} \asymp \lam^2 \big\|\bPhi^*_{k,0}\big\|_{\rm S},
\cr & \tau_{k,r_k} \asymp \lam^2\times \big(
\hbox{the $r_k$-th singular value of }\mat_1(\bPhi_{k,1:h_0})\big),
\eel
by \eqref{new-0}.
In particular, if \eqref{lam-F-U} holds, then \eqref{tau_k-2} holds with ``$\asymp$" replaced by equality as in
\bel{tau_k-3}
\tau_{k,m} = \lam^2\times \big(
\hbox{the $m$-th singular value of }\mat_1(\bPhi_{k,1:h_0})\big).
\eel
As the columns of $\mat_1(\bPhi_{k,1:h_0})$ form an ensemble of mode-$k$ fibers of
$(T-h)^{-1}\sum_{t=h+1}^T\cF_{t-h}\otimes \cF_t$,
$\bPhi_{k,0}, \bPhi^*_{k,0}$ and $\bPhi_{k,1:h_0}$ for fixed $h_0$
can be all replaced by their expectation in \eqref{tau_k-2} under Condition B,
e.g. $\lam_k\asymp \lam$ when the constant matrix $\E[\bPhi_{k,1:h_0}]$ is of rank $r_k$.


The analysis of the TIPUP involves
\bel{new-norm-3c}
\tau^*_{k,m} = \hbox{the $m$-th largest singular value of }\barE\big[{\rm TIPUP}_k\big],
\eel
which is in general different from the $\tau_{k,m}$ for the TOPUP in \eqref{new-norm-3a}.
Similar to \eqref{new-norm-3b},
we characterize the signal strength for the estimation of the projection
$\bP_k = \bA_k\big(\bA_k^\top\bA_k\big)^{-1}\bA_k^\top$ as
\bel{new-norm-3d}
\lam_k^* = \sqrt{h_0^{-1/2}\tau_{k,r_k}^*}.
\eel
Let $\bPhi^*_{k,1:h_0}=\big(\bPhi^*_{k,1},\ldots,\bPhi^*_{k,h_0}\big)\in \R^{r_k\times (r_kh_0)}$ with the $\bPhi^*_{k,h}$ in \eqref{Phi-Theta-star}. Similar to \eqref{tau_k-3}, in model \eqref{lam-F-U}
\bel{tau_k-4}
\tau^*_{k,m} = \lam^2\times \Big(
\hbox{the $m$-th singular value of }\bPhi^*_{k,1:h_0}\Big).
\eel
\chzM{Similar to \eqref{lam_k-relation}, \eqref{Phi-Theta-star}, \eqref{new-norm-2} and Cauchy-Schwarz yield
\bel{lam*_k-relation}
(\lam_k^*)^4
\le \frac{\big\|\barE\big[\mat_1\big(\hbox{\rm TIPUP}_k\big)\big]\big\|_{\rm F}^2}{h_0r_k}
= \sum_{h=1}^{h_0}\frac{\big\|\bTheta^*_{k,h}\big\|_{\rm F}^2}{h_0r_k}
\le \frac{\big\|\bTheta^*_{k,0}\big\|_{\rm S}^2}{(1-h_0/T)^2}.
\eel
For} fixed $h_0$, \eqref{tau_k-4} gives $\lam_k^*\asymp \lam$ under Condition B
when  $\E[\bPhi^*_{k,1:h_0}]$ is of rank $r_k$.
\medskip

\noindent
{\bf Theoretical property of TOPUP:\ }
We present some error bounds for the TOPUP estimator in the following theorem.

\begin{theorem}\label{th-1}
Let $d=\prod_{k=1}^K d_k$ and $r=\prod_{k=1}^K r_k$,
$\lam_k$ be as in \eqref{new-norm-3b},
and
\bes
\Delta_k(\bTheta_{k,0}) &=& \frac{\sigma(2Td)^{1/2}}{T-\chzM{h_0}}
\Big\{\Big(\sqrt{d_k/d}+\sqrt{r/r_k}\Big)\big\|\bTheta^*_{k,0}\big\|_{\rm S}^{1/2}
+ \Big(\sqrt{d_k/d}+\sqrt{r/d_k}\Big)\big\|\bTheta_{k,0}\big\|_{\rm op}^{1/2}\Big\}
\ees
with the norms $\big\|\bTheta_{k,0}\big\|_{\rm op}$ and $\big\|\bTheta^*_{k,0}\big\|_{\rm S}$
in \eqref{new-norm-1} and \eqref{new-norm-2} respectively.
Suppose Condition A holds.
Let $\barE$ be the conditional expectation given $\{\cF_t, 1\le t\le T\}$.
Then, $\barE\big[\bV_{k,h}\big] = \bTheta_{k,h}$ and
\bel{th-1-1}
& \displaystyle
\barE \big\|\mat_1\big(\bV_{k,h}\big) - \mat_1\big(\bTheta_{k,h}\big)\big\|_{\rm S}
\le \Delta_k(\bTheta_{k,0})
+ \chzM{\frac{\sigma^2(1+d_{-k})\sqrt{2d_k}}{\sqrt{T-h}} + \frac{2\sigma^2\sqrt{d_kd}}{T-h},}
\\ \nonumber & \displaystyle
\barE \big\|\mat_1\big({\rm TOPUP}_k\big)
- \barE\big[\mat_1\big({\rm TOPUP}_k\big)\big]\big\|_{\rm S}
\le \sqrt{h_0}\bigg\{\Delta_k(\bTheta_{k,0})
+ \chzM{\frac{\sigma^2(1+d_{-k})\sqrt{2d_k}}{\sqrt{T-h_0}} + \frac{2\sigma^2\sqrt{d_kd}}{T-h_0}}\bigg\}
\eel
for all $k$ and $h_0\le T/4$. Moreover,
\bel{th-1-3}
\barE\Big\|\widehat{\bP_{k, r_k}} - \bP_k\Big\|_{\rm S}
\le 2\lam_k^{-2}\bigg\{\Delta_k(\bTheta_{k,0})
+ \chzM{\frac{\sigma^2(1+d_{-k})\sqrt{2d_k}}{\sqrt{T-h_0}}
+ \frac{2\sigma^2\sqrt{d_kd}}{T-h_0}}\bigg\}
\eel
for the estimator \eqref{TOPUP-P_km} with $m = r_k$, where $\lam_k$ is as in \eqref{new-norm-3b}.
\end{theorem}

\noindent
\chzM{When $\lam_k^2 \le C_1r_{-k}^{-1/2}\big\|\bTheta^*_{k,0}\big\|_{\rm S}/(1-h_0/T)$
as 
discussed below \eqref{lam_k-relation},
$2\sigma^2\sqrt{d_kd}/(T-h_0)$
can be replaced by $C_1\Delta_k(\bTheta_{k,0})/(2r_{-k}^{3/2}\sqrt{d_{-k}})$ in \eqref{th-1-3}
as in the proof of \eqref{th-2-3} of Theorem~\ref{th-2}.}
%
The proof of the theorem is shown in Appendix A.

More explicit error bounds can be given in the one-factor model
with $r_j=1$ for all $j$,
\bel{example-1a}
\cX_t = \lambda f_t\big(\bu_1\otimes \cdots\otimes \bu_K) + \cE_t
\eel
for some unit vectors $\bu_k\in\R^{d_k}$. In this case we have
$\bTheta_{k,h} = \lam^2{\hat\rho}_h\big\{\bu_k
{\rm vec}^\top(\otimes_{j\neq k}\bu_j)\big\}^{\otimes 2}$
and $\bTheta^*_{k,h} = \lam^2{\hat\rho}_h\bu_k\bu_k^\top$
with ${\hat\rho}_h = \sum_{t=h+1}^T f_{t-h}f_t/(T-h)$.
By \eqref{new-norm-1}, \eqref{new-norm-2} and \eqref{new-norm-3b},
we have
\bel{example-1b}
\big\|\bTheta_{k,0}\big\|_{\rm op} = \big\|\bTheta^*_{k,0}\big\|_{\rm S}
= \lam^2 {\hat\rho}_0,\quad
\lam_k^2 = \lam^2\bigg(\frac{1}{h_0}\sum_{h=1}^{h_0}{\hat\rho}_h^2\bigg)^{1/2}.
\eel
Thus, for the TOPUP estimate of $\bu_k$, Theorem \ref{th-1} gives
\bel{example-1c}
\barE\sqrt{1-(\hbu_k^\top\bu_k)^2}
\le \frac{C_1\sigma d^{1/2}\lam {\hat\rho}_0^{1/2}}{\lam_k^2 T^{1/2}}
+ \frac{C_1\sigma^2 d}{\lam_k^2 (d_kT)^{1/2}}
\lesssim \frac{d^{1/2}}{\lam T^{1/2}} + \frac{d}{\lam^2 (d_kT)^{1/2}}.
\eel
for some constant $C_1$.
Here we use the assumption
$\sigma\asymp 1$ and $\sum_{h=1}^{h_0}{\hat\rho}_h^2/h_0\asymp {\hat\rho}_0^2\asymp 1$.
We note that $\{1-(\hbu_k^\top\bu_k)^2\}^{1/2}=\big\|\hbu_k\hbu_k^\top - \bu_k\bu_k^\top\big\|_{\rm S}$
is the absolute value of the sine of the angle between $\hbu_k$ and $\bu_k$, and that
${\hat\rho}_0$ and $\sum_{h=1}^{h_0}{\hat\rho}_h^2/h_0$
can be treated as constants under Condition B.
This analysis is also valid for fixed ranks under Condition B as in the following corollary.

\begin{coro}\label{cor-1}
Suppose $r_1,\cdots,r_K$, $h_0$, and $\sigma$ are fixed, the condition numbers of $\bA_k^\top\bA_k$ are bounded, Conditions A and B hold, and $\E[\mat_1(\bPhi_{k,1:h_0})]$ is of rank $r_k$
for the $\bPhi_{k,1:h_0}$ in \eqref{new-0}. Then,
\bel{cor-1-1}
& \barE \big\|\mat_1\big(\bV_{k,h}\big) - \mat_1\big(\bTheta_{k,h}\big)\big\|_{\rm S}
\lesssim (d/T)^{1/2}\lam + d/(d_kT)^{1/2} \chzM{+ \sqrt{d_kd}/T,}
\\ \nonumber &
\barE \big\|\mat_1\big({\rm TOPUP}_k\big)
- \barE\big[\mat_1\big({\rm TOPUP}_k\big)\big]\big\|_{\rm S}
\lesssim (d/T)^{1/2}\lam + d/(d_kT)^{1/2} \chzM{+ \sqrt{d_kd}/T,}
\eel
with $\lam = \prod_{k=1}^K\|\bA_k\|_{\rm S}$, and $\lam_k\asymp \lam$,
for all $k$ and $h\le h_0$. Moreover,
\bel{cor-1-2}
\barE\Big\|\widehat{\bP_{k, r_k}} - \bP_k\Big\|_{\rm S}
\lesssim  \frac{d^{1/2}}{\lam T^{1/2}} + \frac{d}{\lam^2 (d_kT)^{1/2}}.
\eel
\end{coro}

In the case of $K=2$ where matrix time series $\bX_t = \bA_1\bF_t\bA_2^\top + \bE_t$ is observed,
properties of the TOPUP was studied in \cite{Wang&2019} under the conditions
of Corollary \ref{cor-1} with $\lam\asymp
d_1^{1-\delta_1'}d_2^{1-\delta_2'} = d^{1-\delta_0}$ for some $\delta_0\in [0,1]$.
Their error bounds yield somewhat slower rate
\bes
\big\|\widehat{\bP_{k, r_k}} - \bP_k\big\|_{\rm S} \lesssim d/(\lam_k^2T^{1/2}) \asymp d^{\delta_0}/T^{1/2}.
\ees


For general ranks $r_1,\ldots,r_K$ possibly with slowly diverging
$r = \prod_{k=1}^K r_k$,
we may also characterize the convergence rate in terms of the power of $d$ and $d_k$
as in \cite{Lam&2011} and \cite{Wang&2019} but our error bounds also involve the power of
$r$ and $r_k$ as they are allowed to diverge in our setting.
This is done as follows by relating the norms and singular values in
Theorem~\ref{th-1} and other norms of $\bTheta_{k,h}$
in the scenario where the matrices involved are assumed to have
the fullest rank given ${\rm rank}(\bA_k)=r_k$ and their non-zero singular values are of the same order.
To express such powers of $d, d_k, r$ and $r_k$ scale free, we consider
norms and singular values of $\bTheta_{k,h}/\sigma^2$ which can be viewed as
the signal to noise ratio in the tensor form,
with the $\sigma$ defined in Condition A.

As the elements of $\bTheta_{k,0}\in \R^{d_k\times d_{-k}\times d_k\times d_{-k}}$
are averages of real numbers of the form $(\cM_t)_{\bi}(\cM_t)_{\bj}$
over $t$, we may expect its Hilbert-Schmidt norm to satisfy
\bes
\big\|\bTheta_{k,0}/\sigma^2\big\|_{\rm HS}^2
= \sum_{i_1=1}^{d_k}\sum_{j_1=1}^{d_{-k}}\sum_{i_2=1}^{d_k}\sum_{j_2=1}^{d_{-k}}
\big(\bTheta_{k,0}/\sigma^2\big)_{i_1,j_1,i_2,j_2}^2
\asymp d^{2(1-\delta_0)}
\ees
for some constant $\delta_0$, due to $d_kd_{-k}=d=\prod_{k=1}^K d_k$.
As $\bTheta_{k,0}$ is a nonnegative-definite operator in $\R^{d_k\times d_{-k}}$
with rank $r = \prod_{k=1}^K r_k$, we expect its non-zero eigenvalues to be of the order
\bel{new-cond-1}
\big\|\bTheta_{k,0}/\sigma^2\big\|_{\rm op}
\asymp \Big(\big\|\bTheta_{k,0}/\sigma^2\big\|_{\rm HS}^2/r\Big)^{1/2}
\asymp d^{1-\delta_0}/r^{1/2}.
\eel
Moreover,
\chzM{when the larger quantities in \eqref{norm-relations} are of the same order,}
\bel{new-cond-2}
\big\|\bTheta^*_{k,0}/\sigma^2\big\|_{\rm S}
\asymp r\big(d^{1-\delta_0}/r^{1/2}\big)\big/r_k = d^{1-\delta_0}r^{1/2}/r_k.
\eel
\chzM{In view of \eqref{lam_k-relation}, we}
may also express the singular values $\tau_{k,m}$ and the closely related $\lam_k$
in the same way.
Counting the number of elements in \eqref{TOPUP-k}, we expect
\bes
\big\|\barE\big[\mat_1\big({\rm TOPUP}_k/\sigma^2\big)\big]\big\|_{\rm F}^2
\asymp  h_0 d^{2(1-\delta_1)}
\ees
for some constant $\delta_1$.
\chzM{By \eqref{lam_k-relation} and \eqref{new-cond-1}, we have}
$\delta_1\ge \delta_0$ as TOPUP$_k$ are composed of
auto-covariance elements, whereas $\bTheta_{k,0}$ involves the
covariance (with lag $h=0$).
As the matrix $\barE\big[\mat_1\big({\rm TOPUP}_k\big)\big]$ is of rank $r_k$, we expect
that for some constant $c_1>0$
\bel{new-cond-3}
h_0^{1/2}(\lam_k/\sigma)^2 = \tau_{k,r_k}/\sigma^2
\ge c_1 \sqrt{h_0/r_k}d^{1-\delta_1}>0.
\eel
We summarize the scenario as follows.

\vspace{0.1in}

\noindent
{\bf Condition C-1:} {\it For the quantities given in \eqref{new-norm-1}, \eqref{new-norm-2},
\eqref{new-norm-3a} and \eqref{new-norm-3b},
\bes
{\mathbb{P}}\Big\{\
\hbox{ \eqref{new-cond-1}, \eqref{new-cond-2} and \eqref{new-cond-3}
hold } \Big\} = 1 + o(1).
\ees
Moreover, for certain constants $\delta_2\ge \delta_1$, $c_2>0$ and
$\epsilon_{d,T}=o(1)$,
\bel{new-cond-4}
{\mathbb{P}}\left.\begin{cases}\
\tau_{k,m} - \tau_{k,m+1}
\ge c_2 \sigma^2 \sqrt{h_0/r_k}d^{1-\delta_2}
\cr\ \Big\|\barE\big[\mat_1\big({\rm TOPUP}_k\big)\big]
- \E\big[\mat_1\big({\rm TOPUP}_k\big)\big]\Big\|_{\rm S}
\le \epsilon_{d,T}\sigma^2 \sqrt{h_0/r_k}d^{1-\delta_2}
\end{cases}\right\} = 1+o(1)
\eel
whenever the eigen-gap in \eqref{new-norm-3b} is invoked
for some integer $m\in [1,r_k)$.}

Condition C-1 is more general than Condition B as
$\cF_t$ is not required to be weak stationary and
$r_1,\ldots,r_K, h_0$ are allowed to diverge.
\chzM{Under \eqref{new-cond-1} and \eqref{new-cond-2} of Condition C-1,
the third term $2\sigma^2\sqrt{d_kd}/(T-h_0)$ does not affect the rate in \eqref{th-1-3}
as discussed below Theorem \ref{th-1}.}
To understand the
rates $\delta_0$, $\delta_1$ and $\delta_2$ better, consider
the rank-one case \eqref{example-1a} with $\sigma\asymp 1$. Then
\eqref{example-1b} gives
\bes
\|\bTheta^*_{k,0}\|_{\rm S} = \lam^2{\hat\rho}_0 \asymp d^{1-\delta_0},\quad
\lam_k^2 = \lam^2\big(\hbox{$\sum_{h=1}^{h_0}$}{\hat\rho}_h^2/h_0\big)^{1/2}
\asymp c_1d^{1-\delta_1}.
\ees
We note that $(1-h/T)|{\hat\rho}_h|\le {\hat\rho}_0$ by Cauchy-Schwarz, so that
it would be reasonable to expect $c_1d^{1-\delta_1}\le d^{1-\delta_0}$.
Let $\lam_{k,m}$ be the $m$-th singular value of $\bA_k$,
so that $\lam_{k,1}=\|\bA_k\|_{\rm S}$ and $\lam_{k,1}^2/\lam_{k,r_k}^2$ is the
condition number of $\bA_k^\top\bA_k$.
Let $\lam=\prod_{k=1}^K \lam_{k,1}$ as in Corollary \ref{cor-1}.
For fixed $\{r_1,\ldots,r_K,h_0,\sigma\}$ and under Condition B, \eqref{new-0} gives
$\|\bTheta_{k,0}\|_{\rm op}\vee \|\bTheta^*_{k,0}\|_{\rm S}\lesssim \lam^2$
and $\tau_{k,r_k}\gtrsim \prod_{k=1}^K \lam_{k,r_k}^2$ when $\E[\bPhi_{k,1:h_0}]$ is of rank $r_k$.
These lead to conservative bounds $d^{1-\delta_0}\lesssim \lam^2$
and $d^{1-\delta_1}\gtrsim \prod_{k=1}^K \lam_{k,r_k}^2$, and $\delta_0=\delta_1$ when the condition
numbers of $\bA_k^\top\bA_k$ are bounded.
When $K=1$ and $\delta_0=\delta_1$ our $\delta_0$ is comparable with the $\delta'$ in
\cite{Lam&2011}, and when $K=2$ and $\delta_0=\delta_1$ our $d^{\delta_0}$ is comparable with the $d_1^{\delta'_1}d_2^{\delta'_2}$
in \cite{Wang&2019}.

The rate for the eigen-gap in \eqref{new-cond-4} has the same interpretation as the rate in
\eqref{new-cond-3} as $\tau_{k,r_k+1}=0$.
Condition \eqref{new-cond-4} requires that spectrum distance between
$\barE\big[\mat_1\big({\rm TOPUP}_k\big)\big]$ and its expectation be
within $O(\epsilon_{d,T})$ of the eigen-gap.
It holds in model \eqref{lam-F-U}
when $\|\bPhi_{k,1:h_0} - \E[\bPhi_{k,1:h_0}]\|_{\rm S}$ is within $O(\epsilon_{d,T})$ of the
$m$-th eigen-gap of $\E[\bPhi_{k,1:h_0}]$. 
We leave to the existing literature
for the analysis of the low-dimensional $\bPhi_{k,1:h_0}$
as many options are available.

\begin{coro}\label{cor-2}
Suppose Conditions A and C-1 hold. Then
\bel{th-1-2a}
\Big\|\mat_1\big({\rm TOPUP}_k\big) - \barE\big[\mat_1\big({\rm TOPUP}_k\big)\big]\Big\|_{\rm S}
= O_P(\sigma^2 h_0^{1/2}\eta_k)
\eel
with $\eta_k
=\big(d^{1-\delta_0/2}/T^{1/2}\big)r^{3/4}/r_k + d/(d_k T)^{1/2} \chzM{+ \sqrt{d_kd}/T}$, and
\bel{th-1-3a}
\Big\|\widehat{\bP_{k, r_k}} - \bP_k\Big\|_{\rm S}
= O_P\Big(\frac{d^{\delta_1-\delta_0/2}}{T^{1/2}}\frac{r^{3/4}}{r_k^{1/2}}
+ \frac{r_k^{1/2}d^{\delta_1}}{(d_k T)^{1/2}}\Big)
\eel
for the estimator \eqref{TOPUP-P_km} with $m = r_k$.
\end{coro}




\vspace{0.1in}

The following corollary, a direct consequence of \eqref{th-1-2a} and
condition \eqref{new-cond-4} by Wedin (1972), provides convergence rate
for the estimation of singular-space for the top $m$ singular values.

\begin{coro}\label{cor-3} Suppose Conditions A and C-1 hold.
Let $\bP_{k,m} = \hbox{\rm PLSVD$_m$}\big(\bW_k\big)$
with the $\bW_k$ in \eqref{eq:WA},
and $\widehat{\bP_{k, m}}$ be as in \eqref{TOPUP-P_km} with integer $m\in [1,r_k)$. Then,
\bes
\Big\|\widehat{\bP_{k, m}} - \bP_{k,m}\Big\|_{\rm S}
= O_P\bigg(\frac{d^{\delta_2-\delta_0/2}r^{3/4}}{T^{1/2}r_k^{1/2}}
+ \frac{r_k^{1/2}d^{\delta_2}}{(d_kT)^{1/2}} + \epsilon_{d,T}\bigg)
\ees
with the $\epsilon_{d,T}$ in Condition C-1.
\end{coro}

\noindent

\smallskip

\noindent\medskip
    {\bf Theoretical property of TIPUP:\ }
We summarize our analysis of the TIPUP procedure in the following theorem.

\begin{theorem}\label{th-2}
Let $d=\prod_{k=1}^K d_k$ and $r=\prod_{k=1}^K r_k$, and $\big\|\bTheta^*_{k,0}\big\|_{\rm S}$
be as in \eqref{new-norm-2}.
Suppose Condition A holds.
Then, $\barE\big[\bV^*_{k,h}\big] = \bTheta^*_{k,h}$ and
\bel{th-2-1}
& \displaystyle \barE\big\|\bV^*_{k,h} - \bTheta^*_{k,h}\big\|_{\rm S}
\le \frac{2\sigma(8Td_k)^{1/2}}{T-h}\big\|\bTheta^*_{k,0}\big\|_{\rm S}^{1/2}
+ \chzM{\frac{\sigma^2\sqrt{8d}}{\sqrt{T-h}}+\frac{2\sigma^2d_k}{T-h}}
\\ \nonumber
& \displaystyle  \barE\big\|{\rm TIPUP}_k
- \barE\big[{\rm TIPUP}_k\big]\big\|_{\rm S}
 \le h_0^{1/2}\bigg\{ \frac{2\sigma(8Td_k)^{1/2}}{T-\chzM{h_0}}\big\|\bTheta^*_{k,0}\big\|_{\rm S}^{1/2}
+ \chzM{\frac{\sigma^2\sqrt{8d}}{\sqrt{T-h_0}}+\frac{2\sigma^2d_k}{T-h_0}}\bigg\}
\eel
for all $k$ \chzM{and $h_0\le T/4$.} 
Moreover,
\bel{th-2-3}
\barE\Big\|\widehat{\bP_{k, r_k}} - \bP_k\Big\|_{\rm S}
\le 2\big(\lam^*_k\big)^{-2}
\chzM{\bigg\{\frac{(2+1/16)\sigma(8Td_k)^{1/2}}{T-h_0}\big\|\bTheta^*_{k,0}\big\|_{\rm S}^{1/2}
+ \frac{\sigma^2\sqrt{8d}}{\sqrt{T-h_0}}\bigg\}}
\eel
for the estimator \eqref{TIPUP-P_km} with $m = r_k$, where $\lam^*_k$ is as in \eqref{new-norm-3d}.
\end{theorem}

\noindent
The proof of the theorem is shown in Appendix A.

Again consider the one-factor model \eqref{example-1a} with $r=r_k=1$,
\bes
\cX_t = \lambda f_t\big(\bu_1\otimes \cdots\otimes \bu_K) + \cE_t,\quad
\bTheta^*_{k,h} = \lam^2{\hat\rho}_h\bu_k\bu_k^\top,
\ees
for some unit vectors $\bu_k\in\R^{d_k}$ and ${\hat\rho}_h = \sum_{t=h+1}^T f_{t-h}f_t/(T-h)$.
By \eqref{new-norm-3c} and \eqref{new-norm-3d},
we have
\bel{example-1d}
\big\|\bTheta^*_{k,0}\big\|_{\rm S}
= \lam^2 {\hat\rho}_0,\quad
(\lam^*_k)^2 = \lam_k^2 = \lam^2\bigg(\frac{1}{h_0}\sum_{h=1}^{h_0}{\hat\rho}_h^2\bigg)^{1/2}
\eel
as in \eqref{example-1b}.
Thus, for the TIPUP estimator of $\bu_k$ \eqref{TIPUP-P_km}, Theorem \ref{th-2} gives
\bel{example-1e}
\barE\sqrt{1-(\hbu_k^\top\bu_k)^2}
\le \frac{C_1\sigma d_k^{1/2}\lam{\hat\rho}_0^{1/2}}{\lam_k^2 T^{1/2}}
+ \frac{C_1\sigma^2d^{1/2}}{\lam_k^2 T^{1/2}}
\lesssim \frac{d_k^{1/2}}{\lam T^{1/2}} + \frac{d^{1/2}}{\lam^2 T^{1/2}}
\eel
when $\sigma\asymp 1$ and $\sum_{h=1}^{h_0}{\hat\rho}_h^2/h_0\asymp {\hat\rho}_0^2\asymp 1$.
We note that the convergence rate in \eqref{example-1e} 
is faster than the rate for the TOPUP in \eqref{example-1c} 
since there is no signal cancellation in TIPUP in the one-factor model and
$d_k$ is typically much smaller than $d$.
For general fixed $r$, we have the following corollary.

\begin{coro}\label{cor-4}
Let $\bPhi^*_{k,1:h_0}=\big(\bPhi^*_{k,1},\ldots,\bPhi^*_{k,h_0}\big)\in \R^{r_k\times (r_k h_0)}$
with the $\bPhi^*_{k,h}$ in \eqref{Phi-Theta-star}. Let $\cX_t$ be as in \eqref{lam-F-U}.
Suppose $\sigma, r_1,\ldots,r_K$ are fixed, Conditions A and B hold,
and $\E[\bPhi^*_{k,1:h_0}]$ is of rank $r_k$. Then,
\bel{cor-4-1}
& \barE \big\|\bV^*_{k,h} - \bTheta^*_{k,h}\big\|_{\rm S}
\lesssim (d_k/T)^{1/2}\lam + (d/T)^{1/2} \chzM{+ d_k/T,}
\\ \nonumber &
\barE \big\|{\rm TIPUP}_k
- \barE\big[{\rm TIPUP}_k\big]\big\|_{\rm S}
\lesssim (d_k/T)^{1/2}\lam + (d/T)^{1/2} \chzM{+ d_k/T,}
\eel
for all $k$ and $h\le h_0$. Moreover,
\bel{cor-4-2}
\barE\Big\|\widehat{\bP_{k, r_k}} - \bP_k\Big\|_{\rm S}
\lesssim \frac{d_k^{1/2}}{\lam T^{1/2}}+\frac{d^{1/2}}{\lam^2 T^{1/2}}.
\eel
\end{coro}

We may also count the dimensions and sum 
in the same way as in \eqref{new-cond-3}. This leads to
\bes
\Big\|\barE\big[ {\rm TIPUP}_k\big]\Big\|_{HS}^2 \asymp \sigma^4 h_0 (d_{-k}d_k^2)^{1-\delta_3}
\ees
for some $\delta_3\ge \delta_0$ and rank$({\rm TIPUP}_k)=r_k$, so that
\bel{new-cond-5a}
h_0^{1/2}\big(\lam^*_k\big)^2 = \tau^*_{k,r_k}
\ge c_3\sqrt{\sigma^4 h_0 (d_{-k}d_k^2)^{1-\delta_3}/r_k}
\ge c_3\sigma^2\sqrt{h_0/r_k} (d_kd)^{(1-\delta_3)/2}>0
\eel
for some $c_3>0$.
In the one-factor model \eqref{example-1a} with $\sigma\asymp 1$,
\eqref{new-cond-5a} and \eqref{example-1d} are connected via
$\lam^2{\hat\rho}_0 \asymp d^{1-\delta_0}$
and $(\lam^*_k)^2 \asymp c_1d^{1-\delta_1}$ as in TOPUP.
Compared with \eqref{new-cond-3}
we expect $\delta_3 \ge \delta_1$ in general due to possible signal cancellation
in the inner-product, and we may take $\delta_3 = \delta_1$
in the absence of signal cancellation (e.g. one-factor model with $r=1$) or when
the signal cancellation does not change rates (e.g. as in Corollary \ref{cor-4}).
The counterpart of Condition C-1,
summarizing the expected implications of Condition B on the TIPUP in a general scenario,
is given as follows.

\vspace{0.1in}

\noindent
{\bf Condition C-2:} For the norm in \eqref{new-norm-2} and the matrix
$\barE\big[{\rm TIPUP}_k\big]$,
\bes
{\mathbb{P}}\Big\{\
\hbox{ \eqref{new-cond-2} and \eqref{new-cond-5a} hold } \Big\} = 1 + o(1).
\ees
Moreover, for certain constants $\delta_4\ge \delta_3$, $c_4>0$ and
$\epsilon_{d,T}=o(1)$,
\bel{new-cond-6}
{\mathbb{P}}\left.\begin{cases}\
\tau^*_{k,m}-\tau^*_{k,m+1}
\ge c_4\sigma^2\sqrt{h_0/r_k} (d_kd)^{(1-\delta_4)/2}
\cr\ \Big\|\barE\big[{\rm TIPUP}_k\big]
- \E\big[{\rm TIPUP}_k\big]\Big\|_{\rm S}
\le \epsilon_{d,T}\sigma^2\sqrt{h_0/r_k} (d_kd)^{(1-\delta_4)/2}
\end{cases}\right\} \to 1
\eel
whenever the eigen-gap in \eqref{new-cond-6}
is invoked for some integer $m\in [1,r_k)$.

\begin{coro}\label{cor-5}
Suppose Conditions A and C-2 hold. Then
\bel{th-2-2a}
\Big\|{\rm TIPUP}_k - \barE\big[{\rm TIPUP}_k\big]\Big\|_{\rm S}
= O_P(\sigma^2 h_0^{1/2}\eta^*_k)
\eel
with $\eta^*_k = \sigma^2(d_k/T)^{1/2}d^{(1-\delta_0)/2}r^{1/4}/r_k^{1/2}
+ \sigma^2({d}/T)^{1/2} \chzM{+\sigma^2d_k/T}$, and
\bel{th-2-3a}
\Big\|\widehat{\bP_{k, r_k}} - \bP_k\Big\|_{\rm S}
= O_P\bigg(\frac{d_k^{\delta_3/2}d^{(\delta_3-\delta_0)/2}r^{1/4}}{T^{1/2}}
+ \frac{r_k^{1/2}(d_kd)^{\delta_3/2}}{(d_kT)^{1/2}}\bigg)
\eel
for the estimator \eqref{TIPUP-P_km} with $m = r_k$.
\end{coro}

\smallskip
Compared with the error bound \eqref{th-1-3} for TOPUP, we observe that
\eqref{th-2-3} provides sharper error bounds when $\delta_3=\delta_1$
as it turns some fraction power of $d$ and $r$ into that of $d_k$ and $r_k$
in the numerator. However, as discussed below \eqref{new-cond-5a},
$\delta_3=\delta_1$ may not materialize
when signal cancellation in the inner product in \eqref{TIPUP-k}
changes the rates compared with that of the TOPUP in \eqref{TOPUP-k}.

The following corollary, which is a direct consequence of \eqref{th-2-2a} and
condition \eqref{new-cond-6} by Wedin (1972), provides convergence rate
for the estimation of singular-space for the top $m$ singular values.

\begin{coro}\label{cor-6} Suppose Conditions A and C-2 hold.
Let $\bP^*_{k,m} = \hbox{\rm PLSVD$_m$}\big(\bW^*_k\big)$
with the $\bW^*_k$ in \eqref{W_k-star}
and
$\widehat{\bP_{k, m}}$ be as in \eqref{TIPUP-P_km} with integer $m\in [1,r_k)$. Then,
\bes
\Big\|\widehat{\bP_{k, m}} - \bP^*_{k,m}\Big\|_{\rm S}
= O_P\bigg(\frac{d_k^{\delta_4/2}d^{(\delta_4-\delta_0)/2}r^{1/4}}{T^{1/2}}
+ \frac{r_k^{1/2}(d_kd)^{\delta_4/2}}{(d_kT)^{1/2}} + \epsilon_{d,T}\bigg)
\ees
with the $\epsilon_{d,T}$ in Condition C-2.
\end{coro}

We note that while $\bP_{k,m} = \bP_{k,m}^*$ for $m=r_k$,
the two projections are not the same in general for $1\le m< r_k$
as discussed in the paragraphs below \eqref{eq:WA} and \eqref{eq:WA-star2}.

\vspace{0.1in}

\noindent
{\bf A comparison of TOPUP and TIPUP:}
It is worthwhile to mention here that the rates for
the TIPUP and TOPUP do not dominate each other.
This is expected as the methods are constructed in different ways,
with the inner product in \eqref{TIPUP-k} for the TIPUP and
the tensor (outer) product in \eqref{TOPUP-k} for the TOPUP.
The TIPUP, which features noise cancellation in the inner-product operation,
has a clear advantage in the one-factor model
\eqref{example-1a} in view of \eqref{example-1c} and \eqref{example-1e}
as the model does not have enough flexibility to allow signal cancellation.
In the more general setting, the effects of noise cancellation and possible signal cancellation
are expressed in
the rate $(d_kd)^{\delta_3/2}T^{-1/2}(d^{-\delta_0/2} + d_k^{-1/2})$
in \eqref{th-2-3a} for the TIPUP in the case of bounded rank $r$, in comparison with
the rate $d^{\delta_1}T^{-1/2}(d^{-\delta_0/2} + d_k^{-1/2})$
in \eqref{th-1-3a} for the TOPUP.
Writing $(d_kd)^{\delta_3/2} = (d_k/d)^{\delta_3/2}d^{\delta_3-\delta_1}d^{\delta_1}$,
we may think of factors $(d_k/d)^{\delta_3/2}$ and $d^{\delta_3-\delta_1}$ respectively as
quantifications of the benefit of noise cancellation and the impact of signal cancellation,
as $\delta_3\ge\delta_1\ge 0$ by assumption.
We note that our result for the TOPUP is sharper than the rate
$d/(\lam^2T^{1/2}) \asymp d^{\delta_0}/T^{1/2}$ in \cite{Wang&2019}
as $d^{\delta_0}$ is equivalent to their $d_1^{\delta_1'}d_2^{\delta_2'}$
and their condition implies $\delta_0\in [0,1]$.
In the simplest rank one case for third-order tensor,
our error bound compares favorably with those of order $O_p(d^{3/4}/\lambda)$
obtained recently for the PCA by \cite{Hopkins+2015}.


\vspace{0.1in}

\noindent
{\bf Remark: One-step estimator:}
Let $\bU_k\in\R^{d_k\times r_k}$ be
orthonormal matrices satisfying $\bU_k\bU_k^\top = \bP_k$ as a version of
the left singular matrix of $\bA_k$.
A crucial step in our investigation is to estimate the ``loading matrices"
$\bU_k$, akin to PCA. Once a consistent estimator of $\bU_k$ is constructed,
sharper estimate of them
and the factor model itself can be investigated based on the much smaller
tensor times series. For example, $\bU_1$ can be estimated based on
$\bX_t \times_2 \hbU_2^\top \times_3\ldots \times_k\hbU_k^\top$.
This may lead to significant rate improvement, without using
the popular power iteration methods \citep{KoldaBader09}.
Under proper sample size and signal strength conditions as indicated in Theorems \ref{th-1} and \ref{th-2},
the TOPUP and TIPUP also provide consistent initializations to ensure
the convergence of power iteration methods to a correct solution
among potentially exponentially many
local optima \citep{Auffinger+2013}.
More investigation is needed to study the property of such
one-step and power estimators.

\vspace{0.1in}

\noindent
{\bf Remark: Iterative procedures:}
 Although the one-step estimators are already rate-optimal theoretically,
iterative procedures are shown to have better performance numerically.
Our preliminary empirical results show that
the iterative algorithms significantly improve the estimation accuracy over
their non-iterative counterparts.
Again, more investigation is needed to
study the property of such an estimator.
We note that
in the traditional tensor decomposition problem, the contraction
property can be obtained in parallel to those of tensor power method or
alternating least squares \citep{Golub1996,Anandkumar2014}.

\vspace{0.1in}

\noindent
{\bf Remark: Comparison with traditional tensor decomposition:}
It is well known that, for standard PCA with i.i.d. vectors from distribution $N({\bf 0}, \lam^2 \bu\bu^\top + \bI_{d\times d})$ with $\|\bu\|_2=1$, the
convergence rate of the risk of the loading matrix is
$\sqrt{d/T}(1/\lambda+1/\lambda^2)$, which matches the rate \eqref{example-1c} for the TOPUP (for vector time series, i.e. $K=1$, $d=d_1$) and the rate \eqref{example-1e} for the TIPUP (for $K=1$).  This common rate is faster than the rate obtained in \cite{Lam&2011} for vector time series.
However, in the spiked-PCA model the noise has an identity covariance matrix
but in the factor time series model we consider here
the noise, although white,
can have arbitrary contemporary covariance structure.
This arbitrary noise covariance matrix makes the eigenvectors of the sample
covariance matrix inconsistent with the loading matrices. The use
of the auto-covariance matrix solves the problem.

\vspace{0.1in}




\section{Simulation results}
In this section we present some empirical study on the performance of the
estimation procedures, with various experimental configurations.
We also check the performance of a standard tensor decomposition
procedure which incorporates time as an additional tensor dimension,
and treats the factor as deterministic
without temporal structure. The loading
matrices is then estimated using SVD of the mode-1 (or $2, 3$) matricization
of the expanded tensor $\cY$ to estimate the column space of $\bA_1$ (or $\bA_2$, $\bA_3$).
We will call it the unfolding procedure ({\bf UP}). The main difference
between UP and the estimators TIPUP and TOPUP is that UP does not
incorporate the assumption that the noise is white, while the TIPUP and
TOPUP take full advantage of that assumption.


We demonstrate the finite sample performance under a matrix factor model
setting. We start with a simple setting. Let
\[
\bX_t = \lambda \bu_1f_t\bu_2' + \bE_t
\]
where $\bX_t$ and $\bE_t$ are in $\R^{d_1\times d_2}$,
$\bu_1\in\R^{d_1\times 1}, \bu_2\in\R^{d_2\times 1}$ with $\|\bu_1\|_2=\|\bu_2\|_2=1$,
and
the factor $f_t$ is a univariate time series following $f_t\sim AR(1)$
with AR coefficient $\phi=.6$ and standard $N(0,1)$ noise.
The noise $\bE_t$ is white, i.e.
$\bE_t\perp\bE_{t+h},h>0$ and $\bE_t=\Psi_1^{1/2}\bZ_t\Psi_2^{1/2}$
where $\Psi_1 =\Psi_2$ are the column and row covariance matrices with
the diagonal elements being $1$ and all off diagonal elements being $0.2$. All
elements in the $d_1\times d_2$ matrix $\bZ_t$ are i.i.d $N(0,1)$.
 The
elements of the loadings $\bu_1$ and $\bu_2$ are generated from i.i.d $N(0,1)$,
then normalized so $\|\bu_1\|_2=\|\bu_2\|_2=1$. The sample size $T$, the
dimensions $d_1, d_2$ and the factor strength $\lambda$ are chosen to be
$T = 2^\ell$ for $\ell=1,\ldots,15$, $d_1=d_2 = 2^\ell$ for $\ell=1,\ldots, 6$
and $\lambda = 2^\ell$ for $\ell=-7,-6,\ldots,7$.

\begin{figure}
  \centering
  \includegraphics[height=.3\textheight]{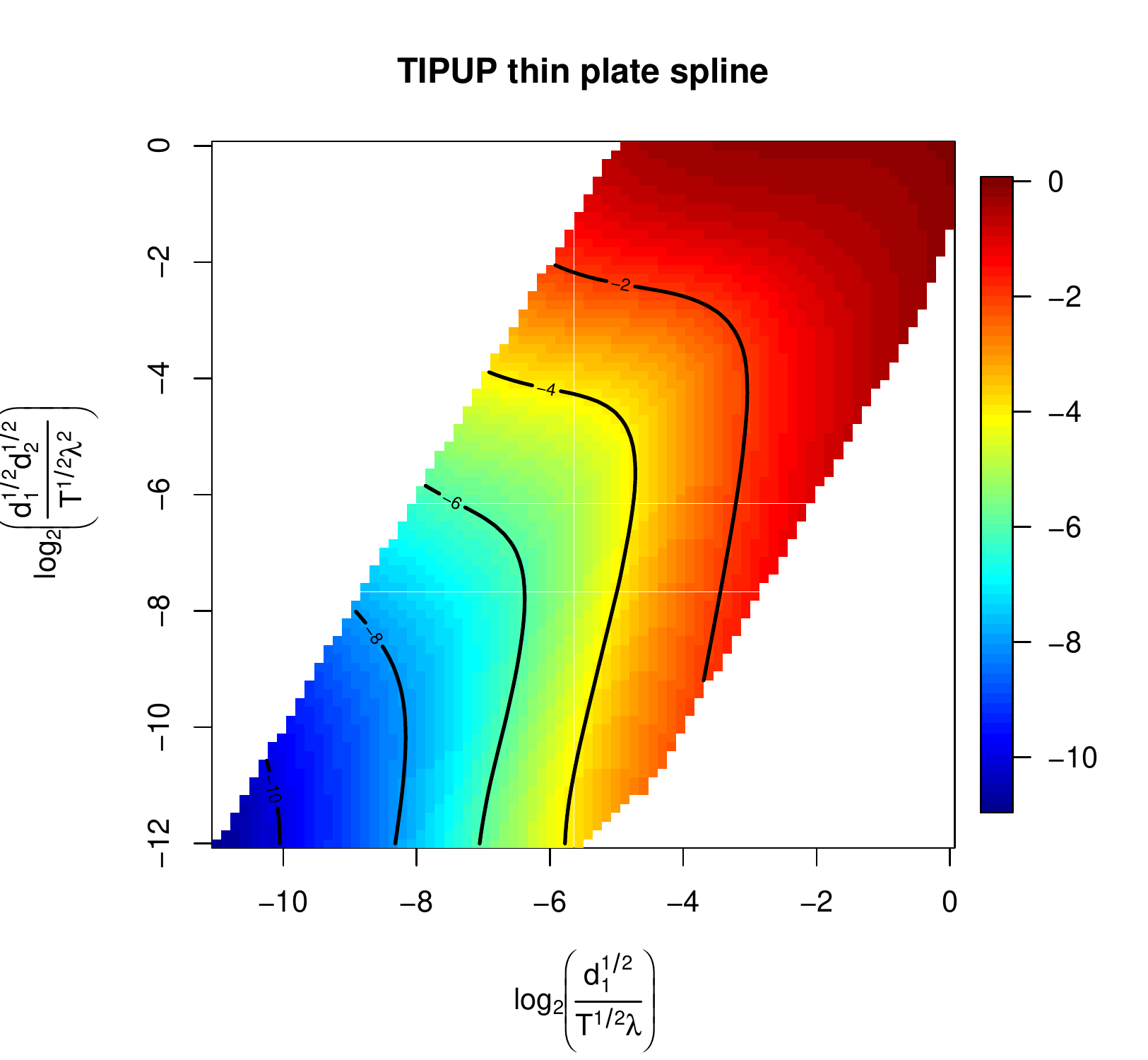}
  \includegraphics[height=.3\textheight]{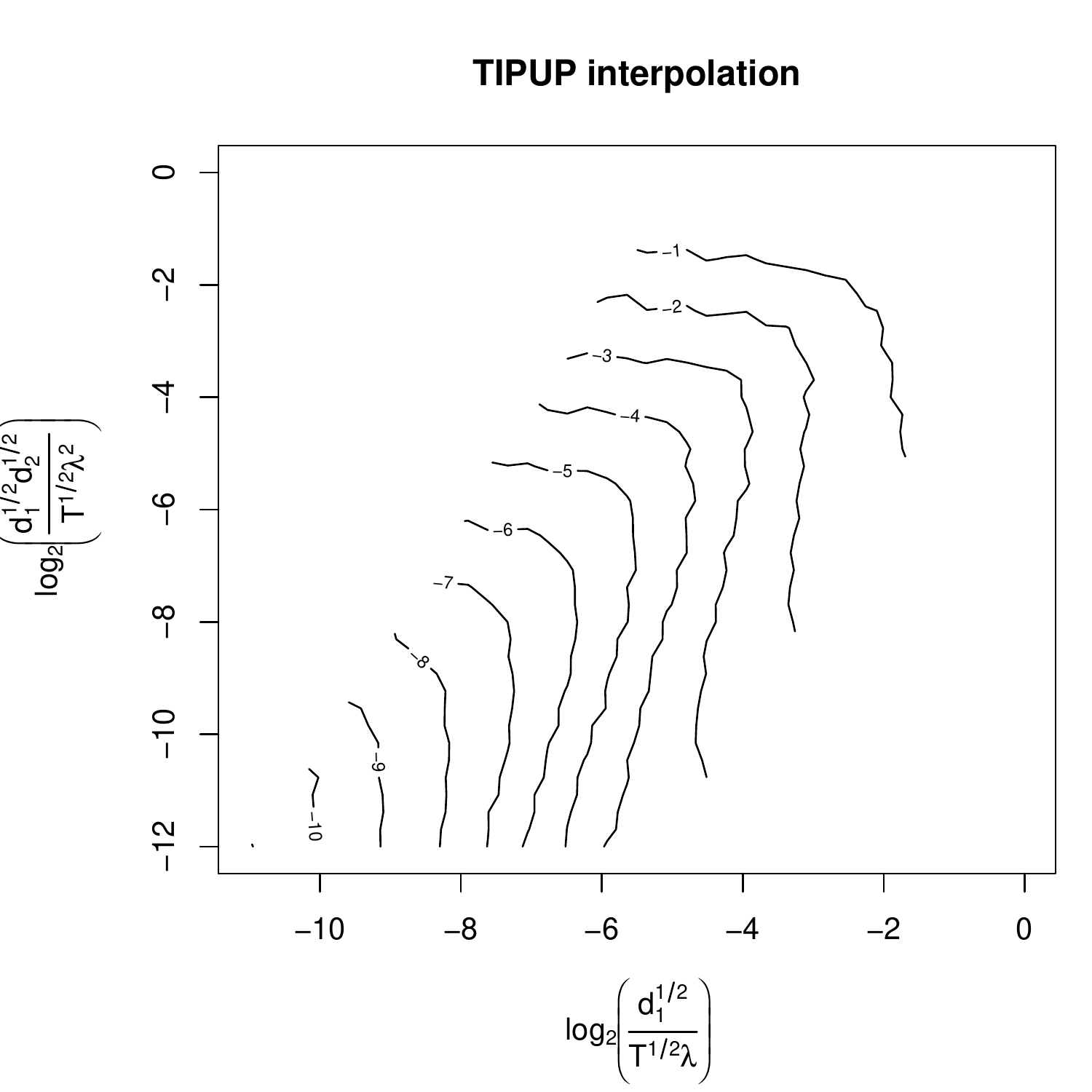}
  \caption{Logarithm of the average loss of estimating $\bu_1$ using the TIPUP
    vs the logarithms of the two components in the convergence rate equation (\ref{simu.rate1})}
    \label{simu.fig1}
\end{figure}

By Theorem 2 and \eqref{example-1e}, the rate for estimating $\bu_1$ via the
TIPUP is
\begin{equation}
    \frac{d_1^{1/2}}{T^{1/2}\lambda}+\frac{(d_1d_2)^{1/2}}{T^{1/2}\lambda^2}.
\label{simu.rate1}
\end{equation}
Let $x=log_2\left(\frac{d_1^{1/2}}{T^{1/2}\lambda}\right)$, $y = log_2\left(\frac{(d_1d_2)^{1/2}}{T^{1/2}\lambda^2}\right)$, and $z$ be the
logarithm of the average of the
corresponding estimation loss, $L=\sqrt{1-(\bu_1^\top\hat\bu_1)^2}=\|\bu_1\bu_1^\top-\hat\bu_1\hat\bu_1^\top\|_{\rm S}$ as in \eqref{example-1e},
of estimating $\bu_1$ over 100
simulation runs. A thin plate
spline fit of $(x,y,z)$ under different $T,d_1,d_2,\lambda$
leads to the left panel of Figure~\ref{simu.fig1} and its
interpolation with the mean leads to the right panel.

The figures clearly confirm the theoretical results. Note that, when the
two terms in (\ref{simu.rate1}) are of different rates (off 45 degree line
in the figure), one of the rates would dominate hence the contour lines of the
error rate should be
either horizontal or vertical in the figure as $(x,y)$ moves away from the
45 degree line.
For negative AR coefficient $\phi=-0.6$ in the factor dynamics,
the results are similar.

We also considered a parametric fit of the TIPUP error rate.
Specifically, we fit the following model to the average loss $L$ of estimating $\bu_1$:
\be \label{simu_fit}
\log_2(L) \sim \log_2(c_12^{\nu_1}+c_62^{\nu_2}),
\ee
where
\begin{eqnarray*}
\nu_1& = & c_2\log_2d_1+c_3\log_2d_2+c_4\log_2\lambda+c_5\log_2T \\
\nu_2& = & c_7\log_2d_1+c_8\log_2d_2+c_9\log_2\lambda+c_{10}\log_2T.
\end{eqnarray*}
and compared the
empirical fit with the theoretical results in Theorem 2. The results are
shown in Table~\ref{simu.table1}.
They are reasonably close.

\begin{table}
\begin{center}
  \begin{tabular}{c|cccccccccc}
  &  $c_1$ & $c_2$ & $c_3$ & $c_5$ & $c_5$ & $c_6$ & $c_7$ & $c_8$ &
        $c_9$ & $c_{10}$ \\ \hline
Thm 2 &  & 0.50 & 0.00 & -1.00 & -0.50 &  & 0.50 & 0.50 & -2.00 & -0.50\\
fitted & 1.19 &0.51 &-0.06 &-0.73 & -0.65 & 0.36 & 0.72 & 0.58 & -1.92 & -0.45
  \end{tabular}
  \caption{Comparison between the theoretical rates and the simulated rates
  for the TIPUP procedure.}
  \label{simu.table1}
\end{center}
\end{table}

\begin{figure}
  \centering
  \includegraphics[height=.3\textheight]{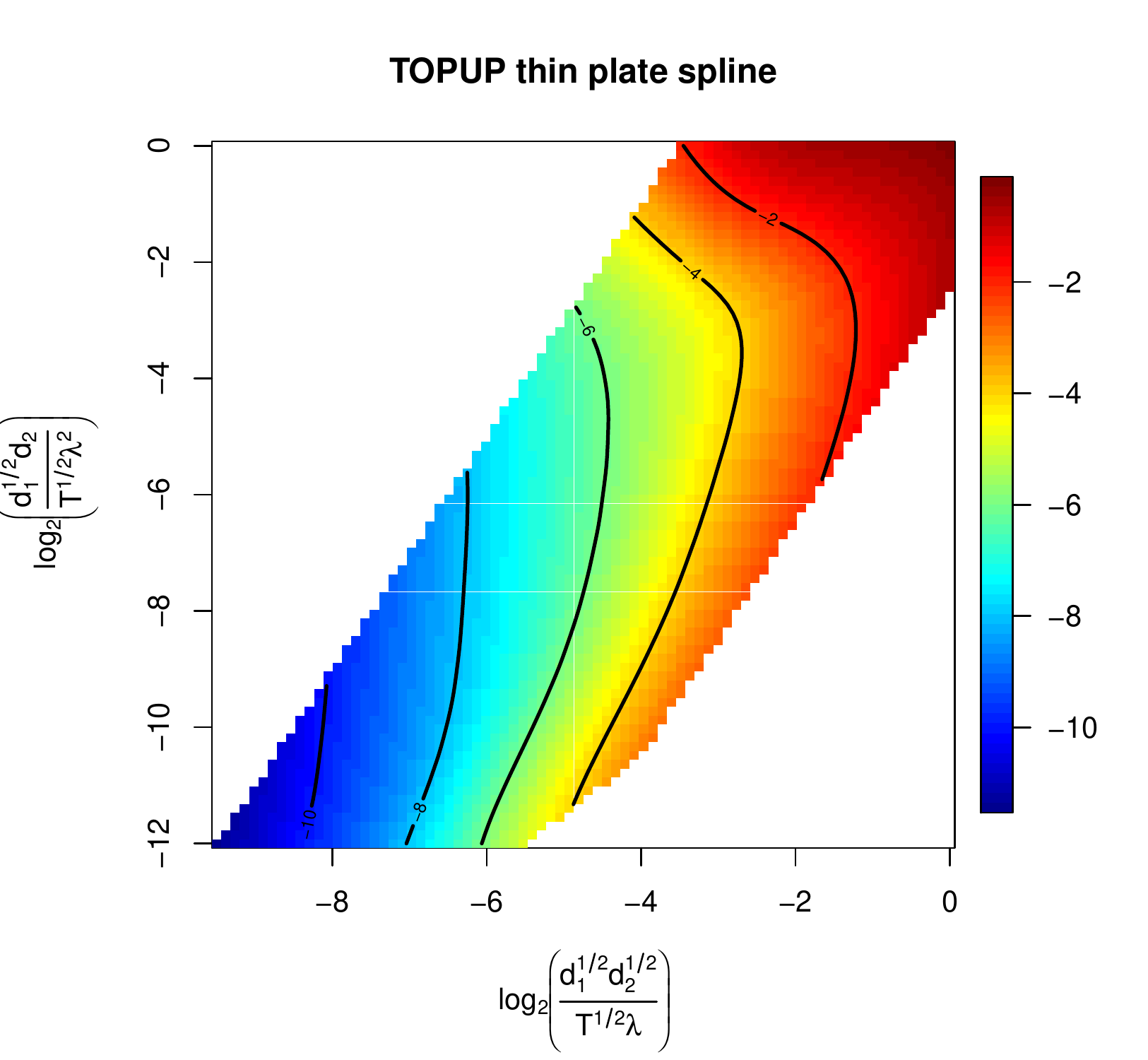}
  \includegraphics[height=.3\textheight]{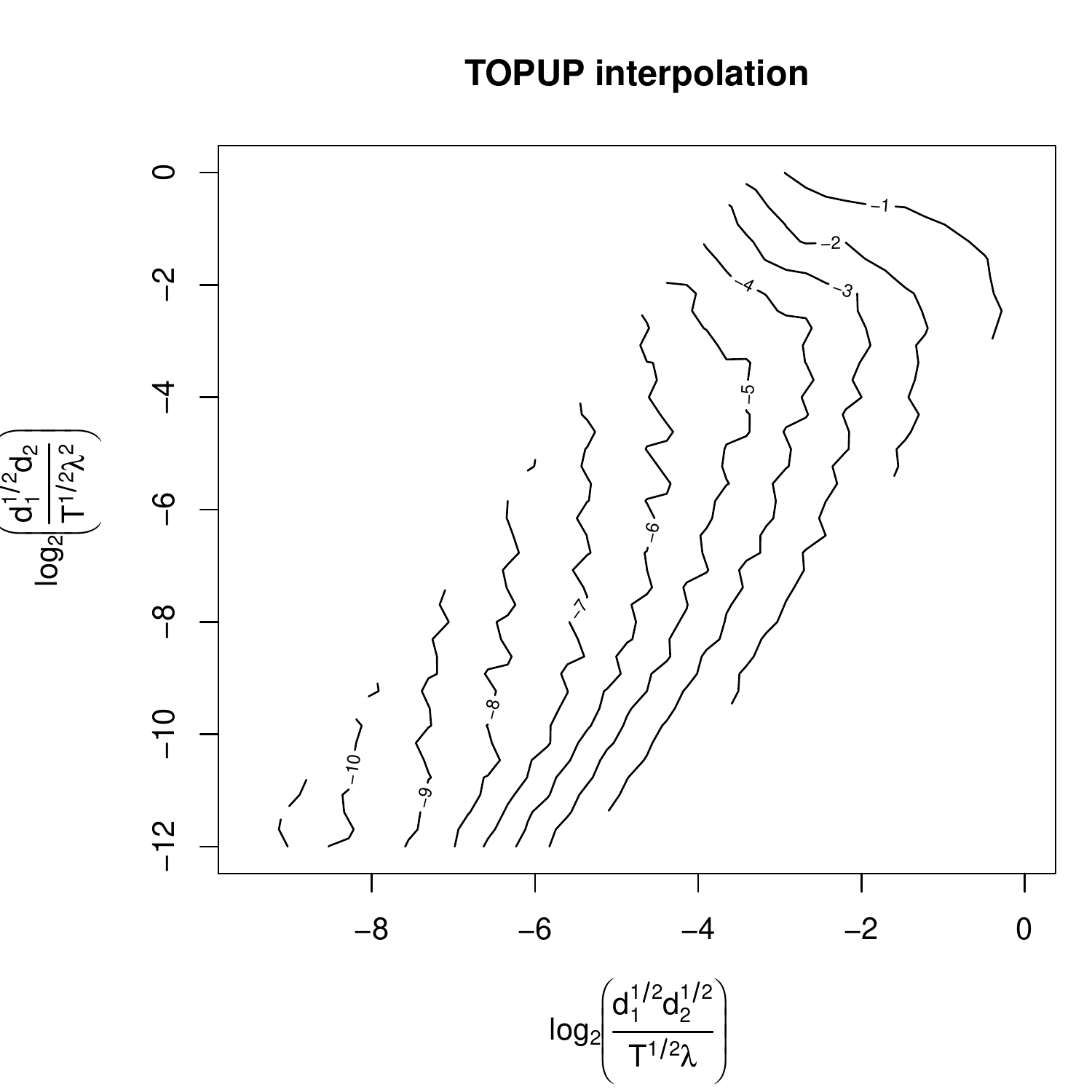}
  \caption{Logarithm of the average loss of estimating $\bu_1$ using the TOPUP
    vs the logarithms of the two components in the convergence rate equation (\ref{simu.rate2})}
    \label{simu.fig2}
 \end{figure}

For TOPUP, the convergence rate based on Theorem 1 and (\ref{example-1c})
in this case is
\be
    \frac{(d_1d_2)^{1/2}}{T^{1/2}\lambda}+\frac{(d_1d_2^2)^{1/2}}{T^{1/2}\lambda^2}.
\label{simu.rate2}
    \ee
Similarly, let $x=log_2\left(\frac{(d_1d_2)^{1/2}}{T^{1/2}\lambda}\right)$
and $y=log_2\left(\frac{(d_1d_2^2)^{1/2}}{T^{1/2}\lambda^2}\right)$
and $z$ be the
logarithm of the average of corresponding estimation error over 100 runs.
Figure~\ref{simu.fig2} shows the results. The picture is not as clean
as that of the TIPUP estimator, but it shows the trend.

Again, we fit the estimation error using (\ref{simu_fit})
and compared it with the theoretical result in (\ref{simu.rate2}), shown
in Table~\ref{simu.table2}.
There is some discrepancy though, possibly due to the limited range of the
simulation setting. The results for multiple rank cases and three dimensional
tensor time series show similar patterns.

\begin{table}
  \begin{center}
    \begin{tabular}{c|cccccccccc}
  &  $c_1$ & $c_2$ & $c_3$ & $c_5$ & $c_5$ & $c_6$ & $c_7$ & $c_8$ &
    $c_9$ & $c_{10}$ \\ \hline
    Thm 1 & & 0.50 & 0.50 &-1.00 & -0.50 & & 0.50 & 1.00 & -2.00 & -0.50\\
    fit & 0.83 & 0.51 & -0.16 & -0.66 & -0.55 & 0.36 & 1.21 & 1.03 & -2.51 &
    -0.70
  \end{tabular}
    \caption{Comparison between the theoretical rates and the simulated rates
    for the TOPUP procedure.}
  \label{simu.table2}
  \end{center}
\end{table}

\begin{figure}
  \centering
  \includegraphics[height=3.0in]{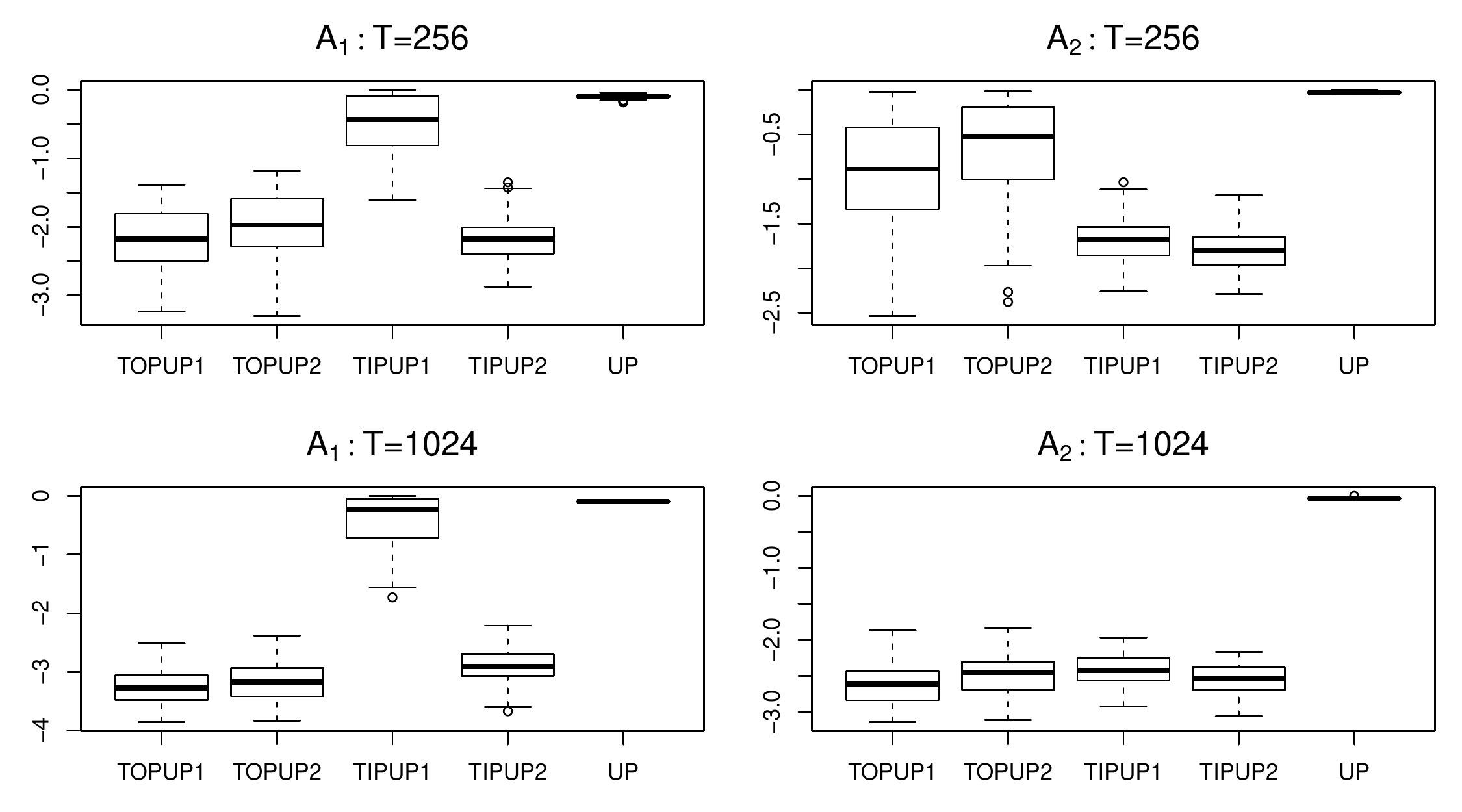}
  \caption{Finite sample comparison between TIPUP, TOPUP and UP with
  different $h_0$. $\phi_1=0.8$, $\phi_2=-0.8$ }
      \label{simu.fig3}
 \end{figure}

\begin{figure}
  \centering
  \includegraphics[height=3.0in]{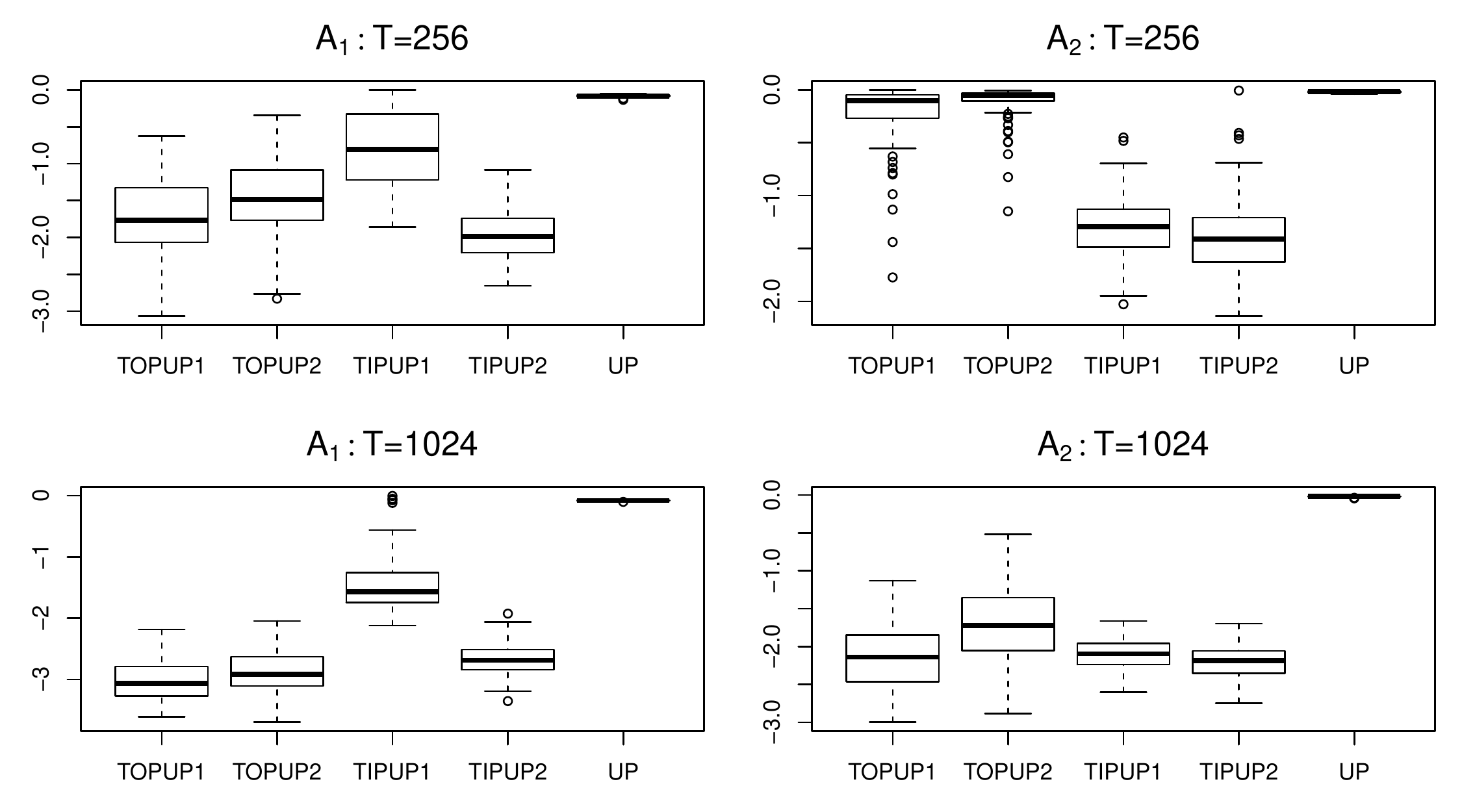}
  \caption{Finite sample comparison between TIPUP, TOPUP and UP with
    different $h_0$. $\phi_1=0.8$, $\phi_2=-0.7$ }
      \label{simu.fig4}
 \end{figure}

To compare the performance of different methods in finite samples, we
generated observations from the following two dimensional model
\[
\bX_t=2\bA_1\bF_t\bA_2^\top+\bE_t
\]
where $\bF_t=[f_{1t},f_{2t}]$ is a $1\times 2$ factor, with two
independent AR(1) processes $f_{it}=\phi_if_{it-1}+e_{it}$. The noise
$\bE_t$ is generated the same way as the simulation in the rank one case.
The elements of the loadings $\bA_1$ (a $d_1\times 1$ matrix) and
$\bA_2$ (a $d_2\times 2$ matrix) are generated from i.i.d N(0,1), then
normalized so that $||\bA_1||=1$ ($\bA_1$ is vector)
and $\bA_2$ is orthonormal through QR
decomposition. We use dimension $d_1=d_2=16$ here.
Figures~\ref{simu.fig3} and \ref{simu.fig4}
show the comparison
of the estimation methods, using boxplots of the logarithm of the estimation error in 100
simulation runs. The estimation error of $\bA_1$ is calculated the same way as that in
the rank one case (since $\bA_1$ is a vector).
The estimation error of $\bA_2$ is the spectral norm of the difference
between $\hat{\bA}_2(\hat{\bA}_2^\top\hat{\bA}_2)^{-1}\hat{\bA}_2^\top$
  and $\bA_2(\bA_2^\top\bA_2)^{-1}\bA_2^\top$, i.e., the difference of the
    two projection matrices.
TOPUP1 and TOPUP2 denote the results using the TOPUP method
with $h_0=1$ and $h_0=2$, respectively.
Similarly, TIPUP1 and TIPUP2 denote the results using the TIPUP method
with $h_0=1$ and $h_0=2$, respectively. UP denotes the results using
simple tensor decomposition. The left panel is for estimating $\bA_1$
and the right panel for $\bA_2$.

Figure~\ref{simu.fig3} shows the results of
using $\phi_1=0.8$ and $\phi_2=-0.8$ in the AR processes of the factors and
sample sizes $T=256$ and $1024$. Note that with
$\phi_1=0.8$ and $\phi_2=-0.8$, we have
\[
\E[\bF_t\bF_{t-1}^\top]=(\phi_1+\phi_2)\sigma_f^2=0.
\]
It violates the condition for the TIPUP in estimating $\bA_1$.
Essentially the signal in
$\sum_{t=h+1}^T\bX_t\bX_{t-h}^\top$, ($h=1$), completely
cancelled out in the TIPUP procedure
in estimating $\bA_1$ when $h_0=1$. Hence the results of TIPUP1 in
the two left panels in
Figure~\ref{simu.fig3} are significantly worse than the respective
TOPUP1. On the other hand, the cancellation does not happen fully with
$h_0=2$, because in this case,
$\E[\bF_t\bF_{t-2}^\top]=(\phi_1^2+\phi_2^2)\sigma_f^2>0$ for the $h=2$ term.
Hence
TIPUP2 is comparable to that of TOPUP2 and TOPUP1. However, notice that
when the sample size $T$ is larger, TIPUP2 is slightly worse that
TOPUP2 and TOPUP1, due to the fact that the cancellation makes the signal
weaker, and lag 2 autocorrelation is also weaker than lag 1. The TOPUP does
not have such a cancellation
problem since it is based on the sum of the squares of column-wise
autocovariance. Note that the cancellation problem should not be very
common in practice. For example,
there is no cancellation for estimating $\bA_2$ when using the
TIPUP in this setting, since
$\E[\bF_t^\top\bF_{t-1}]$ is a full rank matrix. And since the TIPUP in general has a
faster convergence rate, its performance is better than that of the TOPUP,
especially for small sample sizes, as shown in the right panel of Figure~\ref{simu.fig3}.

Figure~\ref{simu.fig4} shows the results of
using $\phi_1=0.8$ and $\phi_2=-0.7$ in the AR processes of the factors.
Here although $\E[\bF_t\bF_{t-1}^\top]$ is not zero, TIPUP1 with $h_0=1$ for estimating $\bA_1$ is still worse than TOPUP due to the partial cancellation, though not as severe as that in the complete cancellation $\phi_2=-0.8$ case.

The UP procedure is always the worst, due to the contemporary
correlation in $\bE_t$. Simulations using other settings show
similar results.

\section{Applications}
\subsection{Tensor factor models for import-export transport
networks} \label{example1}

Here we analyze the multi-category import-export network data
as illustrated in Figure~\ref{fig1}.
The dataset contains the monthly total export among 22 countries in North American and Europe
in 15 product categories from January 2010 to December 2016 (length 84),
so that the original dataset can be viewed as
a 4 way tensor of dimension $22\times 22\times 15\times 84$, with missing
value for the total export from any country to itself. For simpliciy, we treated the missing diagonal values as zero in the analysis. More sophisticated imputation can be implemented. 
The details of the data, countries and product categories
are given in Appendix B.
Following \cite{linnemann1966econometric},
to reduce the effect
of incidental transactions of large trades or unusual shipping
delays, a three-month moving average
of the series is used, so that $\cX_t \in \R^{22\times 23\times 15}$ with $t=1,\ldots, 82$.
Each element $x_{i,j,k,t}$ is the three month moving average
of total export from country $i$ to country $j$ in category $k$ in the
$t$-th month.

\begin{figure}
\centering
\includegraphics[width=3in,height=3in,]{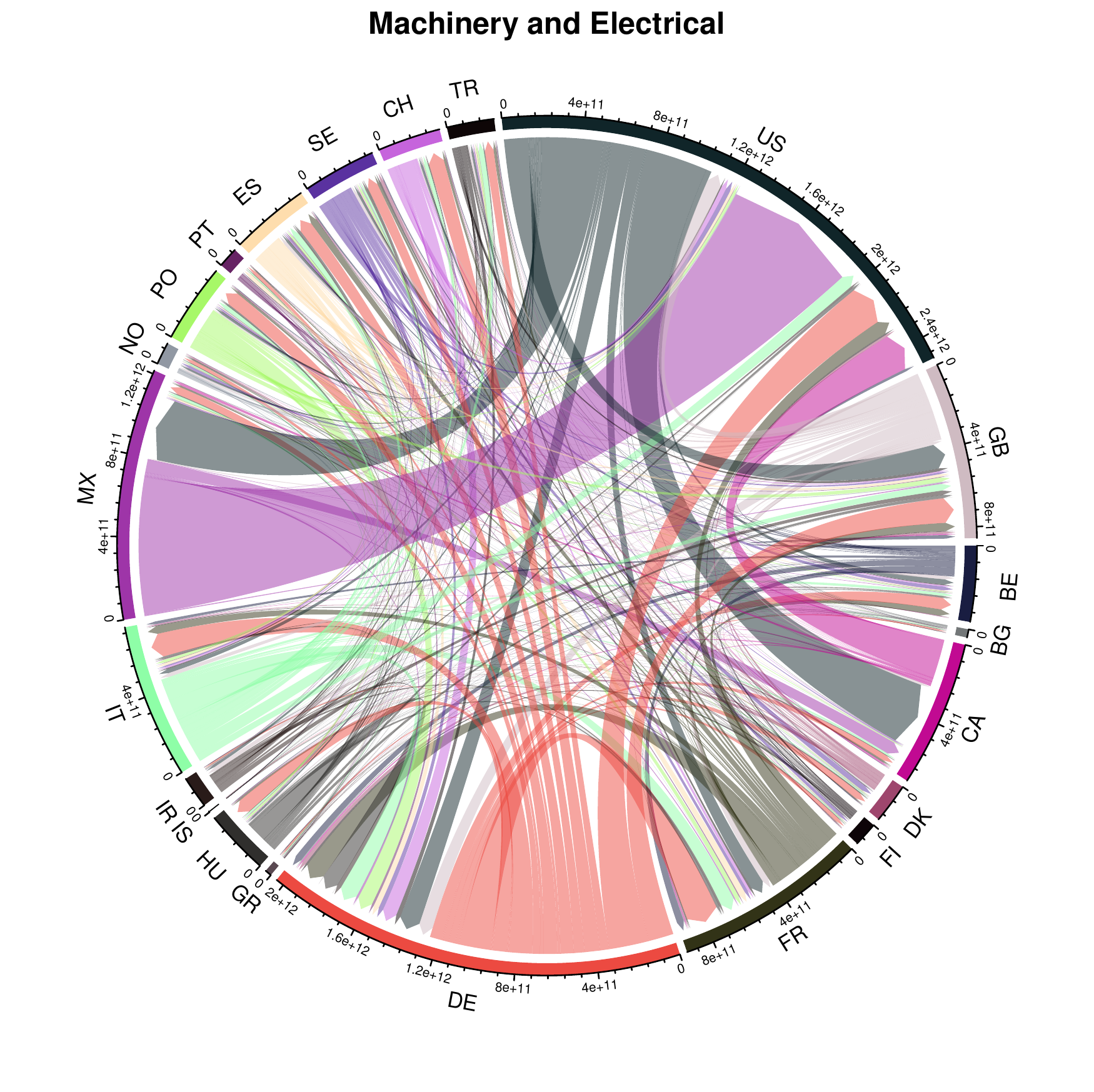}
\includegraphics[width=3in,height=3in,]{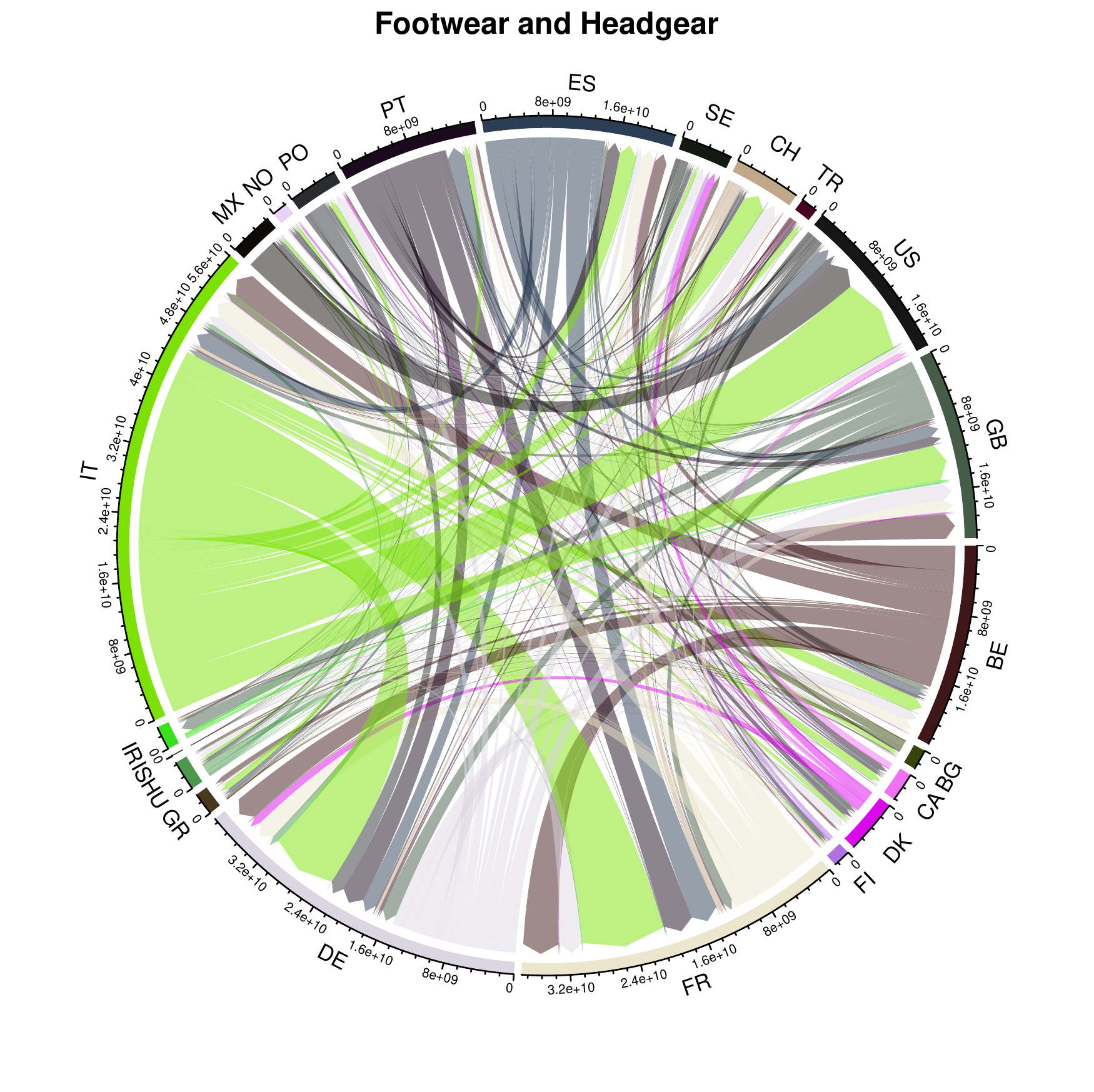}
\caption{Time aggregated import-export volume of
Machinery and Electronic products and
Footwear and Headwear products
among 22 countries in North American and Europe.}
\label{fig2}
\end{figure}

Figure~\ref{fig2} shows the total volume from year 2010 to 2017 in
two categories of products (Machinery and Electronic, and
Footwear and Headwear) among 22 countries in North American and Europe.
The arrows show the trade direction and the width of an arrow reflects
the volume of the trade.
Clearly the networks are quite different for different product
categories. For example,
Mexico is a large importer and exporter of Machinery and Electronic as it
serves as one of the major part suppliers in the product chain of
machinery and
electronics. On the other hand, Italy is
the largest exporter of Footwear and Headwear.

Under our general framework presented in Section 3,
we use the following  model for the dynamic transport networks. Let
$\cX_t$ be the observed tensor at time $t$. The element $x_{i_1i_2i_3,t}$
is the trading volume from country $i_1$ (the exporter) to country $i_2$ (the
importer) of product type $i_3$. Let
\begin{equation}
\cX_t=\cF_t\times_1\bA_1\times_2\bA_2\times_3\bA_3 +\cE_t
\label{trade1}
\end{equation}
where $\cX_t\in \R^{d_1\times d_2\times d_3}$ ($d_1=d_2$),
$\cF_t\in \R^{r_1\times r_2\times r_3}$ ($r\ll d$),
and $\bA_i\in \R^{d_i\times r_i}$. This is similar to the
DEDICOM model
\citep{harshman1978models,kolda2006tophits,kolda2005higher}.
\cite{Chen&Chen2019} provided some interpretations of
the factors in a uni-category import-export network under
the matrix factor model setting of \cite{Wang&2019}.

In the following we provide some interpretation of the model.
Consider
the loading matrix $\bA_3$.
It can be viewed as the loading matrix of
a standard factor model
\[
\bx_{i_1i_2\cdot,t}=\bA_3\bff^{(3)}_{i_1i_2,t}+\bvarepsilon^{(3)}_{i_1i_2\cdot,t}
\]
of the mode-3 fiber $\bx_{i_1i_2\cdot,t}$
for all $(t,i_1,i_2)$. This is essentially unfolding the
four dimensional $d_1\times d_2\times d_3\times T$ tensor $\cY$
into a $d_3\times (d_1d_2T)$ matrix and fit a standard factor model with
$d_1d_2T$ factors, each a vector of dimension $r_3$.
These factors drive the co-moment of all mode-3 fibers
$\cX_{i_1,i_2,\cdot,t}$ at time $t$. The loading matrix reflects how each
element of the mode-3 fiber is related to the factors. Note that
this scheme is only for interpretation. Th estimation
procedure is based on a different set-up.

Table~\ref{loading.3} shows an estimate of $\bA_3$ of the
import-export data under the tensor factor model,
using $r_3=6$ factors. The estimation is based on the TIPUP procedure
with $h_0=2$. The loading matrix is rotated using the varimax procedure
for better interpretation. All numbers are multiplied by 30 then truncated to
integers for clearer viewing.

  \begin{table}
\begin{center}
  \begin{small}
    \begin{tabular}{c|cccccc}
 &	1	& 2	& 3 & 	4 &	5 &	6 \\ \hline
Animal and Animal Products & 0 & 0 & 0 & 0 & 6 & -1 \\
Vegetable Products         & 2 & 1 & -1 & 0 & 5 & 0 \\
Foodstuffs                 & 0 & 0 & 1 & 2 & 6 & 1 \\
Mineral Products           & 0 & {\bf 30} & 0 & 0 & 0 & 0 \\
Chemicals and Allied Industries & -1 & -1 & {\bf 29}  & -1 & 2 & -1 \\
Plastics and Rubbers       & 0 & 0 & 1 & 0 & 16 & -3 \\
Raw Hides, Skins, Leather and Furs & 0 & 0 & 0 & 0 & 1 & 0 \\
Wood and Wood Products     & -2 & 2 & 0 & 2 & 7 & 1 \\
Textiles                   & 1 & -1 & 0 & -1 & 6 & 0 \\
Footwear and Headgear      & 0 & 0 & 0 & 0 & 1 & 0 \\
Stone and Glass            & -1 & 0 & 0 & 0 & 4 & {\bf 29} \\
Metals                     & -1 & 1 & 0 & 1 & 19 & -1 \\
Machinery and Electrical   & {\bf 29} & 0 & -1 & 0 & 3 & -1 \\
Transportation             & 0 & 0 & 0 & {\bf 30} & 0 & 0 \\
Miscellaneous              & 7 & 3 & 8 & 3 & -9 & 6 \\
    \end{tabular}
  \end{small}
  \caption{Estimated loading matrix $\bA_3$ for category fiber. Matrix is
      rotated via varimax. Elements are multiplied by 30 and truncated to
      integer.} \label{loading.3}
\end{center}
  \end{table}

\begin{figure}
\centering
\includegraphics[width=3.5in,height=3.5in]{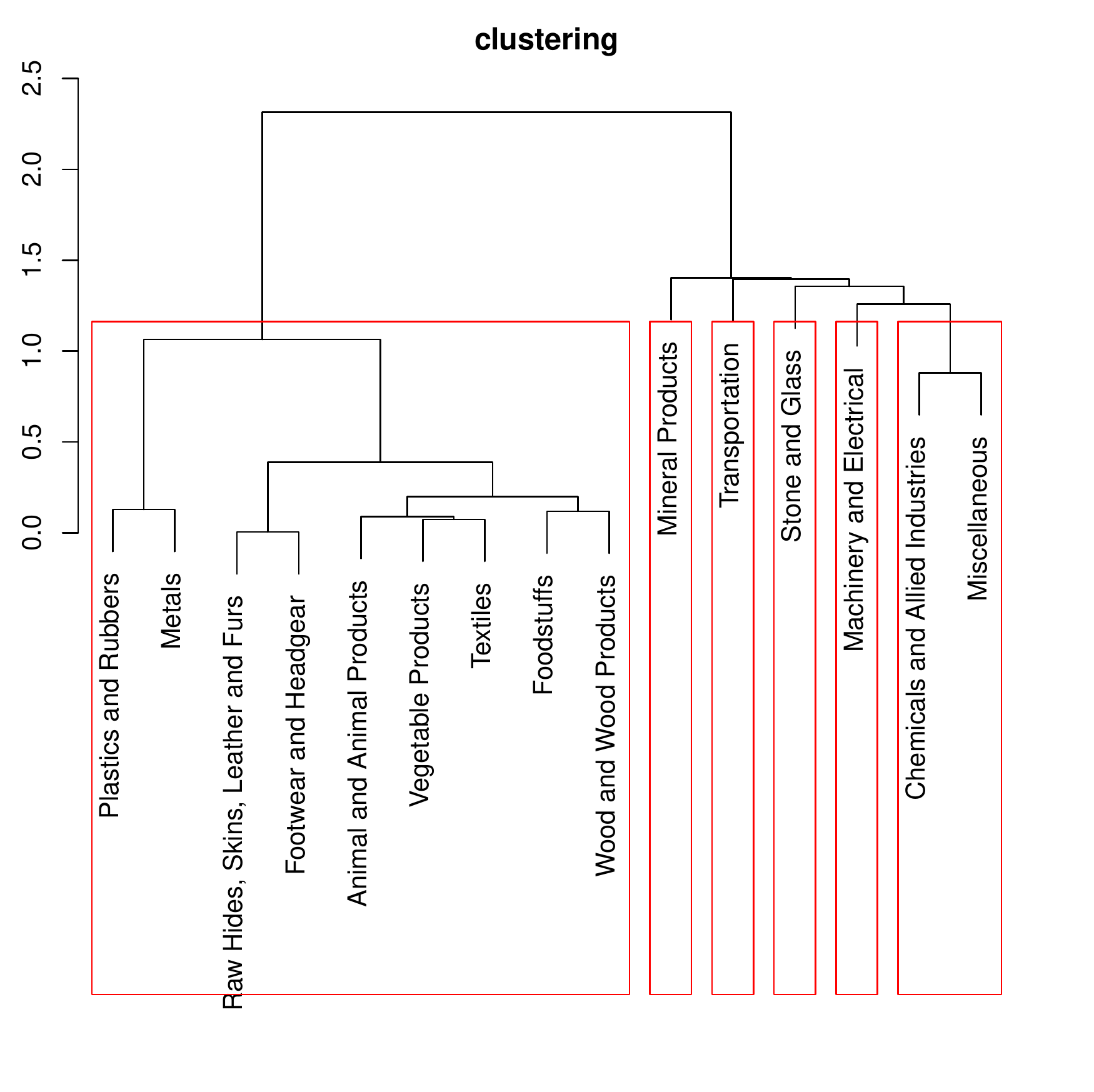}
\caption{Clustering of product categories by their loading coefficients}
\label{fig.clustering.Q3}
\end{figure}

It can be seen that there is a group structure. For example,
Factors 1, 2, 3, 4 and 6 can be interpreted
as the Machinery and Electrical factor, Mineral
factor, Chemicals factor, Transportation factor and Stone and Glass factor,
respectively,
since the corresponding product categories load heavily and almost exclusively
on them. On the other hand, Factor 5 is mixed, with
large loadings by Metals and Plastics/Rubbers, and medium loadings
by Animal, Vegetable and Food products. We will view each
factor as a 'condensed product group'. Figure~\ref{fig.clustering.Q3}
shows the clustering of the product categories according to their loading
vectors.

\begin{figure}
\centering
\includegraphics[width=3.0in,height=3.0in,keepaspectratio=true]{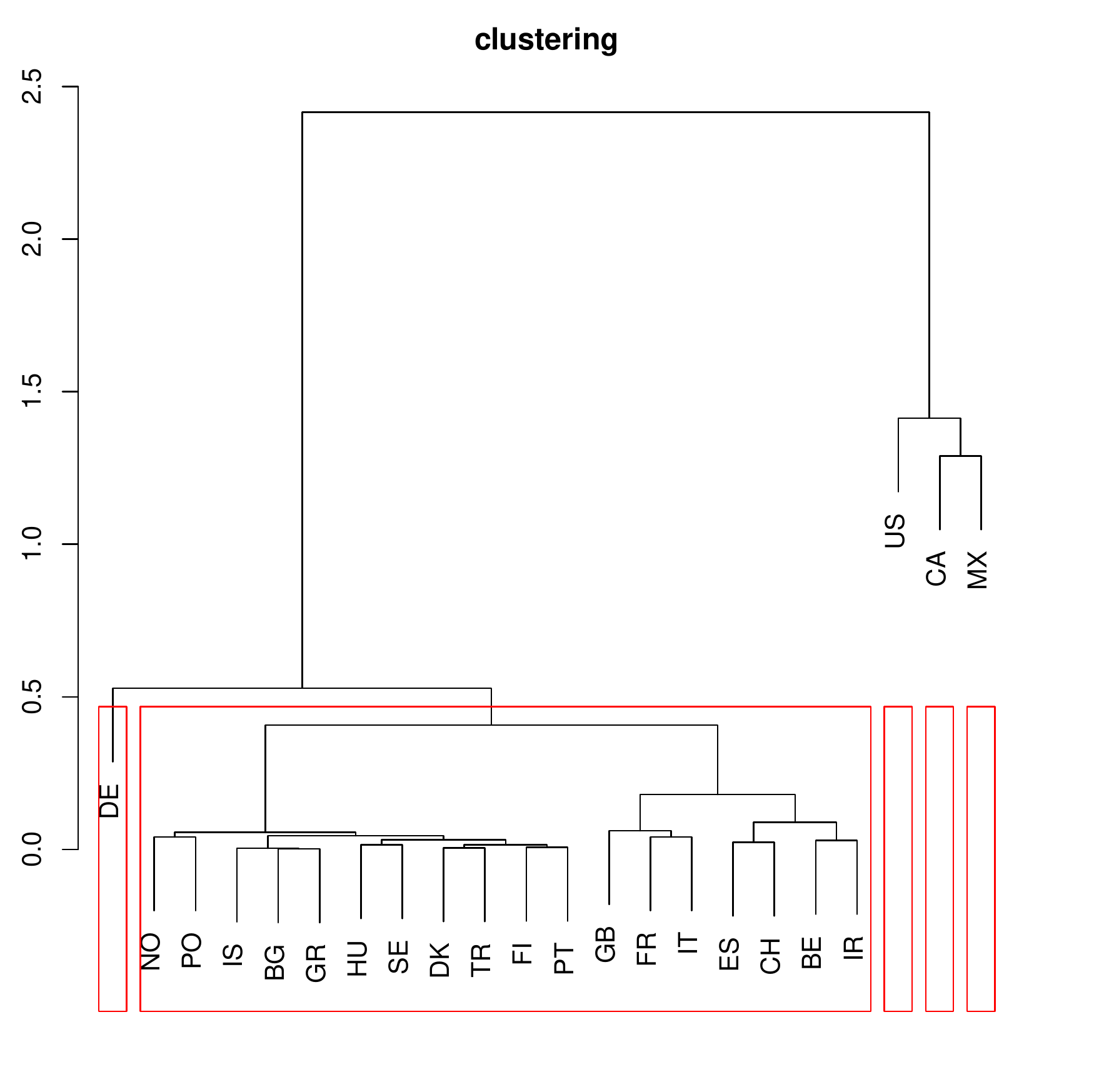}
\includegraphics[width=3.0in,height=3.0in,keepaspectratio=true]{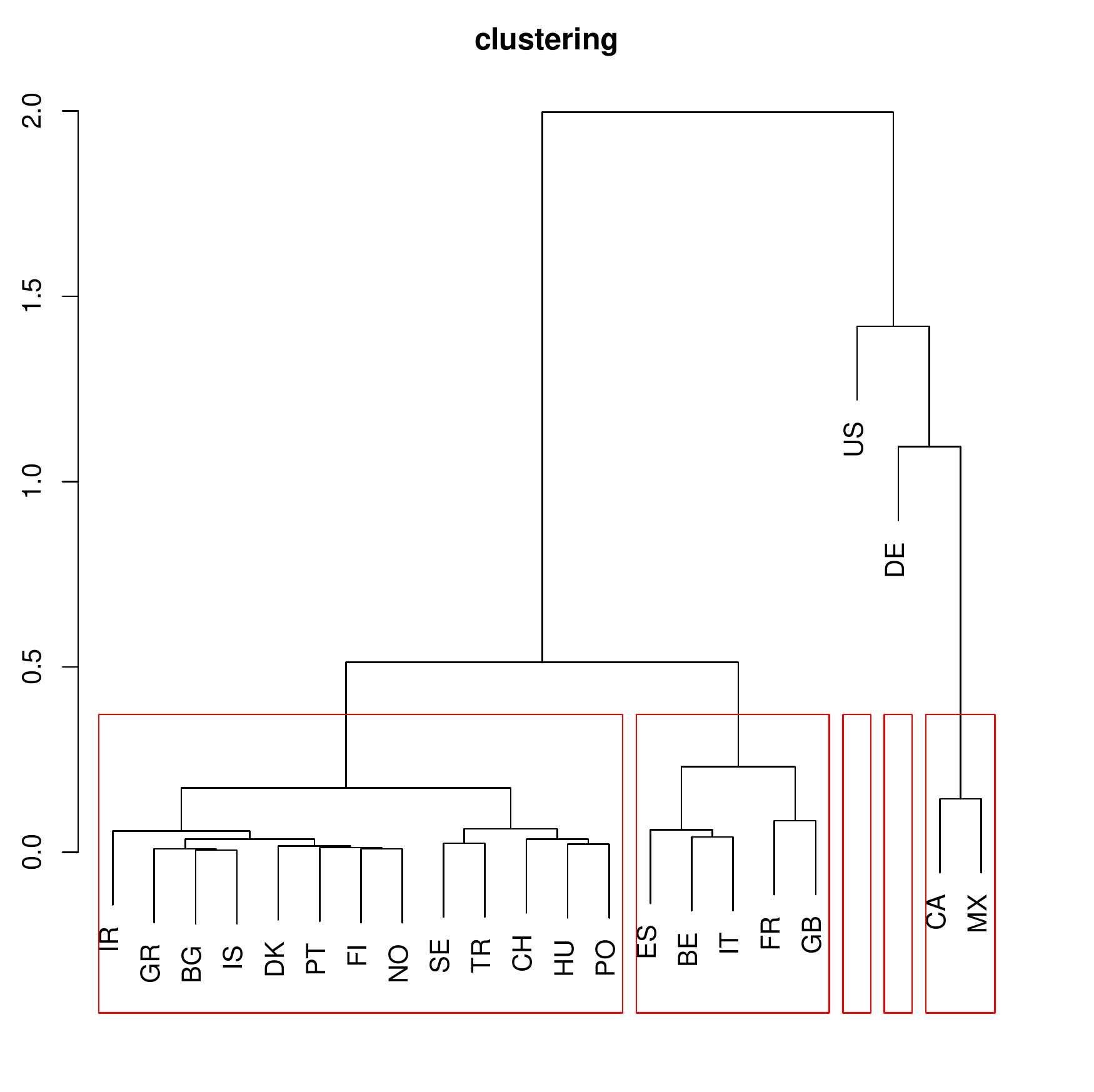}
\caption{Clustering of countries by their export (left) and import (right)
  loading coefficients}
\label{fig.clustering.Q12}
\end{figure}

The factor matrix $\cF_{\cdot,\cdot,i_3,t}$ (for a fixed $i_3$)
can be viewed as the trading pattern
among several {\bf trading hubs} for the $i_3$-th condensed product groups
(product factor).
One can imagine that the export of a product by a country
would first go through a virtual
'export hub', then to a virtual 'import hub', before arriving at the country
that imports the product. Each row of the matrix $\bF_{\cdot,\cdot,i_3,t}$
represents an
export hub and each column represents an import hub. The elements
$\cF_{i_1,i_2,i_3,t}$ can be viewed as the volume of the condensed product group
$i_3$ moved from export hub $i_1$ to import hub $i_2$ at time $t$.
The corresponding loading matrices $\bA_1$ and $\bA_2$ reflects the trading
activities of each country through each of the
export and import hubs, respectively.
We normalize each column of the loading matrices to sum up to one, so the value
can be viewed as the proportion of activities of each country contributes
to the hubs.
Tables~\ref{loading.1} and \ref{loading.2} show the estimated
loading matrices $\bA_1$ and $\bA_2$ after varimax rotation and column
normalization, using four  export hubs (E1 to E4) and four
import hubs (I1 to I4).
All values are in percentage.
There are a few negative values since we do not constrain the loadings
to be positive. The interpretation of the negative values is tricky.
Fortunately there are not many and the values are small. From
Table~\ref{loading.1}, it is seen
 that Canada, US and Mexico  heavily load
on export hubs E1, E2 and E4, respectively, while European countries mainly
load on export hub E3. The clustering based on loading coefficients of $\bA_1$
of each country is shown in the left panel of
Figure~\ref{fig.clustering.Q12}. The three countries in North America
are very different from the European countries. In Europe, Germany behaves
differently from the others as an exporter.
For imports, seen from Table~\ref{loading.2}, US and Germany load heavily
on hubs I1 and I4, respectively, while Canada and Mexico share hub I2. The
European countries other than Germany mainly load on hub I3.
The clustering based on loading coefficients of $\bA_2$
of each country is shown in the right panel of
Figure~\ref{fig.clustering.Q12}.
It seems that
the European countries (other than Germany)
can be divided into two groups of similar import
behavior, mainly based on the size of their economies.

  \begin{table}
\begin{center}
    \scriptsize
                  \addtolength{\tabcolsep}{-3pt}
      \begin{tabular}{c|cccccccccccccccccccccc}
 & BE & BG & CA & DK & FI & FR & DE & GR & HU & IS & IR & IT & MX & NO & PO & PT & ES & SE & CH & TR & US & GB \\ \hline
1 & 4 & 0 & {\bf 80} & 0 & 1 & 2 & -4 & 0 & 0 & 0 & 5 & 0 & -3 & 3 & 0 & 1 & 2 & 0 & 3 & 0 & 3 & 5 \\
2 & -1 & 0 & -4 & 0 & 0 & 1 & -6 & 0 & 1 & 0 & -2 & 2 & 1 & 1 & 1 & 0 & 0 & 0 & 1 & 0 & {\bf 102} & 1 \\
3 & 9 & 0 & -1 & 2 & 1 & 12 & 29 & 0 & 2 & 0 & 7 & 9 & -3 & 1 & 4 & 1 & 7 & 3 & 6 & 2 & 1 & 8 \\
4 & -8 & 0 & 5 & 0 & 0 & 0 & 15 & 0 & 0 & 0 & -7 & 2 & {\bf 104} & -2 & -2 & -1 & -4 & 1 & -3 & -1 & 0 & 2
    \end{tabular}
    \caption{Estimated loading matrix $\bA_1$ for the export fiber (hub). Matrix is
      rotated via varimax and column normalized. Values are in percentage.
      } \label{loading.1}
\end{center}
  \end{table}

  \begin{table}
\begin{center}
    \scriptsize
            \addtolength{\tabcolsep}{-3pt}
  \begin{tabular}{c|cccccccccccccccccccccc}
       & BE & BG & CA & DK & FI & FR & DE & GR & HU & IS & IR & IT & MX & NO & PO & PT & ES & SE & CH & TR & US & GB \\ \hline
1 & 1 & 0 & 2 & 0 & 0 & -1 & 0 & 0 & 0 & 0 & 0 & 0 & -2 & 0 & 0 & 0 & 1 & 0 & 0 & 0 & {\bf 100} & -1 \\
2 & 0 & 0 & 57 & 0 & 0 & 2 & -5 & 0 & 0 & 0 & 1 & -2 & 44 & 0 & -1 & -1 & -2 & -1 & 0 & 1 & 0 & 7 \\
3 & 10 & 1 & -2 & 3 & 2 & 22 & -3 & 1 & 4 & 0 & 1 & 11 & 0 & 2 & 6 & 2 & 6 & 5 & 8 & 4 & 0 & 18 \\
4 & 7 & 0 & 4 & 0 & 0 & 0 & {\bf 68} & 1 & -2 & 0 & 4 & 4 & 1 & 1 & -2 & 1 & 8 & 0 & -2 & 2 & 0 & 5
    \end{tabular}
  \caption{Estimated loading matrix $\bA_2$ for the import fiber (hub).
      Matrix is
      rotated via varimax and column normalized. Values are in percentage.} \label{loading.2}
\end{center}
  \end{table}

\begin{figure}
\centering
\includegraphics[width=3.0in,height=3.0in,keepaspectratio=true]{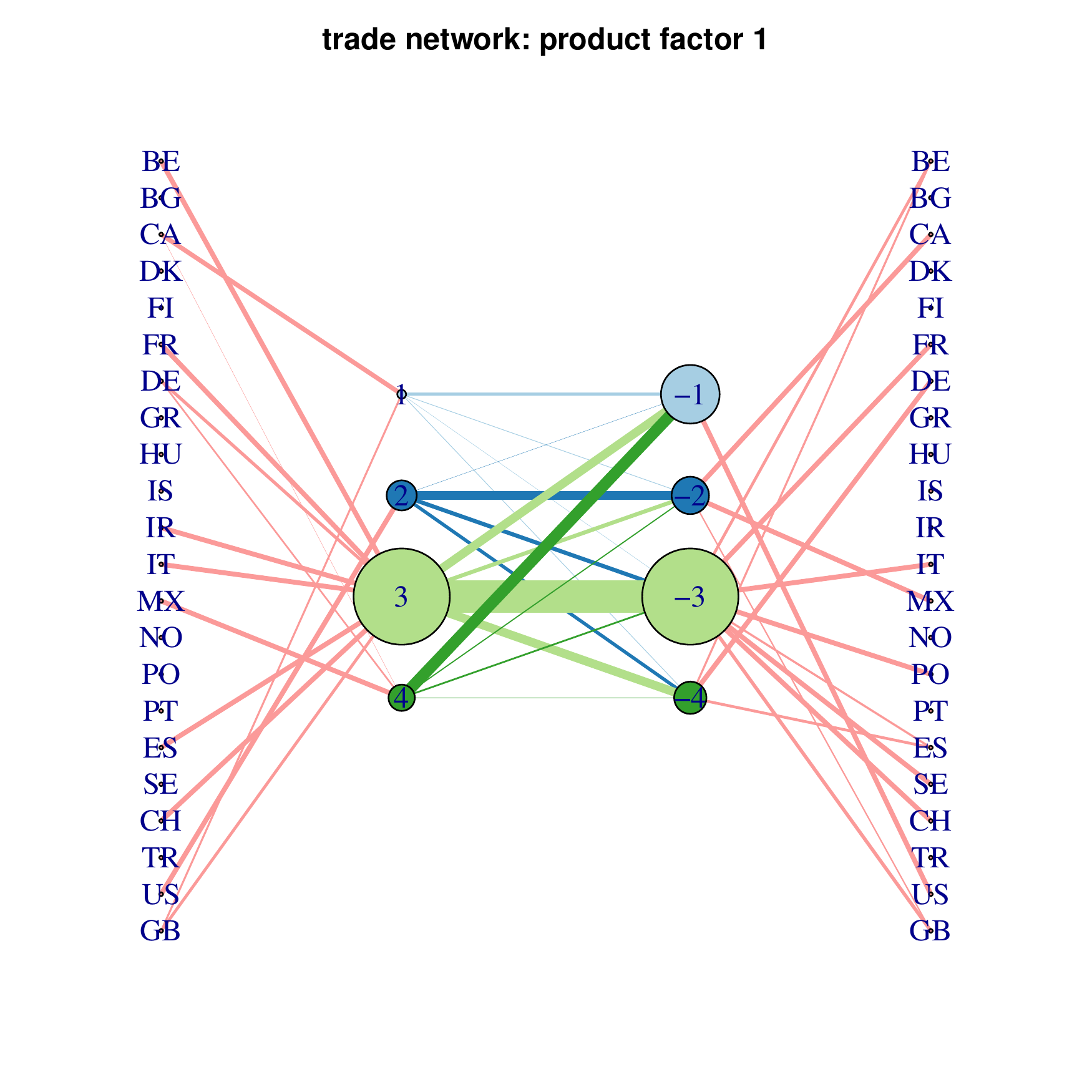}
\includegraphics[width=3.0in,height=3.0in,keepaspectratio=true]{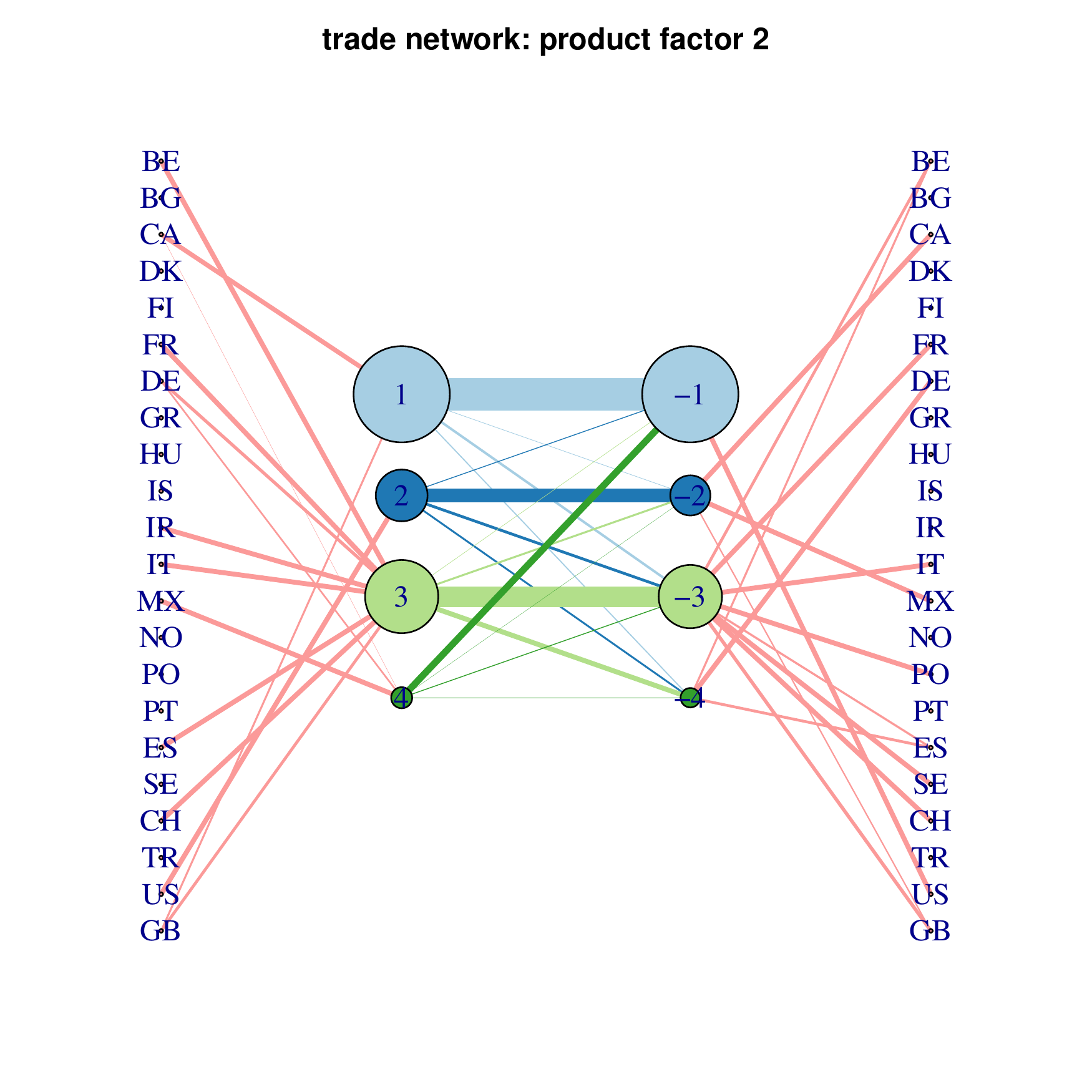}
\caption{Trade network for condensed product group 1 (left) and group 2 (right).
  Export and import hubs are on the left and right of the center network
  respectively. Line
  width is proportional to the total volume of trade between the hubs
  for the last three years (2015 to 2017). Vertex size is proportional to total
  volume of trades through the hub. The line width between the countries
  and the hubs is proportional to the corresponding loading coefficients, for
  coefficients larger than 0.05 only.}
\label{fig.network.1}
\end{figure}

The left panel of
Figure~\ref{fig.network.1} shows the trade transport network for
the condensed product group 1 (mainly Machinery and Electrical). Several
interesting features emerge. Export hub E3 (European hub) has the largest trading volume,
and the goods mainly go to import hub I3 (European hub) and hub I1 (US hub).
This is understandable as trades among the many countries in Europe
accumulate, and US is one of the largest importers. Mexico dominates
export hub E4 and it mainly exports to import hub I1, used by US,
confirming what is shown in the left panel of Figure~\ref{fig2}.
US is also a large exporter
of machinery and electrical, occupying export hub E2, which mainly exports
to import hub I2 used by Mexico and Canada.

On the other hand, for the network of condensed product group 2 (mainly
mineral products) shown in right panel of
Figure~\ref{fig.network.1}, the dynamic is quite different.
Export hub E1, mainly used by Canada, is the largest hub for mineral
products. The import hub I1 is the largest import hub, mainly used by US.
Most of its volume come through export hubs E1 (used mainly by Canada) and
E4 (used mainly by Mexico).
The network plots of other product groups are shown in Appendix B.

\begin{figure}
\centering
\includegraphics[width=4.5in,height=4.0in,keepaspectratio=true]{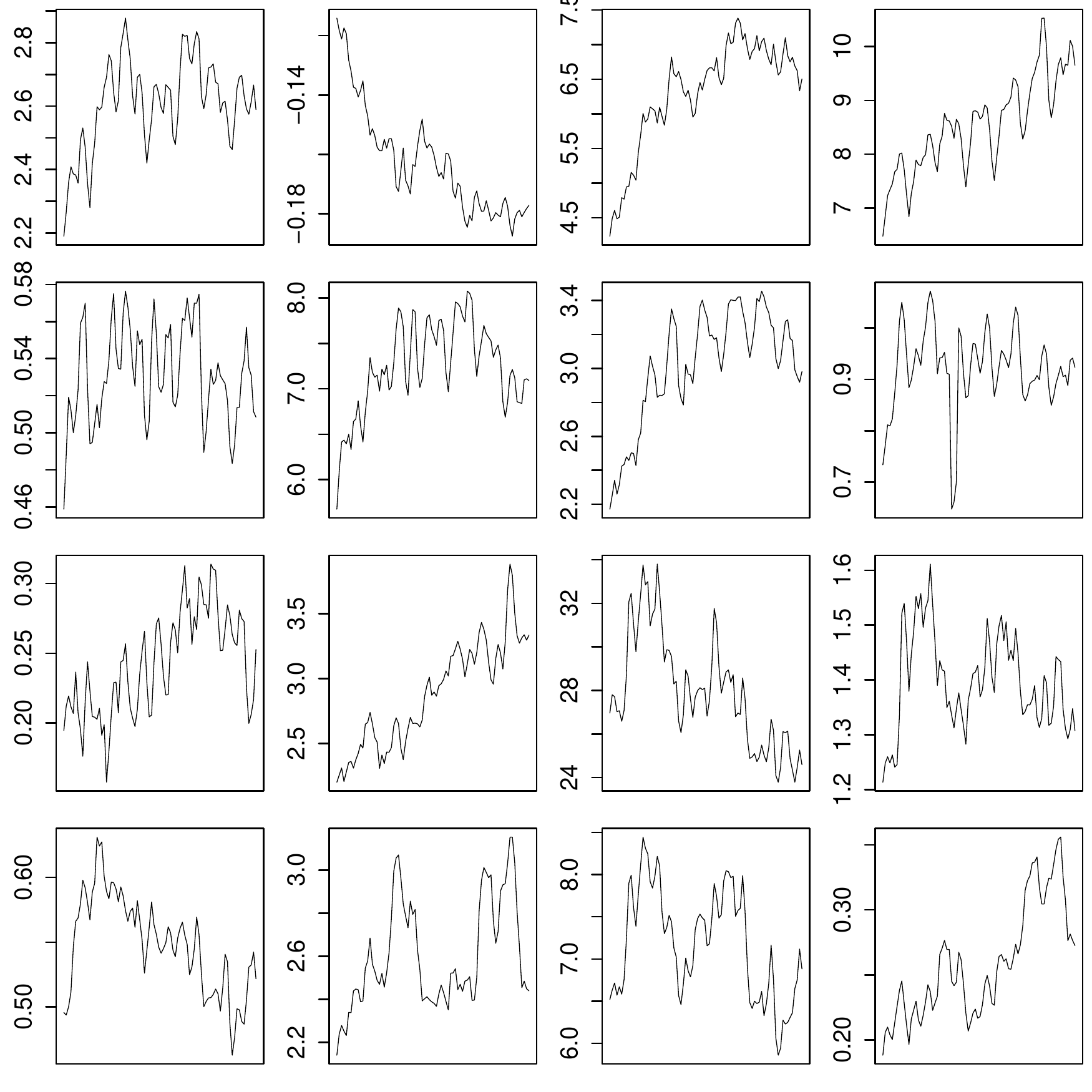}
\caption{Trading volumes among the hubs (factors) of
  condensed product group 1. Rows are for export and columns for import.}
\label{fig.factor.1}
\end{figure}

Figure~\ref{fig.factor.1} shows the normalized trading volumes
among the hubs (factors)
to show the variation in trading through time. Note that
the scales are very different among the figures.

We remark that this analysis is just for illustration and showcasing the
interpretation of the model. A more formal analysis would include the
determination of the number of factors and model comparison procedures.

\subsection{Taxi traffic in New York city}\label{example2}

In this example we analyze taxi traffic pattern in New York city. The data includes
all individual taxi rides operated by Yellow Taxi within New York City, maintained
by the Taxi \& Limousine Commission of New York City and published at

{\tt https://www1.nyc.gov/site/tlc/about/tlc-trip-record-data.page}.

The dataset contains 1.4 billion trip records within the period of January 1, 2009 to December 31, 2017,
among these 1.2 billion are for rides within Manhattan Island.
Each trip record
includes fields capturing pick-up and drop-off dates/times, pick-up and drop-off locations,
trip distances, itemized fares, rate types, payment types, and
driver-reported passenger counts.
As we are interested in the movements of passengers using the taxi service,
our study focuses on the
pick-up and drop-off dates/times, and pick-up and drop-off locations of each ride.
To simplify the discussion, we only consider rides within Manhattan Island.

The pick-up and drop-off location in
Manhattan are coded according to 69 predefined zones in the dataset after 2016 and we will use them to classify the
pick-up and drop-off locations. To
account for time variation during the day, we divide
each day into 24 hourly periods. The first
hourly period is from 0am to 1am. The total
number of rides moving among the zones within each hour is recorded, yielding
a $\cX_t\in\R^{69\times 69\times 24}$ tensor for each day. Here
$x_{i_1,i_2,i_3,t}$ is the number of trips from zone $i_1$ (the pick-up zone)
to zone $i_2$ (the drop-off zone) and the pickup time within the $i_3$-th
hourly period in day $t$. We consider business day and non-business day separately and
ignore the gaps created by the separation. Hence we will analyze two tensor time
series. The business-day series is 2,262 days long, and the non-business-day series
is 1,025 day long, within the period of January 1, 2009 to December 31, 2017.

After some exploratory analysis, we decide to use the tensor factor model with
a $4\times 4\times 4$ core factor tensor and estimate the model using the TIPUP estimator
with $h_0=1$. The TOPUP estimator produces similar results.

\begin{figure}[H]
  \includegraphics[width=.22\textwidth,trim = 12mm 0 12mm 0,clip, page=1]{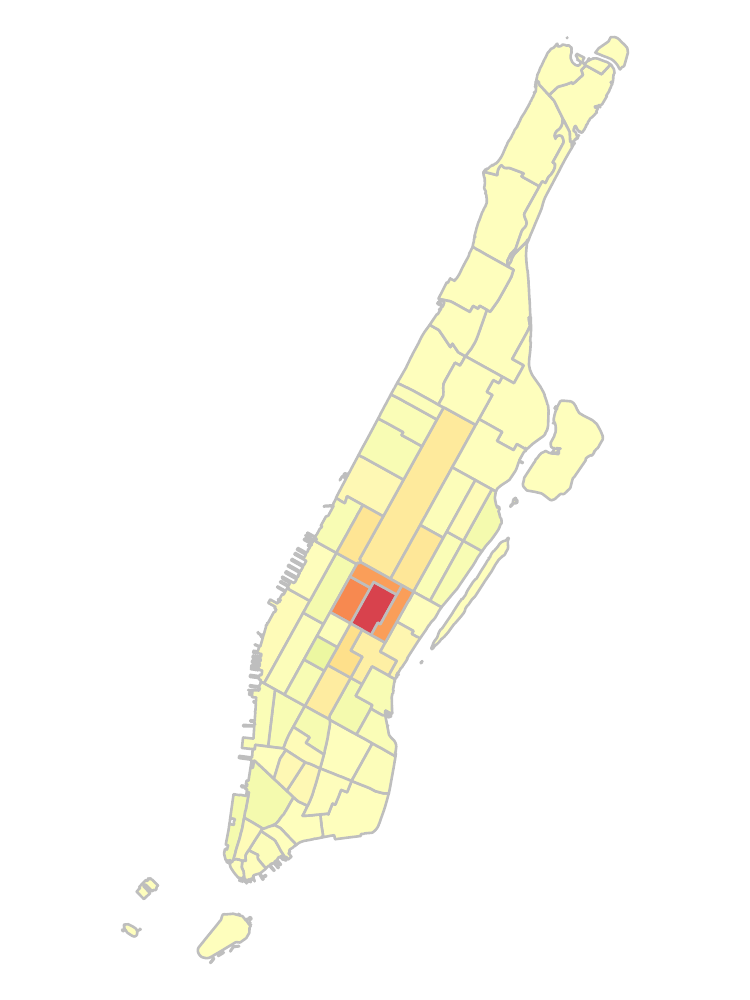}
  \includegraphics[width=.22\textwidth,trim = 12mm 0 12mm 0,clip, page=2]{figure_manhattan_pickup1234_signchange_businessday_separate.pdf}
  \includegraphics[width=.22\textwidth,trim = 12mm 0 12mm 0,clip, page=3]{figure_manhattan_pickup1234_signchange_businessday_separate.pdf}
  \includegraphics[width=.22\textwidth,trim = 12mm 0 12mm 0,clip, page=4]{figure_manhattan_pickup1234_signchange_businessday_separate.pdf}
  \includegraphics[width=.1\textwidth]{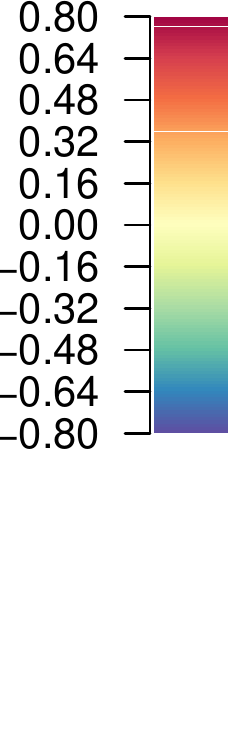}
  \caption{Loadings on four pickup factors for business day series}
  \label{fig:pickup-bus}
\end{figure}

\begin{figure}[H]
  \includegraphics[width=.22\textwidth,trim = 12mm 0 12mm 0,clip, page=1]{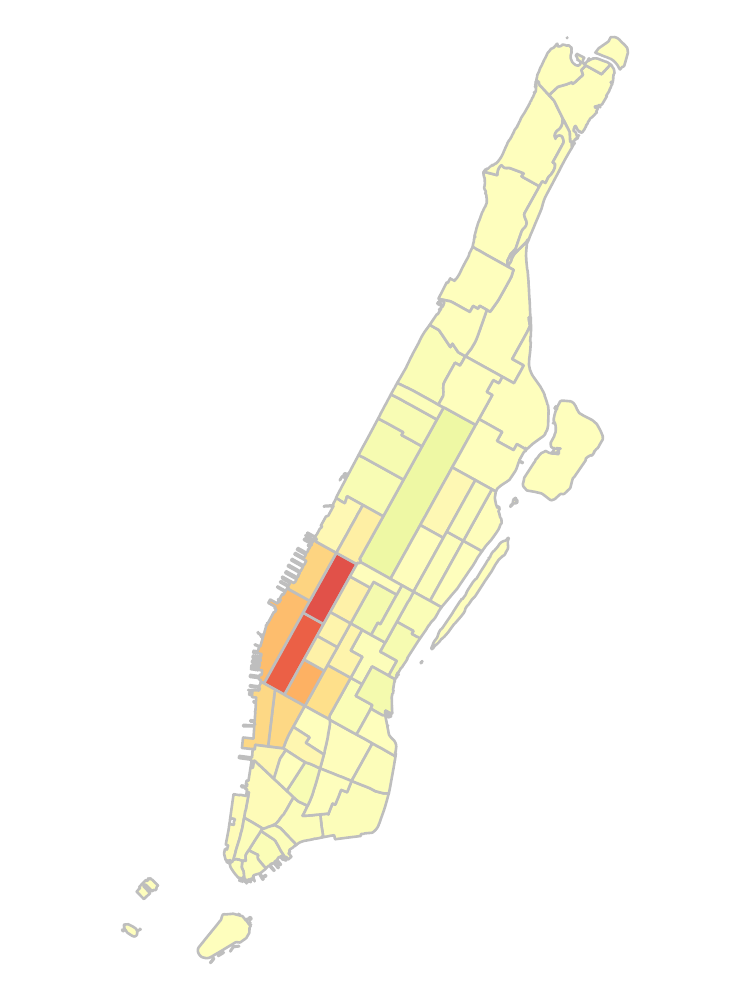}
  \includegraphics[width=.22\textwidth,trim = 12mm 0 12mm 0,clip, page=2]{figure_manhattan_pickup1234_signchange_nonbusinessday_separate.pdf}
  \includegraphics[width=.22\textwidth,trim = 12mm 0 12mm 0,clip, page=3]{figure_manhattan_pickup1234_signchange_nonbusinessday_separate.pdf}
  \includegraphics[width=.22\textwidth,trim = 12mm 0 12mm 0,clip, page=4]{figure_manhattan_pickup1234_signchange_nonbusinessday_separate.pdf}
  \includegraphics[width=.1\textwidth]{figure_manhattan_pickup_dropoff_signchange_separate_color_bar.pdf}
  \caption{Loadings on four pickup factors for non-business day series}
  \label{fig:pickup-non}
\end{figure}

Figure~\ref{fig:pickup-bus} shows the heatmap of
the loading matrix $\bA_1$ (related to
pick-up locations) of the 69 zones in Manhattan.
It is seen that during business days,
the midtown/Times square area is heavily loaded on Factor 1, upper east side on Factor 2, upper west side on Factor 3
and lower east side on Factor 4. For non-business days, the loading matrix is significantly
different, as shown in
Figure~\ref{fig:pickup-non}. The area on the lower west side near Chelsea
(with many restaurants and bars)
that heavily loads on the first
factor is not active for pickups during the business day.

\begin{figure}[H]
  \includegraphics[width=.22\textwidth,trim = 12mm 0 12mm 0,clip, page=1]{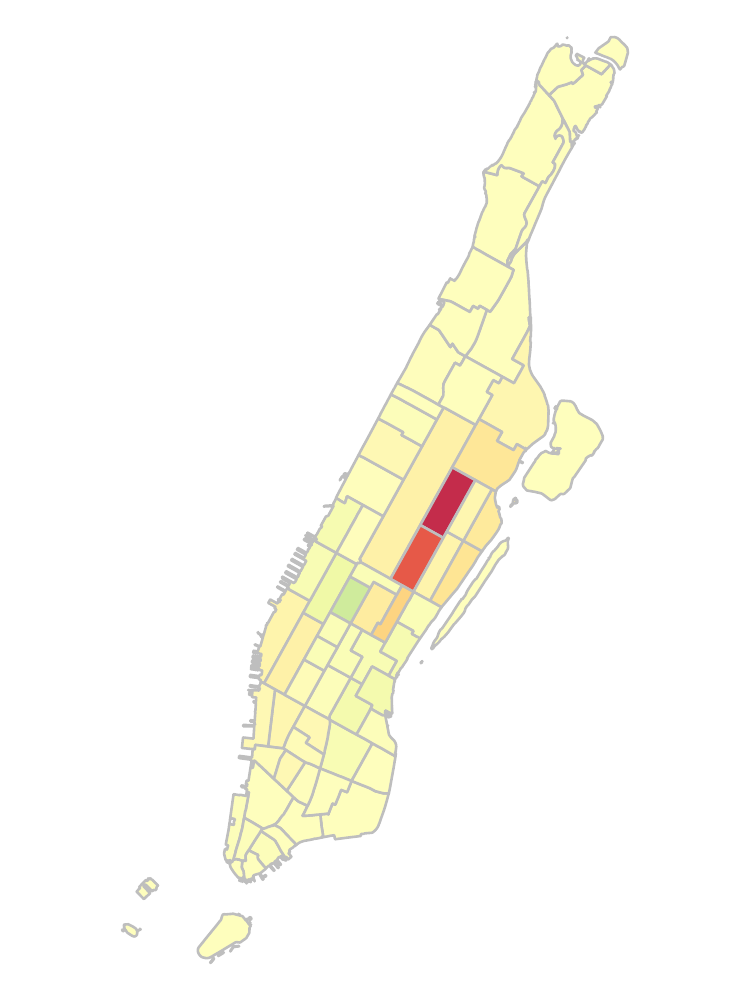}
  \includegraphics[width=.22\textwidth,trim = 12mm 0 12mm 0,clip, page=2]{figure_manhattan_dropoff1234_signchange_businessday_separate.pdf}
  \includegraphics[width=.22\textwidth,trim = 12mm 0 12mm 0,clip, page=3]{figure_manhattan_dropoff1234_signchange_businessday_separate.pdf}
  \includegraphics[width=.22\textwidth,trim = 12mm 0 12mm 0,clip, page=4]{figure_manhattan_dropoff1234_signchange_businessday_separate.pdf}
  \includegraphics[width=.1\textwidth]{figure_manhattan_pickup_dropoff_signchange_separate_color_bar.pdf}
    \caption{Loadings on four dropoff factors for business day series}
  \label{fig:dropoff-bus}
\end{figure}
\begin{figure}[H]
  \includegraphics[width=.22\textwidth,trim = 12mm 0 12mm 0,clip, page=1]{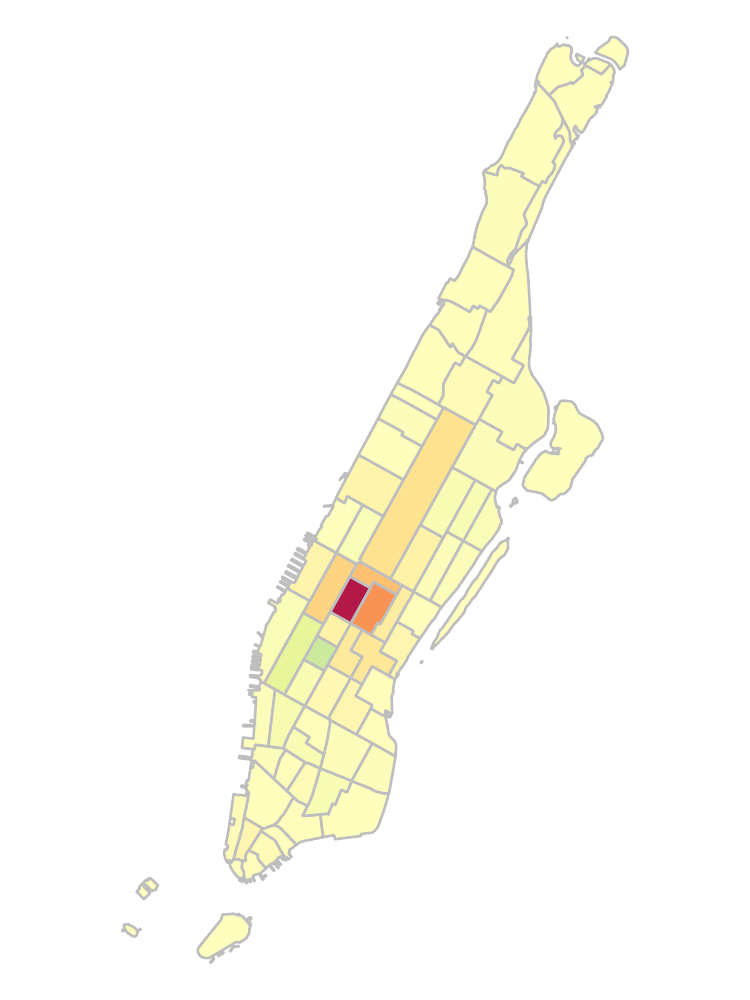}
  \includegraphics[width=.22\textwidth,trim = 12mm 0 12mm 0,clip, page=2]{figure_manhattan_dropoff1234_signchange_nonbusinessday_separate.pdf}
  \includegraphics[width=.22\textwidth,trim = 12mm 0 12mm 0,clip, page=3]{figure_manhattan_dropoff1234_signchange_nonbusinessday_separate.pdf}
  \includegraphics[width=.22\textwidth,trim = 12mm 0 12mm 0,clip, page=4]{figure_manhattan_dropoff1234_signchange_nonbusinessday_separate.pdf}
  \includegraphics[width=.1\textwidth]{figure_manhattan_pickup_dropoff_signchange_separate_color_bar.pdf}
  \caption{Loadings on four dropoff factors for non-business day series}
    \label{fig:dropoff-non}
\end{figure}

Figures~\ref{fig:dropoff-bus} and \ref{fig:dropoff-non}
show the loading matrices
$\bA_2$ (related to dropoff locations)
for business days and non-business days, respectively.
For dropoff during business days, the areas that load heavily on the factors are quite
similar to that for pick-up, except the area that
loads heavily on Factor 2.  This area is around Union Square which is
a big transportation hub servicing the surrounding tri-state area
(New York, Connecticut and New Jersey), and
a heavy shopping/restaurant area.
For non-business days, the dropoff area that heavily loads on Factor 3
(Yorkvill/Lenox hill) is different from all
the areas used for both pickup and dropoff and for both business days and non-business days.
To simplify our presentation and to show comparable results in different settings, we
will roughly match the pickup and dropoff factors by their corresponding heavily loaded
areas, shown in Table~\ref{area} with brief area descriptions.

\begin{table}
\small
  \begin{center}
    \begin{tabular}{l|cccc|l}
      Area  & \multicolumn{4}{c}{source factor} & description \\ \hline
       & \multicolumn{2}{c}{Business} &\multicolumn{2}{c|}{non-Bus} &  \\
 & p & d & p & d & \\ \hline
      1 Upper east                & 2 & 1 & 3 &   & affluent neighborhoods and museums \\
      2 Midtown/Times square      & 1 & 3 &   & 1 & tourism and office buildings\\
      3 Upper west/Lincoln square & 3 & 4 & 4 & 4 & affluent neighborhoods and performing arts \\
      4 East village/Lower east   & 4 &   & 2 &   & historic district with art\\
      5 Union square              &   & 2 &   & 2 & transportation hub with shops and restaurants \\
      6 Clinton east/Chelsea      &   &   & 1 &   & lots of restaurants and bars \\
      7 Yorkvill/Lenox hill       &   &   &   & 3 & a few universities
    \end{tabular}
    \caption{Label of representing areas identified under the tensor factor model with area description. ``p'' stands for pickup and ``d'' for dropoff.}
    \label{area}
  \end{center}
\end{table}

Tables~\ref{loading.3bus} and \ref{loading.3non}
show the loading matrix $\bA_3$
(on the time of day dimension) for business day and non-business day, respectively, after
varimax rotation.
The shaded cells roughly show the dominating periods of
each of the factors, though
the change is more continuously and smooth.
It is seen that, for business days, the morning rush-hours between 6am to 9am
are heavy
and almost exclusively loaded on factor 1 and we will name this factor
the {\it morning
rush-hour} factor. The business hours from 8am to 3pm heavily
load on Factor 2 (the {\it business hour} factor),
the evening rush-hours from 3pm to 8pm load heavily on Factor 3
(the {\it evening rush-hour} factor)
and the night life hours from 8pm to 1am load on Factor 4 (the
{\it night life} factor).
On the other hand, for
nonbusiness days, we have morning activities between 8am to 1pm
(the {\it morning} factor),
afternoon/evening
activities between 12pm to 9pm (the {\it afternoon/evening} factor),
and night activities between 9pm to 12am (the {\it early night} factor)
and 12am to 4am (the {\it late night} factor).

  \begin{table}
\begin{center}
    \scriptsize
      \addtolength{\tabcolsep}{-3pt}
      \begin{tabular}{c|cccccccccccc|cccccccccccc|c}
        \multicolumn{2}{c}{0am} &
        \multicolumn{2}{c}{2} &
        \multicolumn{2}{c}{4} &
        \multicolumn{2}{c}{6} &
        \multicolumn{2}{c}{8} &
        \multicolumn{2}{c}{10} &
        \multicolumn{2}{c}{12pm} &
        \multicolumn{2}{c}{2} &
        \multicolumn{2}{c}{4} &
        \multicolumn{2}{c}{6} &
        \multicolumn{2}{c}{8} &
        \multicolumn{2}{c}{10} &
        \multicolumn{2}{c}{12am} \\ \hline
1&-2&-1&-1&-1&1&10&\cellcolor[gray]{0.6} 47& \cellcolor[gray]{0.6}72& \cellcolor[gray]{0.6} 42&14&2&-6&-11&-9&-8&-5&-1&5&5&4&1&2&0&-2& \\
2&0&0&0&0&-1&-4&-13&-5& \cellcolor[gray]{0.6}32&\cellcolor[gray]{0.6}46&\cellcolor[gray]{0.6}36&\cellcolor[gray]{0.6}35&\cellcolor[gray]{0.6}38&\cellcolor[gray]{0.6}33&\cellcolor[gray]{0.6}29&19&9&1&-2&-5&-3&-3&-1&1& \\
3&-5&-4&-3&-2&-1&1&4&6&-15&-25&-6&4&7&9&19&\cellcolor[gray]{0.6}31&\cellcolor[gray]{0.6}32&\cellcolor[gray]{0.6}43&\cellcolor[gray]{0.6}47&\cellcolor[gray]{0.6}39&22&14&4&-6& \\
4&\cellcolor[gray]{0.6}28&18&11&7&4&1&0&-8&2&14&4&-2&-3&-2&-7&-15&-13&-11&1&19&\cellcolor[gray]{0.6}35&\cellcolor[gray]{0.6}41&\cellcolor[gray]{0.6}46&\cellcolor[gray]{0.6}47&
    \end{tabular}
    \caption{Estimated loading matrix $\bA_3$ for hour of day fiber. Business day.  Matrix is
      rotated via varimax. Values are in percentage.
      } \label{loading.3bus}
\end{center}
  \end{table}

  \begin{table}
\begin{center}
    \scriptsize
      \addtolength{\tabcolsep}{-3pt}
      \begin{tabular}{c|cccccccccccc|cccccccccccc|c}
        \multicolumn{2}{c}{0am} &
        \multicolumn{2}{c}{2} &
        \multicolumn{2}{c}{4} &
        \multicolumn{2}{c}{6} &
        \multicolumn{2}{c}{8} &
        \multicolumn{2}{c}{10} &
        \multicolumn{2}{c}{12pm} &
        \multicolumn{2}{c}{2} &
        \multicolumn{2}{c}{4} &
        \multicolumn{2}{c}{6} &
        \multicolumn{2}{c}{8} &
        \multicolumn{2}{c}{10} &
        \multicolumn{2}{c}{12am} \\ \hline
1&-20&-3&11&10&5&3&9&19&\cellcolor[gray]{0.6}34&\cellcolor[gray]{0.6}47&\cellcolor[gray]{0.6}47&\cellcolor[gray]{0.6}35&\cellcolor[gray]{0.6}23&14&10&12&3&-4&-13&-16&-14&-13&-5&10& \\
2&19&0&-13&-11&-3&0&0&-2&-3&-2&6&17&\cellcolor[gray]{0.6}25&\cellcolor[gray]{0.6}29&\cellcolor[gray]{0.6}30&\cellcolor[gray]{0.6}27&\cellcolor[gray]{0.6}29&\cellcolor[gray]{0.6}34&\cellcolor[gray]{0.6}39&\cellcolor[gray]{0.6}33&\cellcolor[gray]{0.6}22&17&5&-17& \\
3&-11&3&14&7&-2&-4&-4&-3&-2&2&-1&-5&-3&1&0&4&-3&-3&2&17&\cellcolor[gray]{0.6}20&\cellcolor[gray]{0.6}24&\cellcolor[gray]{0.6}45&\cellcolor[gray]{0.6}78& \\
4&\cellcolor[gray]{0.6}53&\cellcolor[gray]{0.6}52&\cellcolor[gray]{0.6}45&\cellcolor[gray]{0.6}37&\cellcolor[gray]{0.6}21&8&6&5&4&2&1&2&-2&-4&-4&-6&-2&0&0&-1&4&6&0&-10&
    \end{tabular}
    \caption{Estimated loading matrix $\bA_3$ for hour of day fiber. Non-Business day.  Matrix is
      rotated via varimax. Values are in percentage.
      } \label{loading.3non}
\end{center}
  \end{table}

\begin{figure}[H]
  \centering
\includegraphics[width=5in,angle=0]{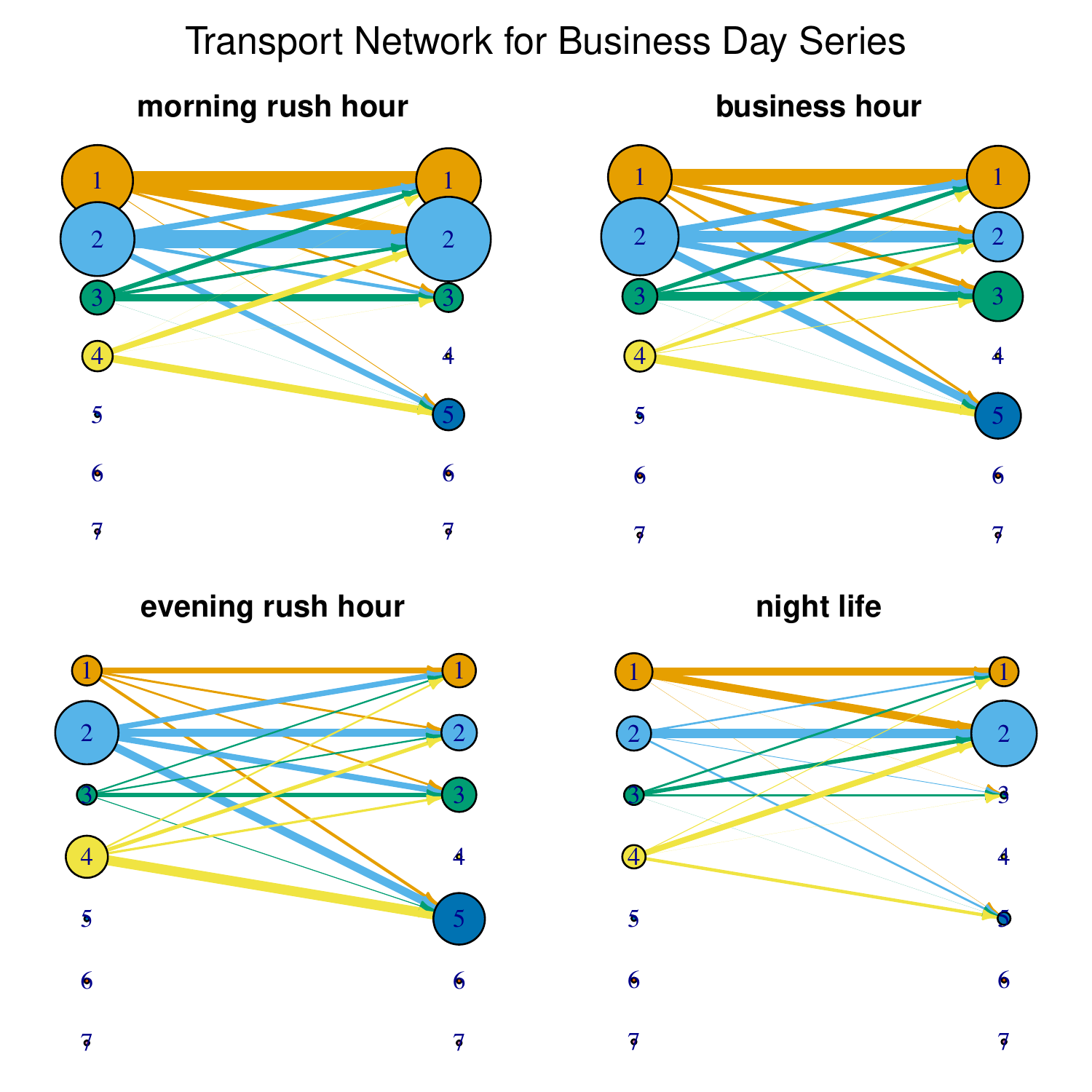}
  \caption{Network Plots during the four time factor periods for business day series}
  \label{fig:network-B}
\end{figure}
\begin{figure}[H]
  \centering
  \includegraphics[width=5in,angle=0]{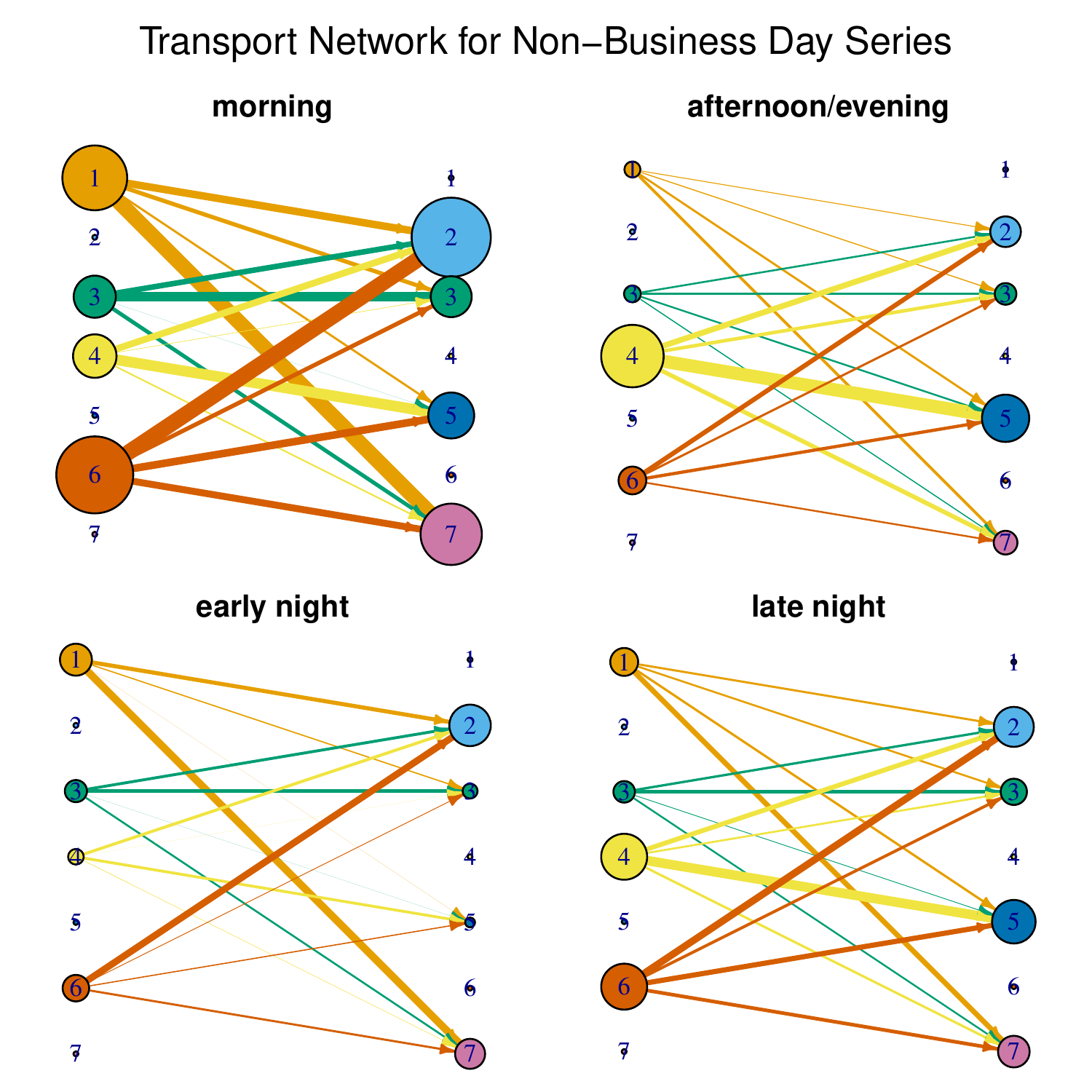}
  \caption{Network Plots during the four time factor periods for non-business day series}
  \label{fig:network-NB}
\end{figure}

Figures~\ref{fig:network-B} and \ref{fig:network-NB}
show the traffic network plots between the areas
defined in Table~\ref{area}
during different time factor periods. The width of the lines reflects
total traffic volume between the major areas
over the entire time series (the sum of the factors $f_{k_1k_2k_3,t}$ over time
$t$.) The size
of the vertices reflects total number of pickups (left vertices) and dropoffs (right vertices)
in the area during the time factor period.

The figures reveal many interesting patterns. For example, during the morning rush-hours
of business days,
traffic mainly goes
from Areas 1 and 2 (upper east and midtown) to Areas 2 (midtown). There is
only a small amount of  traffic to
Area 5. During the business hours and early evening hours,
traffic is mainly within
Areas 1 and 2. During the evening rush-hour, the main pickup area is midtown and
the main dropoff area is the Union square where many people take public transportation to
the surrounding tri-state area. During the night life hours, main traffic is
towards Area 2 (midtown), since Times square is popular among tourists and night-life goers.

For non-business days, the pattern is very different. During morning time from 8am to 12pm,
most traffic takes place from Area 6 (Chelsea) to Area 2 (midtown) and from Area 1
(upper east side) to Area 7 (Yorkvill/Lenox hill); during afternoon/evening from 12pm to 9pm,
many riders take taxi from Area 4 (lower east) to Area 5 (Union square);
during early night (from 8pm to 12am), the traffic volume is much smaller,
mainly from Areas 1 (upper east) and 6 (Chelsea) to Areas 7 (Yorkvill/Lenox hill) and 2 (midtown);
during late night from 12am to 5am, the traffic is heavier than early night, mainly dominated by
pickups from Areas 4 (lower east) and 6 (Chelsea) and dropoffs in Areas 5 (Union square) and
2 (midtown). The late night dropoff to Union square is very plausible since people need to go to
transportation hub to go back home after a long night in New York city after midnight.

Again, this analysis is for demonstration of the tensor factor model only. More thorough and sophisticated analysis may be needed to fully understand the traffic pattern.




\vspace{0.1in}

\newpage

\bibliographystyle{apalike}
\bibliography{paper12}

\newpage

\noindent
{\bf \Large Appendix A: Proof of the theorems}

\vspace{0.1in}

We first prove Theorem \ref{th-2} as the analysis is simpler
and facilitates the proof of Theorem \ref{th-1}.

{\bf Proof of Theorem \ref{th-2}.}
It suffices to consider $k=1$ and $K=2$
as the TIPUP begins with mode-$k$ matrix unfolding in \eqref{TIPUP-k}.
We observe a matrix time series with
$\bX_t = \bA_1\bF_t\bA_2^\top + \bE_t\in \R^{d_1\times d_2}$ and
\bes
\bV^*_{1,h}
= \sum_{t=h+1}^T \frac{\bX_{t-h}\bX_t^\top}{T-h}\in \R^{d_1\times d_1},\quad
\bTheta^*_{1,h}
= \sum_{t=h+1}^T \frac{\bA_1\bF_{t-h}\bA_2^\top\bA_2\bF_t^\top\bA_1^\top}{T-h}\in \R^{d_1\times d_1}.
\ees
Let $\bDelta^*_1$, $\bDelta^*_2$ and $\bDelta^*_3$ be respectively the
three terms on the right-hand side below:
\bel{pf-1}
\bV^*_{1,h} - \bTheta^*_{1,h}
= \sum_{t=h+1}^T \frac{\bA_1\bF_{t-h}\bA_2^\top\bE_t^\top}{T-h}
 +  \sum_{t=h+1}^T \frac{\bE_{t-h}\bA_2\bF_t^\top \bA_1^\top}{T-h}
+ \sum_{t=h+1}^T \frac{\bE_{t-h}\bE_t^\top}{T-h}.
\eel

Let $\sigma_* = \sigma \big\|\bTheta^*_{1,0}\big\|_{\rm S}^{1/2}$
with the norm in \eqref{new-norm-2}. 
By Condition~A, 
for any $\bu$ and $\bv$ in $\R^{d_1}$ we have
\bes
\barE \bigg\{\bu^\top 
\bigg(\sum_{t=h+1}^T \frac{\bA_1\bF_{t-h}\bA_2^\top\bE_t^\top}{T^{1/2}}\bigg)
\bv\bigg\}^2
\le \frac{\sigma^2\|\bv\|_2^2}{T}
\big\|\sum_{t=h+1}^T \bA_2\bF_t^\top \bA_1^\top\bu\big\|_2^2
\le \sigma^2\big\|\bTheta^*_{1,0}\big\|_{\rm S}\|\bu\|_2^2\|\bv\|_2^2.
\ees
Thus for vectors $\bu_i$ and $\bv_i$ with $\|\bu_i\|_2=\|\bv_i\|_2=1$,
\bes
\lefteqn{\{(T-h)^2/(T\sigma_*^2)\}
\barE\big( \bu_1^\top \bDelta^*_1\bv_1 - \bu_2^\top \bDelta^*_1\bv_2\big)^2}
\cr &\le& \Big(\|\bu_1-\bu_2\|_2\|\bv_1\|_2 + \|\bu_2\|_2\|\bv_1 - \bv_2\|_2\Big)^2
\cr &\le& 2\Big(\|\bu_1-\bu_2\|_2^2 + \|\bv_1 - \bv_2\|_2^2\Big)
\cr &=& 2\ \E \big\{(\bu_1-\bu_2)^\top\bxi+ (\bv_1-\bv_2)^\top \bzeta\big\}^2
\ees
where $\bxi$ and $\bzeta$ are iid $N(0,\bI_{d_1})$ vectors. As $\bDelta^*_1$ is a
$d_1\times d_1$ Gaussian matrix under $\barE$,
the Sudakov-Fernique inequality yields
\bes
\Big(\frac{T-h}{T^{1/2}\sigma_*}\Big)\barE \big\|\bDelta^*_1\big\|_{\rm S}
\le \sqrt{2} \E\sup_{\|\bu\|_2=\|\bv\|_2=1}\Big|\bu^\top\bxi+ \bv^\top\bzeta\big|
= \sqrt{2}\E\Big(\|\bxi\|_2+\|\bzeta\|_2\Big).
\ees
As $\E \|\bxi\|_2 = \E \|\bzeta\|_2 \le \sqrt{d_1}$, it follows that
for the first term on the right-hand side of (\ref{pf-1})
\bel{pf-3}
\barE\big\|\bDelta^*_1\big\|_{\rm S}
\le \frac{\sigma_*(8Td_1)^{1/2}}{T-h}
= \frac{\sigma(8Td_1)^{1/2}}{T-h}\big\|\bTheta^*_{1,0}\big\|_{\rm S}^{1/2}.
\eel
Similarly, $\barE\big\|\bDelta^*_2\big\|_{\rm S}
\le \sigma(8Td_1)^{1/2}(T-h)^{-1}\big\|\bTheta^*_{1,0}\big\|_{\rm S}^{1/2}$
for the second term.

Let $h\le T/4$.
For the third term $\bDelta^*_3$ on the right-hand side of (\ref{pf-1}),
we split the sum into two terms
over the index sets, $S_1=\{(h,2h]\cup(3h,4h]\cup\ldots\}\cap (h,T]$ and its complement $S_2$ in $(h,T]$,
so that \chzM{$(\bE_{t-h}, t\in S_a)$ and $(\bE_t, t\in S_a)$ are two independent $d_1\times n_a$
centered Gaussian matrices for each $a=1,2$, with $n_1+n_2 = (T-h)d_2$.
Thus, by Lemma \ref{lm-GH} (i),
\bel{pf-2}
\frac{\barE \big\|\bDelta^*_3\big\|_{\rm S}}{\sigma^2}
\le \sum_{a=1}^2 \barE\bigg\|\sum_{t\in S_a}\frac{\bE_{t-h}\bE_t^\top}{\sigma^2(T-h)}\bigg\|_{\rm S}
\le \sum_{a=1}^2\frac{2\sqrt{d_1n_a}+d_1}{T-h}
\le \frac{\sqrt{8d_1d_2}}{\sqrt{T-h}}+\frac{2d_1}{T-h}.
\eel
We} obtain the first inequality in (\ref{th-2-1})
by applying (\ref{pf-3}) and \eqref{pf-2} to (\ref{pf-1}),
and the second by Cauchy-Schwarz,
${\|(\bDelta_1,\ldots,\bDelta_{h_0})\|_{\rm S}^2} \le h_0\sum_{h=1}^{h_0} \|\bDelta_h\|_{\rm S}^2$.
Finally, 
\chzM{by Wedin (1972),
\bes
\barE\Big\|\widehat{\bP_{1, r_1}} - \bP_1\Big\|_{\rm S}
&\le & \frac{2}{\big(\lam^*_1\big)^{2}}
\bigg\{\frac{2\sigma(8Td_1)^{1/2}}{T-h_0}\big\|\bTheta^*_{1,0}\big\|_{\rm S}^{1/2}
+ \frac{\sigma^2\sqrt{8d_1d_2}}{\sqrt{T-h_0}}+\frac{2\sigma^2d_1}{T-h_0}\bigg\}
\ees
As $\sqrt{T}\big\|\bTheta^*_{1,0}\big\|_{\rm S}^{1/2}\ge \sqrt{T-h}\lam^*_1$ by \eqref{lam*_k-relation}
and $\barE\big\|\widehat{\bP_{1, r_1}} - \bP_1\big\|_{\rm S}\le 1$,
\eqref{th-2-3} holds automatically when
$4\sigma\sqrt{8d_1}\ge \lam^*_1\sqrt{T-h_0}$,
whereas \eqref{th-2-3} follows from
\bes
\frac{2\sigma^2d_1}{T-h_0} \le \frac{\lam^*_1 \sigma \sqrt{d_1}}{2\sqrt{8(T-h_0)}}
\le \frac{\sigma \sqrt{8Td_1}}{16(T-h_0)}\big\|\bTheta^*_{1,0}\big\|_{\rm S}^{1/2}
\ees
otherwise. Thus, \eqref{th-2-3} holds either ways. }
$\hfill\square$

\chzM{\begin{lemma}\label{lm-GH}
(i) Let $G\in \R^{d_1\times n}$ and $H\in \R^{d_2\times n}$ be two centered independent
Gaussian matrices such that $\E(u^\top \text{vec}(G))^2 \le \sigma^2\ \forall\ u\in \R^{d_1n}$ and
$\E(v^\top \text{vec}(H))^2\le \sigma^2\ \forall\ v\in \R^{d_2n}$. Then,
\bes\label{lm-GH-1}
\E\big[\|GH^\top\|_{\rm S}\big] \le \sigma^2\big(\sqrt{d_1d_2}+\sqrt{d_1n} + \sqrt{d_2n}\big).
\ees
(ii) Let $G_i\in \R^{d_1\times d_2}, H_i\in \R^{d_3\times d_4}, i=1,\ldots, n$,
be independent centered Gaussian matrices
such that $\E(u^\top \text{vec}(G_i))^2 \le \sigma^2\ \forall\ u\in \R^{d_1d_2}$ and
$\E(v^\top \text{vec}(H_i))^2\le \sigma^2\ \forall\ v\in \R^{d_3d_4}$. Then,
\bes
\E\bigg[\bigg\|\mat_1\bigg(\sum_{i=1}^n G_i\otimes H_i\bigg)\bigg\|_{\rm S}\bigg]
\le \sigma^2\big(\sqrt{d_1n}+\sqrt{d_1d_3d_4} + \sqrt{nd_2d_3d_4}\big).
\ees
\end{lemma}

{\bf Proof.} Assume $\sigma=1$ without loss of generality. \\
(i) Independent of $G$ and $H$, let $\xi_j\in\R^{d_j}$ and $\zeta_j\in\R^{n}$, $j=1,2$,
be independent standard Gaussian vectors.
For $\|v_1\|_2=\|v_2\|_2=1$ and $\|w_1\|_2\vee\|w_2\|_2\le 1$,
$\E(v_1^\top H w_1 - v_2^\top H w_2)^2
\le \E((v_1-v_2)^\top\xi_2+(w_1-w_2)^\top\zeta_2)^2$.
Thus, by the Sudakov-Fernique inequality
\bes
\E\Big[\|GH^\top\|_{\rm S}\Big|G\Big]
&\le& \|G\|_{\rm S}\E\Big[\max_{\|u\|_2=\|v\|_2=1}(\|G\|_{\rm S}^{-1}u^\top G)H^\top v\Big|G\Big]
\cr &\le& \|G\|_{\rm S}\E\Big[\max_{\|u\|_2=\|v\|_2=1}
(\|G\|_{\rm S}^{-1}u^\top G \zeta_2 + \xi_2^\top v)\Big|G\Big]
\cr &=& \E\Big[\max_{\|u\|_2=1}u^\top G \zeta_2\Big|G\Big] + \|G\|_{\rm S}\sqrt{d_2}.
\ees
Applying the same argument to $G$, we have
\bes
\E\Big[\|GH^\top\|_{\rm S}\Big|G\Big]
&\le & \E\Big[\max_{\|u\|_2=1}u^\top G \zeta_2 + \|G\|_{\rm S}\sqrt{d_2}\Big]
\cr &\le & \E\Big[\|\zeta_2\|_2\max_{\|u\|_2=1}(u^\top\xi_1+\zeta_1^\top\zeta_2/\|\zeta_2\|_2)
+ (\sqrt{d_1}+\sqrt{n})\sqrt{d_2}\Big]
\cr & \le & \sqrt{d_1n} + (\sqrt{d_1}+\sqrt{n})\sqrt{d_2}.
\ees
(ii) We treat $(G_1,\ldots,G_n)\in\R^{d_1\times d_2\times n}$ and
$(H_1,\ldots,H_n)\in\R^{d_3\times d_4\times n}$ as tensors.
Let $\xi_i\in \R^{d_2}$ be additional independent standard Gaussian vectors.
For $u\in \R^{d_1}$ and $V\in \R^{d_2\times (d_3d_4)}$,
\bes
&& \E\bigg[\bigg\|\mat_1\bigg(\sum_{i=1}^n G_i\otimes H_i\bigg)\bigg\|_{\rm S}\bigg]
\cr & = & \E\bigg[\sup_{\|u\|_2=1,\|V\|_F=1} u^\top\mat_1(G_1,\ldots,G_n)
\vec\big(\mat_3(H_1,\ldots,H_n)V^\top\big) \bigg]
\cr & \le & \sqrt{d_1}\E\bigg[\sup_{\|V\|_F=1} \|\mat_3(H_1,\ldots,H_n)V^\top\|_{\rm F} \bigg]
\cr &  & + \E\bigg[\sup_{\|V\|_F=1} (\vec(\xi_1,\ldots,\xi_n))^\top
\vec\big(\mat_3(H_1,\ldots,H_n)V^\top\big) \bigg]
\cr & = & \sqrt{d_1}\E\bigg[\|\mat_3(H_1,\ldots,H_n)\|_{\rm S} \bigg]
 + \E\bigg[\bigg(\sum_{j=1}^{d_2}\sum_{k=1}^{d_3d_4}
\bigg(\sum_{i=1}^n \xi_{i,j}\vec(H_i)_k\bigg)^2\bigg)^{1/2} \bigg]
\cr & \le & \sqrt{d_1}\big(\sqrt{n}+\sqrt{d_3d_4}\big) + \sqrt{nd_2d_3d_4}.
\ees
Again, we apply the Sudakov-Fernique inequality twice above.
$\hfill\square$
}

{\bf Proof of Theorem \ref{th-1}.}
It suffices to consider $k=1$ and $K=2$
as the TOPUP begins with mode-$k$ matrix unfolding in \eqref{TOPUP-k}.
In this case, $\bX_t = \bM_t + \bE_t\in \R^{d_1\times d_2}$
with $\bM_t = \bA_1\bF_t\bA_2^\top$,
\bes
\bV_{1,h} = \sum_{t=h+1}^T \frac{\bX_{t-h}\otimes \bX_t}{T-h},\quad
\bTheta_{1,h} = \sum_{t=h+1}^T \frac{\bM_{t-h}\otimes \bM_t}{T-h}.
\ees
Let $\bDelta_1$, $\bDelta_2$ and $\bDelta_3$ be respectively the
three terms on the right-hand side below:
\bel{new-pf-1}
\lefteqn{\mat_1\big(\bV_{1,h}\big) - \mat_1\big(\bTheta_{1,h}\big)} \nonumber
\\ &=& \sum_{t=h+1}^T \frac{\mat_1(\bM_{t-h}\otimes \bE_t)}{T-h}
 +  \sum_{t=h+1}^T \frac{\mat_1(\bE_{t-h}\otimes \bM_t)}{T-h}
+ \sum_{t=h+1}^T \frac{\mat_1(\bE_{t-h}\otimes \bE_t)}{T-h}.
\eel

For the first term $\bDelta_1$, we notice that $\bM_{t-h}=\bM_{t-h}\bP_2$ for
a fixed orthogonal projection of rank $r_2$.
Let $\bU_2\in \R^{d_2\times r_2}$
with orthonormal columns and $\bU_2\bU_2^\top = \bP_2$.
For $\cV\in \R^{d_2\times d_1\times d_2}$,
$\mat_1(\bM_{t-h}\otimes \bE_t) \vec(\cV)
= \mat_1((\bM_{t-h}\bU_2)\otimes \bE_t) \vec(\cW)$
with $\cW = \cV\times_1\bU_2^\top\in  \R^{r_2\times d_1\times d_2}$
satisfying $\|\vec(\cW)\|_2=\|\vec(\cV)\|_2$, so that
$\|\bDelta_1\|_{\rm S} = \|\bbDelta_1\|_{\rm S}$ with
\bes
\bbDelta_1 = \sum_{t=h+1}^T \frac{\mat_1((\bM_{t-h}\bU_2)\otimes \bE_t)}{T-h}.
\ees
By Condition~A, for any $\bu\in \R^{d_1}$ and $\cW\in \R^{r_2\times d_1\times d_2}$
\bes
\lefteqn{\barE\bigg(T^{-1/2}\sum_t\bu^T \mat_1((\bM_{t-h}\bU_2)\otimes \bE_t) \vec(\cW)\bigg)^2}
\cr &=&  T^{-1}\sum_t \barE \bigg(\sum_{i_1,j_1,i_2, j_2}
 u_{i_1}(\bM_{t-h}\bU_2)_{i_1,j_1}\big(\bE_t\big)_{i_2,j_2}w_{j_1,i_2,j_2}\bigg)^2
\cr &\le& T^{-1}\sum_t \sigma^2\sum_{i_2, j_2}\bigg(\sum_{i_1,j_1}
u_{i_1}(\bM_{t-h}\bU_2)_{i_1,j_1}w_{j_1,i_2,j_2}\bigg)^2
\cr &\le&\sigma^2T^{-1} \sum_t \sum_{i_2, j_2}\sum_{j_1}\bigg(\sum_{i_1}
u_{i_1}(\bM_{t-h}\bU_2)_{i_1,j_1}\bigg)^2 \sum_{j_1}w_{j_1,i_2,j_2}^2
\cr &=& \sigma^2 T^{-1}\sum_t \|\bM_{t-h}^\top\bu\|_2^2 \big\|\vec(\cW)\big\|_2^2.
\cr &\le & \sigma^2 \big\|\bTheta^*_{1,0}\big\|_{\rm S}\|\bu\|_2^2 \big\|\vec(\cW)\big\|_2^2.
\ees
As in the derivation of \eqref{pf-3}, it follows that for $\|\bu_i\|_2=\|\vec(\cW_i)\|_2=1$
\bes
\lefteqn{\{(T-h)^2/(T\sigma_*^2)\}
\barE \big\{\bu_1^\top\bbDelta_1\vec(\cW_1) - \bu_2^\top \bbDelta_1\vec(\cW_2)\big\}^2}
\cr &\le& 2\big\{\|\bu_1-\bu_2\|_2^2 + \big\|\vec(\cW_1-\cW_2)\big\|_2^2\big\}.
\ees
As $\bbDelta_1$ is a Gaussian matrix under $\barE$, the Sudakov-Fernique inequality yields
\bel{new-pf-2}
\barE \big\|\bDelta_1\big\|_{\rm S} = \barE \big\|\bbDelta_1\big\|_{\rm S}
\le \frac{\sigma(2T)^{1/2}(\sqrt{d_1}+\sqrt{r_2d_1d_2})}{T-h}\big\|\bTheta^*_{1,0}\big\|_{\rm S}^{1/2}.
\eel

For the second term $\bDelta_2$,
$\mat_1(\bE_{t-h}\otimes \bM_t) \vec(\cV)
= \mat_1(\bE_{t-h}\otimes (\bU_1^\top \bM_t\bU_2)) \vec(\cW)$
with $\bU_j\bU_j^\top=\bP_j$ and
$\cW = \cV\times_2\bU_1^\top\times_3 \bU_2^\top\in  \R^{d_2\times r_1\times r_2}$, so that
$\|\bDelta_2\|_{\rm S} = \|\bbDelta_2\|_{\rm S}$ with
\bes
\bbDelta_2 = \sum_{t=h+1}^T \frac{\mat_1(\bE_{t-h}\otimes (\bU_1^\top\bM_t\bU_2))}{T-h}
\in \R^{d_1\times d_2r_1r_2}.
\ees
Moreover, for $\bu\in\R^{d_1}$ and $\cW\in \R^{d_2\times r_1\times r_2}$,
\bes
\lefteqn{\barE\bigg(T^{-1/2}\sum_t \bu^T \mat_1(\bE_{t-h}\otimes (\bU_1^\top\bM_t\bU_2))\vec(\cW)\bigg)^2}
\cr &=&  T^{-1}\sum_t \barE \bigg(\sum_{i_1,j_1,i_2, j_2}
 u_{i_1}(\bE_{t-h})_{i_1.j_2}(\bU_1^\top\bM_t\bU_2)_{i_2.j_2}w_{j_1,i_2,j_2}\bigg)^2
\cr &\le&\sigma^2T^{-1}\sum_t \sum_{i_1, j_1}
\bigg(\sum_{i_2, j_2} u_{i_1}(\bU_1^\top\bM_t\bU_2)_{i_2.j_2}w_{j_1,i_2,j_2}\bigg)^2
\cr & = &\sigma^2\|\bu\|_2^2 \sum_{j_1} T^{-1} \sum_t
{\rm trace}\Big((\bU_1^\top\bM_t\bU_2)\cW^{(2,3)}_{j_1}\Big)^2
\cr & \le &\sigma^2\|\bu\|_2^2 \sum_{j_1} \big\|\bTheta_{1,0}\big\|_{\rm op}
\big\|\cW^{(2,3)}_{j_1}\big\|_{\rm F}^2
\cr & = &\sigma^2\big\|\bTheta_{1,0}\big\|_{\rm op}\|\bu\|_2^2 \big\|\vec(\cW)\big\|_2^2.
\ees
Thus, as $\cW$ is of dimension $d_2\times r_1\times r_2$, the derivation of \eqref{new-pf-2} yields
\bel{new-pf-3}
\barE \big\|\bDelta_2\big\|_{\rm S} = \barE \big\|\bbDelta_2\big\|_{\rm S}
\le \frac{\sigma(2T)^{1/2}(\sqrt{d_1}+\sqrt{d_2r_1r_2})}{T-h}\big\|\bTheta_{1,0}\big\|_{\rm op}^{1/2}.
\eel

For the third term $\bDelta_3$, we partition $(h,T]$ as $S_1\cup S_2$ as
in the derivation of \eqref{pf-2},
\chzM{so that by Lemma \ref{lm-GH} (ii)
\bel{new-pf-4}
\barE \big\|\bDelta_3\big\|_{\rm S}
&\le& \sum_{a=1}^2 \E \bigg\|\mat_1\bigg(\sum_{t\in S_a}\frac{\bE_{t-h}\otimes \bE_t}
{T-h}\bigg)\bigg\|_{\rm S}
\cr &\le& \sum_{a=1}^2\frac{\sigma^2\big(\sqrt{d_1|S_a|}+d_1\sqrt{d_2} + d_2\sqrt{|S_a|d_1}\big)}{T-h}
\cr &\le& \frac{\sigma^2(1+d_2)\sqrt{2d_1}}{\sqrt{T-h}} + \frac{2\sigma^2d_1\sqrt{d_2}}{T-h}.
\eel
We} obtain (\ref{th-1-1})
by applying \eqref{new-pf-2}, \eqref{new-pf-3} and \eqref{new-pf-4} to the three terms in (\ref{new-pf-1})
and Cauchy-Schwarz.
Finally \eqref{th-1-3} follows from \eqref{th-1-1} via Wedin (1972). $\hfill\square$

\newpage

\noindent
{\bf \Large Appendix B: Import-Export network example}

\vspace{0.1in}

In this appendix we provide the detailed data description of the import
export data used in the example, as well as some additional figures.

The data is obtained from UN Comtrade Database at
{\tt https://comtrade.un.org}. In this study we
use the
monthly observations of 22 large economies in North America and Europe
from January 2010 to December 2016. The countries used are
Belgium (BE), Bulgaria (BU), Canada (CA), Denmark (DK), Finland (FI),
France (FR),
Germany (DE), Greece (GR), Hungary (HU), Iceland (IS), Ireland (IR), Italy (IT),
Mexico (MX), Norway (NO), Poland (PO), Portugal (PT), Spain (ES), Sweden (SE),
Switzerland (CH), Turkey (TR), United States (US) and United Kingdom (UK).

The trade data includes commodity classifier (2 digit Hamonized System codes).
Following the classification shown at
{\tt https://www.foreign-trade.com/reference/hscode.htm}, we divide all
products into 15 categories, including Animal \& Animal Products
(HS code 01-05),
Vegetable Products (06-15),
Foodstuffs (16-24),
Mineral Products (25-27),
Chemicals \& Allied Industries (28-38),
Plastics \& Rubbers (39-40),
Raw Hides, Skins, Leather \& Furs (41-43),
Wood \& Wood Products (44-49),
Textiles (50-63),
Footwear \& Headgear (64-67),
Stone \& Glass (68-71),
Metals (72-83),
Machinery \& Electrical (84-85),
Transportation (86-89), and
Miscellaneous (90-97).

The following two figures are the network figures for condensed product
groups 3 to 6.

\begin{figure}
\centering
\includegraphics[width=3.0in,height=3.0in,keepaspectratio=true]{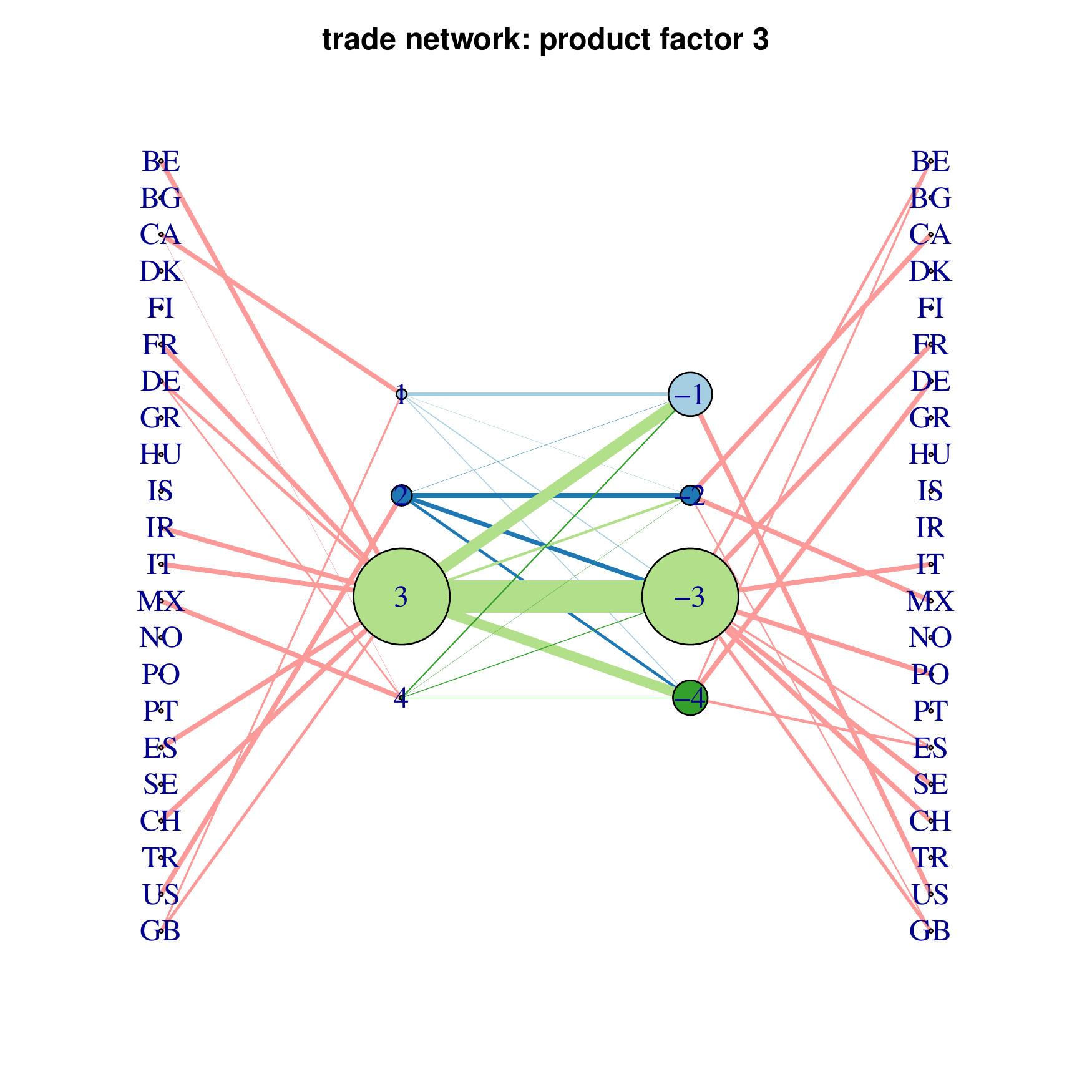}
\includegraphics[width=3.0in,height=3.0in,keepaspectratio=true]{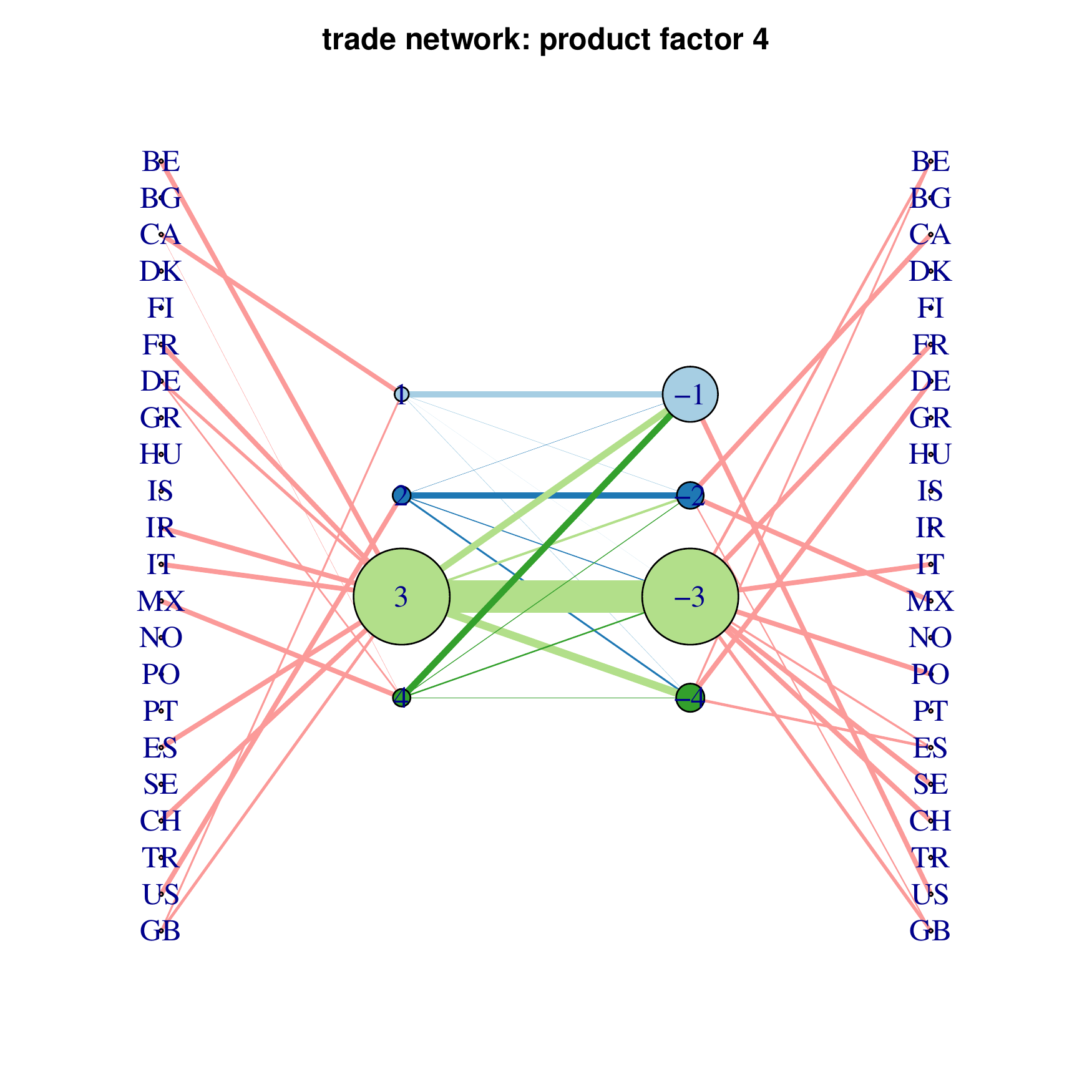}
\caption{Trade network for condensed product group 3 (left) and group 4 (right).
    Export and import hubs on the left and right of the center network. Line
  width is proportional to total volume of trade between the hubs
  for the last three years (2015 to 2017). Vertex size is proportional to total
  volume of trades through the vertex. The line width between the countries
  and the hubs is proportional to the corresponding loading coefficients, for
  coefficients larger than 0.05 only.}
\label{fig.network.3}
\end{figure}

\begin{figure}
\centering
\includegraphics[width=3.0in,height=3.0in]{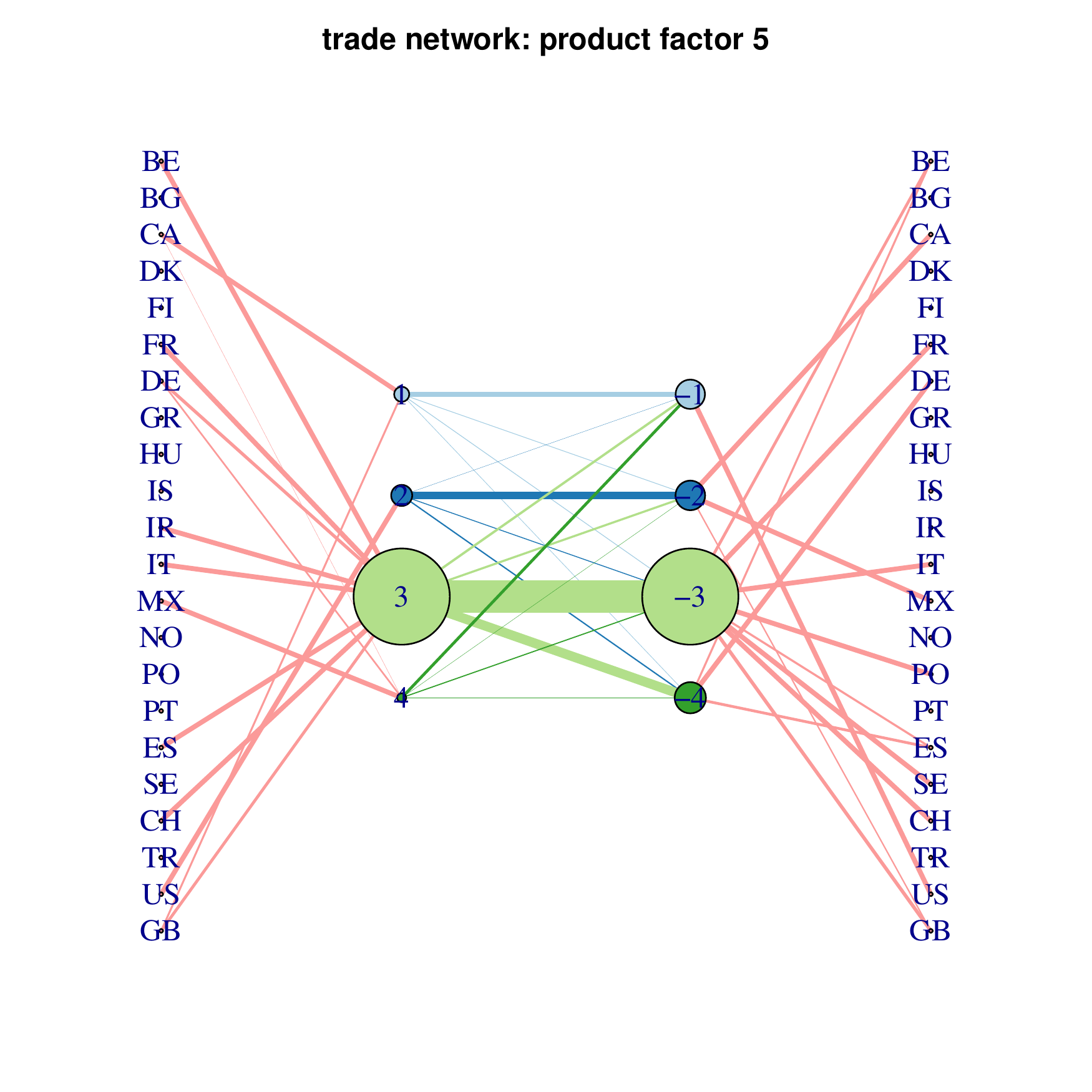}
\includegraphics[width=3.0in,height=3.0in]{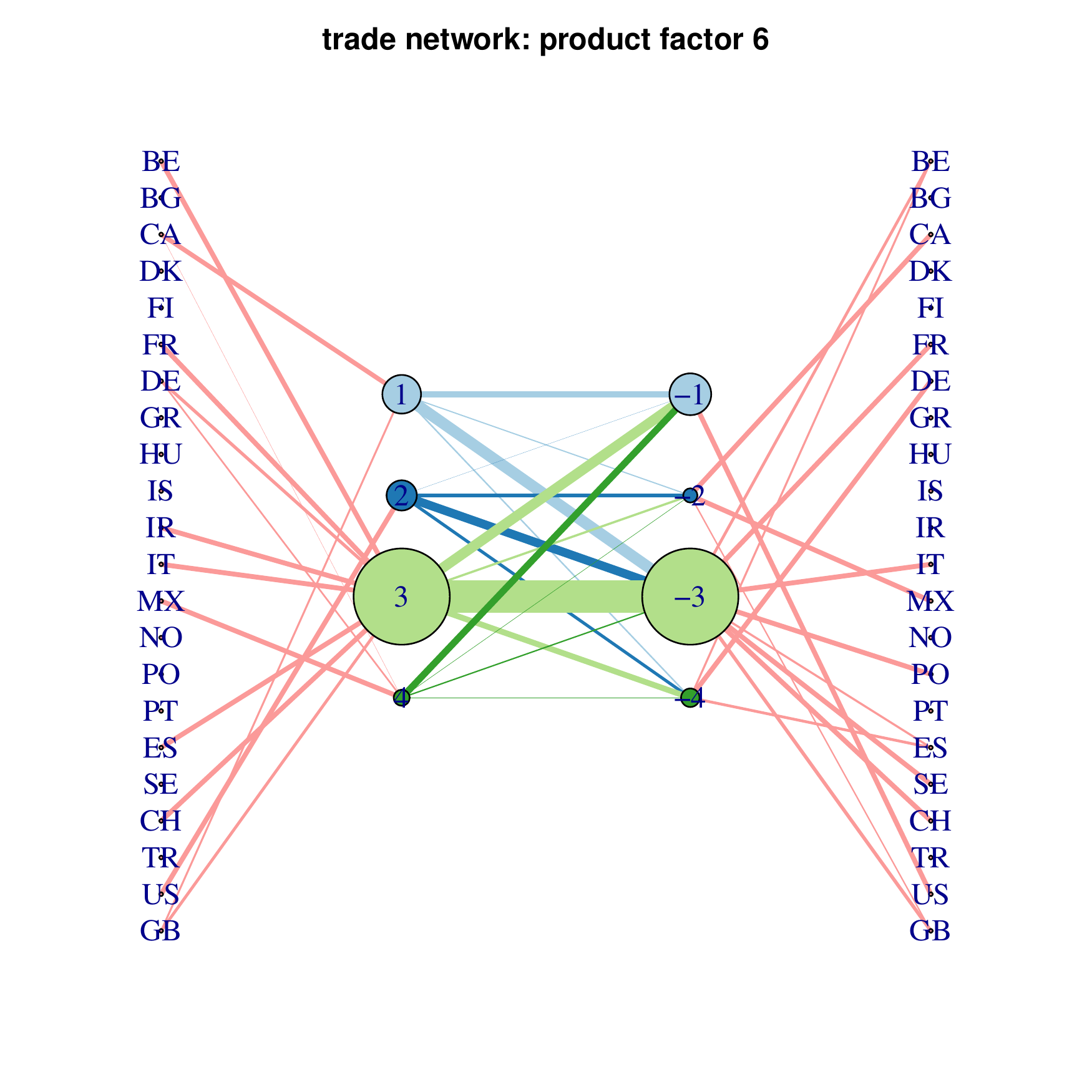}
\caption{Trade network for condensed product group 5 (left) and group 6 (right).
    Export and import hubs on the left and right of the center network. Line
  width is proportional to total volume of trade between the hubs
  for the last three years (2015 to 2017). Vertex size is proportional to total
  volume of trades through the vertex. The line width between the countries
  and the hubs is proportional to the corresponding loading coefficients, for
  coefficients larger than 0.05 only.}
\label{fig.network.5}
\end{figure}

\end{document}